\documentclass[iop,apj,numberedappendix,onecolumn]{my_emulateapj}
\usepackage{amsmath}
\usepackage{bm}
\usepackage{natbib}
\usepackage{xcolor}
\usepackage[normalem]{ulem} 
\usepackage{calligra} 
\usepackage[version=3]{mhchem}
\usepackage[colorlinks=true,linkcolor=blue,citecolor=blue,%
urlcolor=blue,linktocpage=true]{hyperref}
\usepackage[capitalize]{cleveref}

\makeatletter
\renewcommand\tableofcontents{\@starttoc{toc}}
\makeatother

\DeclareMathAlphabet{\mathcalligra}{T1}{calligra}{m}{n}
\DeclareFontShape{T1}{calligra}{m}{n}{<->s*[2.2]callig15}{}

\newcommand{\newacronym}[3]{\newcommand{#1}[1]{#3##1 (#2##1)%
\renewcommand{#1}[1]{#2####1}}}

\newacronym{\NSE}{NSE}{nuclear statistical equilibrium}
\newacronym{\ODE}{ODE}{ordinary differential equation}
\newacronym{\NR}{NR}{Newton--Raphson}
\newacronym{\EOS}{EOS}{equation of state}

\newcommand{\skynet}{\emph{SkyNet}}
\newcommand{\winnet}{\emph{WinNet}}
\newcommand{\xnet}{\emph{XNet}}
\newcommand{\skyneturl}{\url{https://bitbucket.org/jlippuner/skynet}}
\newcommand{\code}[1]{\texttt{#1}}

\crefformat{footnote}{#2\footnotemark[#1]#3}
\crefname{equation}{Equation}{Equations}
\crefname{figure}{Figure}{Figures}

\newcommand{\sref}[1]{\S\ref{#1}}
\newcommand{\aref}[1]{Appendix~\ref{#1}}
\newcommand{\crefp}[1]{(\namecref{#1}~\ref{#1})}
\newcommand{\crefalt}[1]{\namecref{#1}~\ref{#1}}

\renewcommand{\vec}{\bm}
\newcommand{\smce}[1]{\smash{\ce{#1}}}
\newcommand{\scriptr}{\mathcalligra{r}}
\newcommand{\nn}{\nonumber}

\newcommand{\kb}{\smash{k_B\,\text{baryon}^{-1}}}

\newlength{\singlecol}
\shorttitle{\skynet}
\shortauthors{Jonas Lippuner and Luke F.\ Roberts}

\hypersetup{pdftitle={SkyNet: A modular nuclear reaction network library}}
\hypersetup{pdfauthor={Jonas Lippuner and Luke F. Roberts}}

\begin{document}

\title{\textbf{\skynet: A modular nuclear reaction network
library\vspace{1.1pc}}}

\author{Jonas Lippuner}
\affil{TAPIR, Walter Burke Institute for Theoretical Physics, California
Institute of Technology, Pasadena, CA 91125, USA}
\affil{CCS-2, Los Alamos National Laboratory, P.O.\ Box 1663, Los Alamos,
NM 87545, USA;
\href{mailto:jlippuner@lanl.gov}{jlippuner@lanl.gov}}
\affil{Center for Nonlinear Studies, Los Alamos National Laboratory,
P.O.\ Box 1663, Los Alamos, NM 87545, USA}
\affil{Center for Theoretical Astrophysics, Los Alamos National Laboratory,
P.O.\ Box 1663, Los Alamos, NM, 87545, USA}
\author{Luke F.\ Roberts}
\affil{National Superconducting Cyclotron Laboratory, Michigan State
University, East Lansing, MI 48824, USA}
\affil{Department of Physics and Astronomy, Michigan State University, East
Lansing, MI 48824, USA}

\begin{abstract}
Almost all of the elements heavier than hydrogen that are present in our solar
system were produced by nuclear burning processes either in the early universe
or at some point in the life cycle of stars. In all of these environments, there
are dozens to thousands of nuclear species that interact with each other to
produce successively heavier elements. In this paper, we present \skynet, a new
general-purpose nuclear reaction network that evolves the abundances of nuclear
species under the influence of nuclear reactions.  \skynet\ can be used to
compute the nucleosynthesis evolution in all astrophysical scenarios where
nucleosynthesis occurs. \skynet\ is free and open source and aims to be easy to
use and flexible. Any list of isotopes can be evolved, and \skynet\ supports
different types of nuclear reactions.  \skynet\ is modular so that new
or existing physics, like nuclear reactions or equations of state, can easily be
added or modified. Here, we present in detail the physics implemented in
\skynet\ with a focus on a self-consistent transition to and from nuclear
statistical equilibrium to non-equilibrium nuclear burning, our
implementation of electron screening, and coupling of the network to an equation
of state. We also present comprehensive code tests and comparisons with existing
nuclear reaction networks. We find that \skynet\ agrees with published results
and other codes to an accuracy of a few percent.  Discrepancies, where they
exist, can be traced to differences in the physics implementations.
\end{abstract}


\section*{Contents}
\twocolumngrid
\tableofcontents
\vspace{2\baselineskip}
\onecolumngrid

\section{Introduction}

Nuclear and weak reactions play a crucial role in many astrophysical scenarios.
Nuclear reactions typically occur at high
temperatures and densities, because a large amount of energy is required to
overcome the Coulomb repulsion between positively charged nuclei. Inside
the cores of main-sequence stars, nuclear fusion converts hydrogen into helium, which
releases nuclear binding energy as heat, keeping the star from collapsing
\citep{bethe:39}. When massive stars undergo core collapse at the end of their
lives, nuclear and weak reactions serve as important energy sinks and sources (neutrino
cooling and heating). In core-collapse and Type Ia supernovae (SNe Ia), explosive nuclear
burning mainly creates iron-group elements that are ejected into the
interstellar medium \citep{nomoto:97, woosley:02}. According to our current
understanding, a weak version of the rapid neutron capture process (r-process)
can also occur
\citep{wanajo:13} in core-collapse supernovae. However, it appears that the
full r-process that can synthesize all of the heavy elements predominantly happens
during neutron star mergers \citep{freiburghaus:99}. Heavy
elements up to bismuth can also be created in stars via the slow neutron
capture process \citep[s-process;][]{burbidge:57}. Finally, when hydrogen and
helium gas accretes onto a white dwarf, the accumulated material can undergo a
thermonuclear explosion that creates a short-lived bright flash of light called
a nova. If the accretor is a neutron star instead, the thermonuclear
explosion results in an X-ray burst \citep[ch.~6]{jose:16} or a superburst
\citep{strohmayer:02}.

To adequately understand these astrophysical objects and phenomena, one needs
to account for the nuclear reactions that drive them. In some cases, reaction
networks are mainly used to track the nuclear energy generation
\citep[e.g.,][]{weaver:78,mueller:86,timmes:00b}. But in most
cases, the evolution of the entire composition due to nuclear reactions is
of interest. Because of the many ways that nuclei can react with each other to
form other nuclides, the number of nuclear species that are relevant for many
astrophysical processes ranges from dozens to thousands, and the number of
nuclear reactions involved is hundreds to tens of thousands. For this reason, a
large number of variables (i.e., nuclide abundances) that are
coupled together by non-linear nuclear reaction rates (see \sref{sec:rates}) need to be evolved.
Mathematically and computationally, the ensemble of coupled nuclear reactions
is described by a \emph{nuclear reaction network}.

Large-scale (several dozen species or more) nuclear reaction networks were
first developed in the late 1960s and early 1970s \citep[e.g.,][]{truran:66b, truran:67, arnett:69,
woosley:73}. These first networks were mainly for explosive nuclear
burning in massive stellar evolution and supernovae, although earlier stellar evolution models also took nuclear
reactions into account and evolved a handful of nuclear species
\citep[e.g.,][]{hayashi:62, hofmeister:64, iben:67}.  Early networks
ranged in size from a few dozen to around a hundred species, with up to a few
hundred reactions connecting the nuclei. Since then, nuclear reaction networks of different sizes have
been used to study various astrophysical scenarios. Big bang nucleosynthesis
calculations require the smallest networks with typically fewer than a dozen
nuclear species, although some authors utilize much bigger networks of up to
several dozen species \citep[e.g.,][]{wagoner:73, nollett:00, orlov:00, coc:12,
cyburt:16}. Networks with dozens to hundreds of species are employed in
stellar evolution codes \citep[e.g.,][]{arnett:77, weaver:78, meader:00, paxton:11,
bressan:12, jones:15}. Similar-sized or larger (up to hundreds of species)
networks are also used to compute explosive nucleosynthesis in SNe Ia
\citep[e.g.,][]{thielemann:86, iwamoto:99, hillebrandt:13,
seitenzahl:13, leung:15}, core-collapse supernovae
\citep[e.g.,][]{thielemann:96, limongi:03, heger:10, harris:14},
novae \citep[e.g.,][]{weiss:90, jose:98, iliadis:02, starrfield:16}, and X-ray
bursts \citep[e.g.,][]{schatz:01, woosley:04, cyburt:10, parikh:13}.

The largest nuclear networks are needed to simulate neutron capture processes. For
the s-process in massive stars, it may be sufficient to use a few hundred to
about a
thousand nuclei \citep[e.g.,][]{prantzos:99, kaeppeler:11, nishimura:17}.
Larger nuclear reaction networks (typically thousands of isotopes) have been
used for r-process nucleosynthesis calculations in neutrino-driven winds from
core-collapse supernovae \citep[e.g.,][]{woosley:92, hoffman:97,
freiburghaus:99b, arcones:10, farouqi:10, roberts:10, wanajo:13},
in the jets of magnetorotational core-collapse supernovae
\citep[e.g.,][]{winteler:12, nishimura:15}, in the dynamical ejecta of neutron
star mergers \citep[e.g.,][]{freiburghaus:99, goriely:11, bauswein:13,
wanajo:14, eichler:15, just:15,fernandez:17a},
in accretion disk ejecta following neutron star mergers
\citep[e.g.,][]{surman:08, perego:14, martin:15, lippuner:17a}, and in broader
astrophysical contexts \citep[e.g.,][]{panov:95, panov:01, blinnikov:96,
mumpower:12}.

To evolve a nuclear reaction network, the rates of all reactions in the network
are required. Most reaction rates, e.g.\ interactions between two or more
nuclides, depend strongly on the energies of the incoming particles, due to
Coulomb barrier penetration, resonances in the compound nuclear system, and
other effects \citep[e.g.,][\S4]{clayton:68}. The rates of reactions only
involving a single particle in the entrance channel, like $\beta$-decays and
spontaneous fission, are constant.\footnote{Strictly speaking, $\beta$-decay
rates are only constant in vacuum. In the medium, electron phase-space blocking
introduces a dependency on the electron chemical potential and temperature
\citep[e.g.,][\S11.3]{arnett:96}.}
Some reaction rates involving nuclides sufficiently close to the valley of
stability can be measured experimentally as a function of energy, although in
many cases astrophysical reactions occur at energies that are much lower than
the experimentally accessible energy ranges \citep[e.g.,][\S4]{rolfs:88}.
Furthermore, most astrophysical processes involve unstable nuclei that may be
very far away from stability and are not experimentally accessible for rate
measurements. Therefore, theoretical models are necessary to compute the reaction
rates needed by the reaction network \citep[e.g.,][]{cyburt:10}. The
Hauser--Feshbach approach, which assumes that the reactants form a single
compound nucleus that subsequently decays into the reaction products, has been
used extensively to compute nuclear reaction rates for astrophysics applications
\citep[e.g.,][]{hauser:52, holmes:76, woosley:78, thielemann:86b, rauscher:00,
goriely:08}.

Nuclear reaction networks also require the properties, such as masses and internal
nuclear
partition functions \citep[e.g.,][]{arcones:11, brett:12, mendoza:15,
mumpower:16}, of all nuclides evolved in the
network. These properties are needed to compute \NSE{} and inverse reaction
rates (see \sref{sec:inverse_rates}), as well as $\beta$-decay rates. Some of
these nuclear properties, such as the masses, also enter the theoretical
reaction rate calculations. Since many of the nuclides of interest are
extremely unstable, special radioactive ion beam facilities are needed to
produce and measure these exotic nuclei \citep[see, e.g.,][and references
therein]{lunney:03, schatz:13, mumpower:16}. Current radioactive beam
facilities have made great
progress in measuring unstable nuclei and new facilities or upgrades to current
facilities are being built and planned. These new facilities will extend the
reach to more
exotic nuclei that are highly relevant to nuclear astrophysics scenarios
\citep[e.g.,][]{schatz:13, schatz:16, mumpower:16}. For the foreseeable future,
however, it is necessary to use theoretical models to compute nuclear masses
and $\beta$-decay properties for a large fraction of the nuclear species present
in r-process networks \citep[e.g.,][and references therein]{lunney:03,
moller:03, mumpower:16}.

Many authors who use nuclear reaction networks do not make
the code of these networks publicly available. This makes it hard to reproduce
and verify published results and also presents a barrier to new researchers
joining the field since they first have to write their own nuclear
reaction network. Notable exceptions of nuclear reaction networks that are
publicly available are the various networks by
\citet{timmes:99b},\footnote{%
\url{http://cococubed.asu.edu/code_pages/burn.shtml}} \xnet\ by
\citet{hix:99},\footnote{\url{http://eagle.phys.utk.edu/xnet/trac}}
and \emph{NucNet} by
\citet{meyer:07}.\footnote{\url{https://sourceforge.net/projects/nucnet-tools}}
In this paper, we present a new nuclear reaction network called \skynet\ that
is publicly available as an open-source software at \skyneturl\
\citep{skynet_ascl}. This paper is based on version 1.0 of \skynet\
\citep{skynet_v1.0}.

\skynet\ was initially designed for evolving large reaction networks for
r-process nucleosynthesis calculations, but thanks to its modularity and
flexibility, \skynet\ can easily be used for nucleosynthesis computations
in many other astrophysical situations.  Besides correctness, the main design
goals behind \skynet\ are usability and flexibility, making
\skynet\ an easy to use and versatile nuclear reaction network that is
available for anyone to use. \skynet\ can evolve an arbitrary set of nuclear
species under various different types of nuclear reactions
(\sref{sec:reac_types}). \skynet\ can also compute \NSE{} compositions
(\sref{sec:inverse_rates}) and switch between evolving \NSE{} and the full
network in an automated and self-consistent way (\sref{sec:NSE_evolution}).
\skynet\ contains electron screening corrections (\sref{sec:screening}) and an
\EOS{} that takes the entire composition into account (\sref{sec:eos}). For
ease of use, \skynet\ can be used from within Python (\sref{sec:impl}), and
\skynet\ can make movies of the nucleosynthesis evolution (see examples
at \url{http://stellarcollapse.org/lippunerroberts2015}).

\skynet\ has been used for
r-process nucleosynthesis calculations in different scenarios by various
authors: \citet{lippuner:15}, \citet{radice:16b}, \citet{roberts:16b}, \citet{lippuner:17a},
\citet{siegel:17}, \citet{vlasov:17}, and \citet{fernandez:17a}. Here, we discuss the physics used in
\skynet, provide details on how it is implemented, and perform code tests and
comparisons with other nuclear reaction networks.

This paper is organized as follows. In \sref{sec:rates}, we derive the
equations that govern nuclear abundance evolution and equilibrium.
\sref{sec:net_evo} deals with the numerical implementation of the reaction
network. We discuss in detail the electron screening corrections implemented in
\skynet\ in \sref{sec:screening}. In \sref{sec:impl}, we describe code
implementation details. The code tests and comparisons are the subject of
\sref{sec:tests}. We summarize in \sref{sec:summary}. In \aref{app:eos}, we
briefly present the physics of an ideal Boltzmann gas and the \EOS{}
implemented in \skynet. We show how \skynet\ computes \NSE{} in
\aref{app:calc_NSE}, and in \aref{app:nu_reac} we discuss neutrino interaction
reactions.

Throughout this paper, we set the Boltzmann constant $k_B = 1$ (i.e., all
temperatures are measured in energy), the speed of light $c = 1$, and the
reduced Planck constant $\hbar = 1$.

\section{Nuclear reaction network basics}
\label{sec:rates}

Astrophysical nuclear reaction networks track the composition of a system
containing many species of nuclei, electrons, positrons, photons, and sometimes
neutrinos. Essentially, they evolve the numbers of different nuclei in a system
given a set of reactions and rates for those reactions that transmute nuclei
into other nuclei. Although it is straightforward to heuristically write down
a system of rate equations \citep{hix:06}, it is useful to start from kinetic
theory to tie the rate equations to the microscopic processes driving the
nuclear transmutations.

\subsection{Kinetic theory}
\label{sec:kinetic}

\newcommand{\nc}{\smce{^{12}C}}
\newcommand{\nhe}{\smce{^4He}}
\newcommand{\no}{\smce{^{16}O}}
\newcommand{\nne}{\smce{^{20}Ne}}

Consider a homogeneous system of different species of particles (including nuclei,
electrons, etc.) connected by a set of interactions, a subset of which
changes particles of one type into another. A reaction indexed by $n$
converts a set of reactants into a set of products and vice versa. We write
reaction $n$ as
\begin{align}
\sum_{\alpha\in\mathcal{\tilde R}_n} [\alpha] \rightleftharpoons
\sum_{\beta\in\mathcal{\tilde P}_n} [\beta],
\end{align}
where $[\alpha]$ is a reactant, $[\beta]$ is a product, and $\mathcal{\tilde
R}_n$ and $\mathcal{\tilde P}_n$ are the sets of all reactants and products,
respectively. We emphasize that all particles are individually labeled, even if
they are of the same species. For example, for the reaction $\nc + \nhe
\rightleftharpoons \no + \gamma$, we have $\mathcal{\tilde R}_n = \{0,1\}$ and
$\mathcal{\tilde P}_n = \{2,3\}$ with $[0] = \nc$, $[1] = \nhe$, $[2] = \no$,
and $[3] = \gamma$. For $\nc + \nc \rightleftharpoons \nne + \nhe + \gamma$,
we have $\mathcal{\tilde R}_n = \{0,1\}$ and $\mathcal{\tilde P}_n = \{2,3, 4\}$
with $[0] = [1] = \nc$, $[2] = \nne$, $[3] = \nhe$, and $[5] = \gamma$. Note
that reaction $n$ includes both the forward process (going from reactants to
products) and the inverse process (going from products to reactants). Of
course, it is arbitrary which set we call reactants and which we call products.
In the following, we use the convention that if we consider particle
$[\epsilon]$, then we choose the reactants and products such that $\epsilon \in
\mathcal{\tilde R}_n$.

If the
particles are uncorrelated, the system can be described in terms of the
individual particle distribution functions $f_\epsilon$. Kinetic theory then
gives the time evolution of the distribution functions as
\citep[e.g.,][]{danielewicz:91, buss:12}
\begin{align}
\label{eq:kinetic}
& \left(\partial_t + \frac{\partial k_\epsilon^0}{\partial \vec{k}_\epsilon}
\cdot \nabla \right) f_\epsilon(x^\mu, k_\epsilon^\mu) \nn\\
= & -f_\epsilon \sum_n \mathcal{N}^\text{forward}_n
\left[\prod_{\alpha \in \mathcal{\tilde R}_n,\alpha\neq\epsilon}
\int_{[\alpha]} f_\alpha \right]
\left[\prod_{\beta \in \mathcal{\tilde P}_n} \int_{[\beta]} (1 \pm
f_\beta) \right]
\delta^4\left(\sum_{\alpha \in \mathcal{\tilde R}_n} k^\mu_{\alpha}
-\sum_{\beta \in \mathcal{\tilde P}_n} k^\mu_\beta \right)
\scriptr_n\left(k_{\{\alpha\}}^\mu, k_{\{\beta\}}^\mu\right) \nn\\
&+ (1 \pm f_\epsilon) \sum_n \mathcal{N}^\text{inverse}_n
\left[\prod_{\alpha \in \mathcal{\tilde R}_n,\alpha\neq\epsilon} \int_{[\alpha]}
(1 \pm f_\alpha) \right]
\left[\prod_{\beta \in \mathcal{\tilde P}_n} \int_{[\beta]} f_\beta
\right]
\delta^4 \left(\sum_{\alpha \in \mathcal{\tilde R}_n} k^\mu_{\alpha}
-\sum_{\beta \in \mathcal{\tilde P}_n} k^\mu_\beta \right)
\scriptr_n\left(k_{\{\alpha\}}^\mu, k_{\{\beta\}}^\mu\right),
\end{align}
where the sum over $n$ only includes interactions that have $\epsilon \in
\mathcal{\tilde R}_n$. $\mathcal{N}^\text{forward}_n$ and
$\mathcal{N}^\text{inverse}_n$ are factors that avoid double counting if the
interaction involves multiple particles of the same species. These will be
defined later after introducing some additional notation. $k_\epsilon^\mu =
(k^0_\epsilon, \vec k_\epsilon)$ is the four-momentum of particle $[\epsilon]$
and $\delta^4$ is the 4-dimensional delta function that enforces conservation of
momentum. $\scriptr_n$ denotes the differential rate of reaction $n$. The upper
($+$) signs are for bosons and the lower ($-$) signs are for fermions. We use
the shorthand
\begin{align}
\int_{[\epsilon]} = g_\epsilon \int \frac{d^3k_\epsilon}{(2\pi)^3}
\label{eq:intk}
\end{align}
for the phase-space integral of particle $[\epsilon]$, where $g_\epsilon$ is
the spin degeneracy factor. Note that the differential rate $\scriptr_n$
depends on the momenta of all particles (reactants and products). The second
line in \cref{eq:kinetic} is due to the forward process of interaction $n$, and
the third line is due to the inverse process. The third line is required by
the assumption of reversibility of interactions. Note that the differential
rate is the same for the forward and inverse processes, and the delta function is
also identical since it is an even function.

For simple interactions, e.g., weak interactions between nucleons and
neutrinos, $\scriptr$ is given by
\begin{align}
\scriptr\left(k_{\{\alpha\}}^\mu, k_{\{\beta\}}^\mu\right) = (2\pi)^4
\frac{\langle|\mathcal{T}^2|\rangle}{2 \prod_{\alpha\in\mathcal{\tilde R}} 2
k^0_\alpha \prod_{\beta\in\mathcal{\tilde P}} 2 k^0_\beta},
\end{align}
where $\langle|\mathcal{T}^2|\rangle$ is the spin-averaged reduced matrix
element (averaged over the spins of both the initial and final states) of the
interaction (see \citet{brown:97} for a discussion in the
non-relativistic context). For more complicated interactions between nuclei,
$\scriptr_n$ could include transition probabilities between multiple internal
states.

Generally, a reaction network only evolves some subset of the particles present
in the system. For instance, photons are assumed to be in equilibrium at all
times, and the electron and positron densities are determined by charge
neutrality, so their number evolution does not need to be tracked explicitly.
Therefore, it is useful to define the part of a reaction that only includes
particles that will be tracked by the network as
\begin{align}
\sum_{j\in\mathcal{R}_n} N_j^n [j] \rightleftharpoons
\sum_{l\in\mathcal{P}_n} N_l^n [l], \label{eq:reac_species}
\end{align}
where $[j]$ is a reactant species, $[l]$ is a product species, and $\mathcal{R}_n$
and $\mathcal{P}_n$ (without tildes) are the sets of distinct reactant species
and product species including only species that are present in the network,
respectively. The positive integers \smash{$N_j^n$} and \smash{$N_l^n$} are
the numbers of particles of reactant species $[j]$ destroyed and the number
of particles of product species $[l]$ created, respectively. Note that we use
Latin indices to refer to particle species, and we use Greek indices to refer to
individual particles. In the earlier example of $\nc + \nhe \rightleftharpoons
\no + \gamma$, we now have $\mathcal{R}_n = \{0,1\}$ and $\mathcal{P}_n = \{2\}$
with $[0] = \nc$, $[1] = \nhe$, $[2] = \no$, and $N_0^n = N_1^n = N_2^n = 1$.
The photon that is contained in $\mathcal{\tilde P}_n$ is not in
$\mathcal{P}_n$, because it is not explicitly tracked in the network. Similarly,
for $\nc + \nc \rightleftharpoons \nne + \nhe + \gamma$, we get $\mathcal{R}_n =
\{0\}$ and $\mathcal{P}_n = \{1,2\}$ with $[0] = \nc$, $[1] = \nne$, and $[2] =
\nhe$. But in this case, we have $N_0^n = 2$ and $N_1^n = N_2^n = 1$.

Since \cref{eq:kinetic} essentially counts the pairs (or triplets, etc.) of
reactants that can interact with each other (or pairs of products for the
inverse process), we need to be careful to avoid double counting if the
interaction involves multiple particles of the same species. If two distinct
particles $[0]$ and $[1]$ interact with each other, then there are $N_0 N_1$
distinct pairs, where $N_0$ and $N_1$ are the numbers of particles $[0]$ and
$[1]$, respectively. But if we have $N$ particles of the same species where two
react with each other, then the total number of distinct pairs is $N^2/2$ and
not $N^2$. If it is three identical particles that react with each other, the
number of distinct triplets is $N^3/6$, since there are $6 = 3!$ ways to order
a set of three items. Thus, we need to divide by the product of factorials of the
multiplicities of the interacting particle species. With the notation
introduced in \cref{eq:reac_species}, we can write this as
\begin{align}
\mathcal{N}_n^\text{forward} = \prod_{j\in\mathcal{R}_n} \frac{1}{N_j^n!}
\qquad \text{and} \qquad
\mathcal{N}_n^\text{inverse} = \prod_{l\in\mathcal{P}_n} \frac{1}{N_l^n!}.
\label{eq:double_counting}
\end{align}

We can now define the reaction rate of the forward process as
\begin{equation}
\lambda_n = n_B^{-1} \mathcal{N}^\text{forward}_n \left[\prod_{j
\in \mathcal{R}_n} \left(\frac{n_B}{n_j}\right)^{N^n_j}  \right]
\left[\prod_{\alpha \in \mathcal{\tilde R}_n} \int_{[\alpha]} f_\alpha \right]
\left[\prod_{\beta \in \mathcal{\tilde P}_n} \int_{[\beta]} (1 \pm f_\beta)
\right]
\delta^4 \left(\sum_{\alpha\in\mathcal{\tilde R}_n} k^\mu_{\alpha}
-\sum_{\beta\in\mathcal{\tilde P}_n} k^\mu_\beta \right)
\scriptr_n\left(k_{\{\alpha\}}^\mu, k_{\{\beta\}}^\mu\right),
\label{eq:kinetic_rates}
\end{equation}
where
\begin{align}
n_m = \int_{[m]} f_m = g_m \int \frac{d^3 k_m}{(2\pi)^3} f_m
\end{align}
is the number density of species $[m]$ and $n_B$ is the total baryon number
density. $\lambda_n$ is the forward process term of reaction $n$ on the
right-hand side of \cref{eq:kinetic} integrated over the phase space of particle
$[\epsilon]$ and normalized by the number densities of the particles in the
entrance channel. The reaction rate is just the rate at which a reaction
proceeds per particle in the entrance channel. These reaction rates are only
non-zero when the particles in the entrance channel differ from those in the
exit channel. The other interactions included in \cref{eq:kinetic} may change
the momentum space distribution of the particles in the system, but they cannot
change the total number of particles of any species. The reaction rate of the
inverse process $\lambda_n'$ is defined analogously to \cref{eq:kinetic_rates}
with the reactant and product sets switched.

Now, the standard reaction network equations follow from integrating over the
phase space of particle $[\epsilon]$ in \cref{eq:kinetic} to find
\begin{equation}
\partial_t n_\epsilon + \nabla \cdot \left(\langle {\vec v}_\epsilon
\rangle n_\epsilon\right) =
\sum_n\left[-\lambda_n n_B^{1 - N^n_\mathcal{R}} \prod_{j\in\mathcal{R}_n}
n_j^{N_j^n}
+ \lambda_n' n_B^{1-N^n_\mathcal{P}} \prod_{l\in\mathcal{P}_n}
n_l^{N_l^n}\right]. \label{eq:dn_kinetic}
\end{equation}
Here, $\langle {\vec v}_\epsilon \rangle = (2 \pi)^{-3} n_\epsilon^{-1} \int d^3
k_\epsilon f_\epsilon \partial_{\vec k_\epsilon} k^0_\epsilon$ is the average
velocity of particle $[\epsilon]$, and we define
\begin{align}
N_\mathcal{R}^n = \sum_{j\in\mathcal{R}_n} N_j^n
\qquad \text{and} \qquad
N_\mathcal{P}^n = \sum_{l\in\mathcal{P}_n} N_l^n.
\end{align}
Note that the left-hand side of \cref{eq:dn_kinetic} is for an individual
particle, not a particle species, but we need the evolution equations for the
particle species. A reaction $n$ that involves $N_i^n$ particles of species
$[i]$ will contribute the right-hand side in \cref{eq:dn_kinetic} $N_i^n$ times
to the derivative of $n_i$ and so we multiply the right-hand side by $N_i^n$.
Furthermore, due to the symmetry between the forward and inverse processes in
\cref{eq:dn_kinetic}, it makes sense to treat the forward and inverse processes
separately. So far, we have indexed reactions with $n$ and each reaction
consisted of the forward and inverse directions. Let us now index reactions with
$\alpha$, where the forward and inverse processes are counted separately. The
set of reactions $\{\alpha\}$ is thus twice as big as the set of reactions
$\{n\}$, although some inverse reactions may be ignored since they are
extremely unlikely to occur. \cref{eq:dn_kinetic} thus becomes
\begin{align}
\partial_t n_i + \nabla \cdot\left(\langle {\vec v}_i\rangle
n_i\right) = \sum_\alpha \lambda_\alpha (-R_i^\alpha + P^\alpha_i) N_i^\alpha
n_B^{1-N_\mathcal{R}^\alpha} \prod_{m\in \mathcal{R}_\alpha} n_m^{N_m^\alpha},
\label{eq:dn}
\end{align}
where
\begin{align}
R^\alpha_i = \left\{\begin{array}{@{}l@{\ \ }l@{}}
1 & \text{if $i \in \mathcal{R}_\alpha$} \\
0 & \text{otherwise}
\end{array}\right. \qquad \text{and} \qquad
P^\alpha_i = \left\{\begin{array}{@{}l@{\ \ }l@{}}
1 & \text{if $i \in \mathcal{P}_\alpha$} \\
0 & \text{otherwise.}
\end{array}\right.
\end{align}
For every interaction $n$, there is a forward reaction $\alpha$ that has
$\lambda_\alpha = \lambda_n$, $\mathcal{R}_\alpha = \mathcal{R}_n$, and
$N_\mathcal{R}^\alpha = N_\mathcal{R}^n$. For that reaction, \cref{eq:dn}
contributes the forward part of \cref{eq:dn_kinetic} with a multiplicative
factor $(-R_i^\alpha + P^\alpha_i) N_i^\alpha = -N_i^\alpha$, since
$N_i^\alpha$ particles of species $[i]$ are destroyed. Similarly, there is an
inverse reaction $\alpha'$ for the same interaction $n$ that has
$\lambda_{\alpha'} = \lambda_n'$, $\mathcal{R}_{\alpha'} = \mathcal{P}_n$, and
$N_\mathcal{R}^{\alpha'} = N_\mathcal{P}^n$. This reaction contributes the
inverse part of \cref{eq:dn_kinetic} with a factor of $(-R_i^\alpha +
P^\alpha_i) N_i^\alpha = N_i^\alpha$, since $N_i^\alpha$ particles of species
$[i]$ are created in the inverse reaction. Note that $N_m^{\alpha} =
N_m^{\alpha'} = N_m^n$ for all reactants and products $[m]$.

Finally, it is useful to define the \emph{abundance} $Y_i$ as
\begin{align}
Y_i \equiv \frac{n_i}{n_B} = \frac{N_i/V}{N_B/V} = \frac{N_i}{N_B}, \label{eq:Y}
\end{align}
where $V$ is the volume of the fluid element, and $N_i$ and $N_B$ are the total
numbers of particles of species $[i]$ and baryons, respectively. Since the
number density $n_i$ of species $[i]$ changes with both the number of particles
of species $[i]$ and the total volume of the system, which is often changing in
astrophysical systems undergoing nuclear burning, it is convenient to evolve the
abundances $Y_i$ instead of the number densities $n_i$. Assuming that all of the
species move as a single fluid, i.e., $\langle {\vec v}_i \rangle = {\vec v}$,
and using the Lagrangian time derivative, $d/dt = \partial_t + {\vec v} \cdot
\nabla$, we find
\begin{align}
\frac{dY_i}{dt} &= (\partial_t + \vec v \cdot
\nabla)\left(\frac{n_i}{n_B}\right) \nn\\
&= \frac{1}{n_B} \left[\partial_t n_i
-\frac{n_i}{n_B}\partial_tn_B + \nabla\cdot(\vec v n_i) - n_i\nabla\cdot\vec v
-\frac{n_i}{n_B}\left(\nabla\cdot(\vec v n_B) - n_B\nabla\cdot\vec v\right)
\right] \nn\\
&= \frac{1}{n_B}\left[\partial_t n_i + \nabla\cdot(\vec v n_i) -
\frac{n_i}{n_B}\left(\partial_tn_B + \nabla\cdot(\vec v n_B)\right)\right] \nn\\
&= \frac{1}{n_B}\left[\partial_t n_i + \nabla\cdot(\vec v n_i)\right],
\end{align}
where we used the identity $\vec v \cdot \nabla f = \nabla\cdot(\vec v f) -
f\nabla\cdot \vec v$ and the baryon number continuity equation $\partial_t n_B +
\nabla \cdot (\vec v n_B) = 0$. But since $\langle \vec v_i\rangle = \vec v$,
the above is the left-hand side of \cref{eq:dn} and so we get
\begin{align}
\frac{dY_i}{dt} = \sum_\alpha \lambda_\alpha (- R_i^\alpha + P^\alpha_i)
N^\alpha_i \prod_{m \in \mathcal{R}_\alpha} Y_m^{N^\alpha_m}, \label{eq:dotYi}
\end{align}
which is the familiar abundance evolution equation \citep[e.g.,][]{hix:99}.
Essentially, for a given set of rates \smash{$\lambda_\alpha$}, \skynet\ solves
this coupled, first-order, non-linear system of equations.

Even though it might look somewhat complicated, \cref{eq:dotYi} is very easy to
understand. It says that the total time derivative of a species $[i]$ is the
sum over all reactions that involve that species, and each reaction contributes
a term consisting of the reaction rate multiplied by the abundances of the
reactants and a factor that gives the number of particles of species $[i]$
destroyed or created in the reaction. For example, for the forward and inverse
reactions $\nc + \nhe \rightleftharpoons \no$,
\cref{eq:dotYi} says
\begin{align}
\frac{dY_{\nc}}{dt} &= -\lambda_\alpha Y_{\nc} Y_{\nhe} + \lambda_{\alpha'}
Y_{\no} + \cdots, \\
\frac{dY_{\nhe}}{dt} &= -\lambda_\alpha Y_{\nc} Y_{\nhe} + \lambda_{\alpha'}
Y_{\no} + \cdots, \\
\frac{dY_{\no}}{dt} &= \lambda_\alpha Y_{\nc} Y_{\nhe} - \lambda_{\alpha'}
Y_{\no} + \cdots.
\end{align}
For the reactions $\nc + \nc \rightleftharpoons \nne + \nhe$, we get
\begin{align}
\frac{dY_{\nc}}{dt} &= -2\lambda_\alpha Y_{\nc}^2 + 2\lambda_{\alpha'} Y_{\nne}
Y_{\nhe} + \cdots, \\
\frac{dY_{\nne}}{dt} &= \lambda_\alpha Y_{\nc}^2 - \lambda_{\alpha'} Y_{\nne}
Y_{\nhe} + \cdots, \\
\frac{dY_{\nhe}}{dt} &= \lambda_\alpha Y_{\nc}^2 - \lambda_{\alpha'} Y_{\nne}
Y_{\nhe} + \cdots.
\end{align}

\subsection{Reaction rates and velocity-averaged cross-sections}

Specializing to astrophysical systems consisting of a range of nuclear species,
scattering reactions mediated by the nuclear and electromagnetic forces bring
particles into thermal equilibrium at temperature $T$ on a much shorter
timescale than nuclear reactions bring particles into chemical equilibrium.
In that case, the distribution functions only depend on temperature and the
chemical potentials, i.e., $f_i = f_i(T, \mu_i)$.
As written, the rates defined in \cref{eq:kinetic_rates} depend on the momentum
space distribution functions of the particles involved in the reaction and may
be quite complicated. Nevertheless, in thermal equilibrium, the reaction rates
depend only on the parameters of the distribution functions, i.e.,
$\lambda_\alpha = \lambda_\alpha(T,n_B,\mu_{\{m\}})$, where the index $m$ ranges
over reactant and products.

If all of the particles involved in a reaction obey Boltzmann statistics or are
photons with chemical potential zero, then we find that $\lambda_\alpha =
\lambda_\alpha(T, n_B)$ does not depend on the chemical potentials of the
particles. This is because we can set the blocking factors $(1 \pm f_i)$ of
Boltzmann particles to 1, since quantum effects for Boltzmann particles are
negligible. For Boltzmann particles, we have $n_i \propto \exp(\mu_i/T)$ and
$f_i \propto \exp(\mu_i/T)$, which means that the dependence on $\mu_i$ cancels for
Boltzmann particles because every $f_\alpha$ in \cref{eq:kinetic_rates} is
divided by an $n_\alpha$.\footnote{Actually, the $f_\alpha$'s are divided by
$n_j^{N_j^n}$, but this is just an $n_\alpha$ for every particle of species
$[j]$.} In most
astrophysical scenarios, we can assume that the distribution functions $f_i$
follow a thermal Maxwell--Boltzmann distribution, so the rates of reactions
involving only nuclei and photons only depend on the temperature and density.

Nuclei can also undergo weak
interactions that may involve leptons with non-zero chemical potentials. The
leptons are generally not evolved in the network, but rather the electron (and
positron) chemical potential is determined by the requirement of charge
neutrality or by the number of electrons per baryon $Y_e$. Neutrinos are also
not evolved in reaction networks since their distribution functions are often
non-thermal in astrophysical scenarios in which they play a role. Therefore,
weak interaction rates have a dependence $\lambda_{\alpha, \text{weak}}
= \lambda_{\alpha, \text{weak}}(T, n_B, Y_e, f_{\{\nu\}})$ where $f_{\{\nu\}}$
are the externally specified neutrino distribution functions of the relevant
neutrino species (see \aref{app:nu_reac}). However, weak decay rates of nuclei
are just constants when final state blocking by leptons can be safely ignored.

For two particle reactions, it is common to define the cross-section as
\cite[][\S4]{peskin:95}
\begin{equation}
\sigma_\alpha(k^\mu_1 - k^\mu_2) = \frac{1}{v_\text{rel}}
\left[\prod_{l \in \mathcal{P}_\alpha} \int_{[l]}
\right]
\delta^4\left(k_1^\mu + k_2^\mu - \sum_{l\in\mathcal{P}_\alpha} k^\mu_l\right)
\scriptr_\alpha\left(k_1^\mu, k_2^\mu, k^\mu_{\{l\}}\right),
\end{equation}
where $v_\text{rel}$ is the relative velocity between particles [1] and [2].
Adopting the viewpoint that the $[i]$ are stationary targets and the $[j]$
are incoming projectiles impinging on the targets, the cross-section
$\sigma_{\alpha}$ can be interpreted as \citep[e.g.,][\S4]{clayton:68}
\begin{align}
\sigma_\alpha &= \frac{\text{number of reactions per second per target
$[i]$}}{\text{flux of incoming projectiles $[j]$}} = \frac{R_{i,j}/(n_i
V)}{v_\text{rel} n_j} = \frac{r_{i,j}}{v_\text{rel}n_in_j}, \label{eq:sigma}
\end{align}
where $R_{i,j}$ is the number of reactions per second, $r_{i,j} = R_{i,j} /V$
is the number of reactions per second per volume, and $n_i$, $n_j$ are the
number densities
of $[i]$ and $[j]$. Assuming Boltzmann statistics so
that $(1 \pm f_l) \to 1$ for the products, \cref{eq:kinetic_rates} gives
\begin{align}
\lambda_\alpha &= \mathcal{N}_\alpha n_B \int_{[1]} \frac{f_1}{n_1}
\int_{[2]} \frac{f_2}{n_2} v_\text{rel} \sigma_\alpha = \mathcal{N}_\alpha n_B
\langle \sigma_\alpha v_\text{rel}\rangle,
\end{align}
where $\mathcal{N}_\alpha$ is the double counting factor from
\cref{eq:double_counting}. Since the distribution functions are normalized by
the densities of species [1] and [2], this expression shows that
$\lambda_\alpha$ is proportional to the cross-section
averaged over the relative velocities between the two particles (after
transforming to the center of mass frame; see e.g., \citealt{clayton:68}).
Therefore, using \cref{eq:sigma}, one arrives at the standard relation between
the reaction rate and the velocity-averaged cross-section (e.g.,
\citealt[\S4]{clayton:68}; \citealt[\S3]{rolfs:88}),
\begin{equation}
r_{i,j} = n_B^{-1} \lambda_\alpha n_1 n_2
= n_1 n_2 \mathcal{N}_\alpha \langle \sigma_\alpha v_\text{rel} \rangle.
\end{equation}

\subsection{Nuclear Statistical Equilibrium (NSE) and inverse reaction rates}
\label{sec:inverse_rates}

\Cref{eq:kinetic} shows that for every reaction, there is an inverse
reaction.  The relationship between the forward and reverse rates, which
only depends on the density, temperature, and the internal properties of the
nuclei, is generally called detailed balance.  In some cases, for example for
$\beta$-decays or fission reactions, the inverse reactions are extremely
unlikely to occur and can be ignored. For other reactions, e.g., a neutron
capture reaction, the inverse reaction can occur very frequently and sometimes
even more often than the forward reaction. Thus, it is important to take inverse
reactions into account.  At temperatures of about 5~GK and above, inverse strong
reactions such as photodissociation of nuclides can be in equilibrium with
their forward reactions. In that case, the reaction is said to be in
\emph{chemical equilibrium}. For example, the reactions $\smce{^{196}Au} +
\text{n} \rightleftharpoons \smce{^{197}Au} + \gamma$ and $\nne + \gamma
\rightleftharpoons \no + \nhe$ can be in chemical equilibrium at
sufficiently high temperatures. The situation of all strong reactions being in
equilibrium is called \NSE{}. This situation can
also be thought of as an equilibrium between the reaction of forming a nucleus
$(Z,N)$ from $Z$ free protons and $N$ free neutrons, and its inverse reaction,
namely, completely dissociating a nucleus $(Z,N)$ into $Z$ protons and $N$
neutrons. In other words, if \NSE{} holds, then the forward and inverse
reactions,
\begin{align}
(Z,N) \rightleftharpoons Z[\text{p}] + N[\text{n}], \label{eq:nse:reac}
\end{align}
are in equilibrium for all nuclides that are part of the composition. Of
course, there are no reactions that directly create a nuclide $(Z,N)$ out of
$Z$ protons and $N$ neutrons. But there is a chain of strong reactions that
connects $(Z,N)$ to free neutrons and protons. So if all strong reactions are
in equilibrium, then we effectively have the reactions shown above and they are
also in equilibrium. When nucleons are in chemical equilibrium with all other
nuclear species, the energetic cost of turning $Z_i$ protons and $N_i$ neutrons
into a single nucleus must be zero, which requires
\begin{equation}
\mu_i = Z_i \mu_\text{p} + N_i \mu_\text{n}, \label{eq:NSE}
\end{equation}
where $\mu_i$ is the chemical potential of species $[i]$.

When the composition moves into \NSE{}, the
forward and inverse strong reactions approach equilibrium. In order to ensure
that the equilibrium composition determined by the forward and inverse reaction
rates is the same as the \NSE{} composition computed from the equality of the
chemical potentials, we need to compute the inverse reaction rates directly from
the forward rates and nuclide properties. Consider the reaction $\alpha$ and its
inverse reaction $\alpha'$. In equilibrium, each set of terms on the right-hand
side of \cref{eq:kinetic} must be zero. Then, by the symmetry of the
differential rate $\scriptr_\alpha = \scriptr_{\alpha'}$ and casting
\cref{eq:kinetic} into \cref{eq:dn_kinetic}, we have
\begin{equation}
\lambda_\alpha \prod_{j \in \mathcal{R}_\alpha} Y_{j,\text{eq}}^{N_j^\alpha} =
\lambda_{\alpha'} \prod_{l \in \mathcal{P}_\alpha} Y_{l,\text{eq}}^{N_l^\alpha},
\label{eq:lambda_balance}
\end{equation}
where $Y_{i,\text{eq}}$ is the abundance of species $[i]$ in chemical
equilibrium. For a Boltzmann gas, the abundance is given by \crefp{eq:mu}
\begin{align}
Y_i = \frac{G_i(T)}{n_B} \left(\frac{m_i T}{2 \pi}\right)^{3/2}
\exp[(\mu_i-m_i)/T],
\end{align}
where $G_i(T)$ is the internal partition function of species $[i]$ (see
\crefalt{eq:part_func}) and $m_i$ is its rest mass. Substituting the above into
\cref{eq:lambda_balance} yields
\begin{equation}
\frac{\lambda_{\alpha'}}{\lambda_\alpha} =
\frac{\prod_{j \in \mathcal{R}_\alpha} [G_j/n_B (m_j T/
2\pi)^{3/2}]^{N_j^\alpha}}
{\prod_{l \in \mathcal{P}_\alpha} [G_l/n_B (m_l T/ 2\pi)^{3/2}]^{N_l^\alpha}}
\exp\left[\frac{1}{T}\sum_{j\in\mathcal{R}_\alpha} N_j^\alpha(\mu_{j,\text{eq}}
- m_j) - \frac{1}{T}\sum_{l\in\mathcal{P}_\alpha}
N_l^\alpha(\mu_{l,\text{eq}} - m_l) \right]. \label{eq:basic_balance}
\end{equation}

Since the forward and reverse reactions are in equilibrium, the chemical
potentials on both sides are equal, hence
\begin{align}
\sum_{j\in\mathcal{R}_\alpha} N_j^\alpha \mu_{j,\text{eq}} =
\sum_{l\in\mathcal{P}_\alpha} N_l^\alpha \mu_{l,\text{eq}},
\label{eq:mu_balance}
\end{align}
and the chemical potentials in the exponential of \cref{eq:basic_balance}
cancel. Then, the inverse reaction rate $\lambda_{\alpha'}$ is
\begin{align}
\lambda_{\alpha'}(T,\rho) &= \lambda_\alpha(T,\rho)
e^{-Q_\alpha/T}\Gamma_\alpha(T) M_\alpha^{3/2}
\left(\frac{T}{2\pi} \right)^{ 3\Delta N_\alpha/2}
n_B^{-\Delta N_\alpha}, \label{eq:detailed_balance}
\end{align}
where we define
\begin{align}
Q_\alpha &= \sum_{j\in\mathcal{R}_\alpha} N_j^\alpha m_j -
\sum_{l\in\mathcal{P}_\alpha} N_l^\alpha m_l, \label{eq:Qa} \\
\Gamma_\alpha(T) &= \frac{\prod_{j\in\mathcal{R}_\alpha}G_j(T)^{N_j^\alpha}}
{\prod_{l\in\mathcal{P}_\alpha}G_l(T)^{N_l^\alpha}}, \\
M_\alpha &= \frac{\prod_{j\in\mathcal{R}_\alpha}m_j^{N_j^\alpha}}
{\prod_{l\in\mathcal{P} _\alpha }m_l^{N_l^\alpha}}, \\
\Delta N_\alpha &= \sum_{j\in\mathcal{R}_\alpha} N_j^\alpha -
\sum_{l\in\mathcal{P}_\alpha} N_l^\alpha = N_\mathcal{R}^\alpha -
N_\mathcal{P}^\alpha.
\end{align}
Although the above is derived under the assumption that the abundances are
such that the two reactions are in equilibrium, it still holds for any
abundances because the reaction rates only depend on the temperature and
density. In \aref{app:calc_NSE}, we show how calculating \NSE{} is implemented
in \skynet.

\section{Network evolution}
\label{sec:net_evo}

In this section, we focus on the specific implementation of the physics
described in the previous section in \skynet.
In essence, evolving the reaction network means solving the large coupled system
of first-order, non-linear \ODE{s} given in \cref{eq:dotYi}. But there are
various other pieces that are needed to make the evolution robust and
efficient, which are also discussed this section.

\subsection{Implicit integration method} \label{sec:NR}

The system of \ODE{s} \crefp{eq:dotYi} we need to solve is extremely stiff
because of the enormous range of reaction rates, which can span many orders of
magnitude \citep[e.g.,][]{timmes:99b, hix:06}. Thus, an explicit integration
method
would be constrained to taking extremely small time steps. This is why nuclear
reaction networks are typically integrated with an implicit method
\citep[e.g.,][]{arnett:69, woosley:73, timmes:99b, winteler:13, longland:14}.
However, some authors
have proposed various explicit methods specifically tuned to integrate stiff
reaction networks \citep[e.g.,][]{feger:11, guidry:12, guidry:13b,
guidry:13c, guidry:13a, brock:15}. Currently, \skynet\ uses the first-order
implicit backward Euler method \citep[e.g.,][]{gear:71, hix:99}, but it is
straightforward to implement higher-order implicit integration methods as well.
We plan to do this in the future, since \citet{timmes:99b} recommends using
higher-order methods, such as the variable-order Bader--Deuflhard method
\citep{bader:83}. Let the vector $\vec Y(t) = Y_i(t)$ denote
the composition at time $t$. If we want to take a time step $\Delta t$ using
the first-order implicit backward Euler method, we write
\begin{align}
& \dot{\vec Y}(t+\Delta t) = \frac{\vec Y(t+\Delta t) - \vec Y(t)}{\Delta t}
\nonumber\\
\Leftrightarrow \ \ & \vec 0 = \dot{\vec Y}(\vec x, T(t+\Delta t),
\rho(t+\Delta t)) -
\frac{\vec x - \vec Y(t)}{\Delta t} \nonumber\\
&\phantom{\vec 0{}}{}= \vec F(\vec x, T(t+\Delta t), \rho(t+\Delta t)),
\label{eq:root}
\end{align}
where $\vec x = \vec Y(t+\Delta t)$ are the unknown abundances at the end
of the time step we are trying to find and $\vec Y(t)$ are the known
abundances at the beginning of the time step. $\dot{\vec Y}(\vec Y, T, \rho)$ is
the function defined in \cref{eq:dotYi} that gives the time derivatives of the
abundances as a function of a given set of abundances $\vec Y$, a temperature
$T$, and density $\rho$. Note that we need to know the temperature $T$ and
density $\rho$ as a function of time. The function $\vec F$ is simply the
right-hand side of
the above equation. We thus have a multidimensional root-finding problem $\vec
0 = \vec F (\vec x, T, \rho)$, where $T$ and $\rho$ are known functions of time.
\skynet\ uses the \NR{} method to find the root. For every time step, the
following iteration is performed to find the solution $\vec x$ to
\cref{eq:root}:
\begin{align}
\vec x_{n+1} &= \vec x_n - [J_{\vec F}(\vec x_n)]^{-1} \vec F(\vec x_n,
T(t+\Delta t), \rho(t+\Delta t)),
\label{eq:NR}
\end{align}
where the iteration starts with $\vec x_0 = \vec Y(t)$, the abundances from the
previous time step, and
\begin{align}
(J_{\vec F})_{ij} = \frac{\partial F_i}{\partial Y_j} = \frac{\partial \dot
Y_i}{\partial Y_j} - \frac{\delta_{ij}}{\Delta t},
\end{align}
is the Jacobian matrix and $\delta_{ij}$ is the Kronecker delta. The partial
derivatives $\partial\dot Y_i/\partial Y_j$ are computed from \cref{eq:dotYi}
as
\begin{align}
\frac{\partial \dot Y_i}{\partial Y_j} &= \sum_\alpha \lambda_\alpha(-R_i^\alpha
+ P_i^\alpha) N_i^\alpha \sum_{n\in\mathcal{R}_\alpha} \delta_{nj}
\frac{N_n^\alpha}{Y_n}\prod_{m\in\mathcal{R}_\alpha} Y_m^{N_m^\alpha}
\nonumber\\
&= \sum_\alpha \lambda_\alpha (-R_i^\alpha + P_i^\alpha) N_i^\alpha
R_j^\alpha N_j^\alpha Y_j^{N_j^\alpha-1}
\prod_{m\in\mathcal{R}_\alpha,m\neq j} Y_m^{N_m^\alpha}.
\label{eq:dydy}
\end{align}

The size of the Jacobian matrix $J_{\vec F}$ is $N\times N$,
where $N$ is the number of nuclear species in the network. For large networks
($N\sim 8000$), inverting this large matrix can be quite costly, because the
linear system in \cref{eq:root} has to be solved for every \NR{} iteration.
Since the Jacobian matrix depends on the unknown abundances at the end of the
time step, we have to recompute the Jacobian matrix after every \NR{} iteration
that updates our guess for the abundances at the end of the time step.
Fortunately, however, the Jacobian matrix is very sparse (only 0.24\% of the
$N^2$ entries are non-zero for large networks), because most nuclear species
are not directly connected by a single reaction. By exploiting the sparseness of
the Jacobian, we drastically reduce the memory requirement to store the
Jacobian and we can also use matrix solver packages that are specifically
designed for sparse systems (see \sref{sec:impl}). The most expensive parts of
the evolution are computing the Jacobian entries from \cref{eq:dydy} and then
solving the sparse linear system given by \cref{eq:root}. These two operations
have to be performed at every \NR{} iteration. The method for choosing the time
step $\Delta t$ and when to terminate the \NR{} iteration will be discussed in
\sref{sec:dt}.

\subsection{Self-heating evolution} \label{sec:self-heating}

In the previous section, we showed how \skynet\ integrates the nuclear
abundances
forward in time if both the temperature and density are
given as functions of time. In most applications, the density history $\rho(t)$
is given (for example, when \skynet\ is used to post-process nucleosynthesis for
tracer particles from a hydrodynamical simulation), but the temperature is not
necessarily known as a function of time. Even if we do have a temperature
history available, it most likely would not include heating due to the nuclear
reactions that \skynet\ evolves. But based on the kinetic theory description of
the reaction network equations in \sref{sec:kinetic}, as well as the discussion
of detailed balance in \sref{sec:inverse_rates}, it is clear that the reaction
network and the thermodynamic state of the fluid are intimately related and
need to be treated consistently. Therefore, we want the temperature to be
evolved in the network under the influence of nuclear reactions, which
release nuclear binding energy as heat. This is referred to
as a self-heating network evolution \citep[e.g.,][]{freiburghaus:99}. We still
require knowing the density as a function of time, though.

Recall the first law of thermodynamics,
\begin{align}
dU = \delta Q - \delta W = \delta Q - PdV, \label{eq:1st_law}
\end{align}
where $dU$ is the infinitesimal change in internal energy, $\delta Q$ is the
infinitesimal heat added to the system from the surroundings, and $\delta W$ is
the infinitesimal mechanical work performed by the system. If the system
expands or contracts, the work performed is $\delta W = PdV$, where $P$ is
the pressure and $dV$ is the infinitesimal change in volume. We use the entropy
$S$, volume $V$, and composition $\{N_k\}$ as our independent thermodynamic
variables. Note that the index $k$ ranges over all particles in the system, not
just nuclides. The total differential of the internal energy is thus
\begin{align}
dU &= \left(\frac{\partial U}{\partial S}\right)_{V,\{N_k\}}dS +
\left(\frac{\partial U}{\partial V}\right)_{S,\{N_k\}}dV +
\sum_k\left(\frac{\partial U}{\partial N_k}\right)_{S,V,\{N_{l\neq k}\}}dN_k
\nn\\
&= TdS - PdV + \sum_k \mu_k dN_k, \label{eq:dU}
\end{align}
where $T$ is the temperature and $\mu_k$ is the chemical potential of particle
$k$.
Equating \cref{eq:1st_law} and \cref{eq:dU} yields
\begin{align}
\delta Q = TdS + \sum_k \mu_k dN_k.
\end{align}
If we divide the above by $N_B$, the total number of baryons, and replace
the infinitesimal changes by the differences of the quantities from one time
step to the next (i.e., $\Delta X = X(t + \Delta t) - X(t)$), we find
\begin{align}
\Delta q = T\Delta s + \sum_k \mu_k \Delta Y_k, \label{eq:dq1}
\end{align}
since $Y_k = N_k/N_B$ \crefp{eq:Y}. Note that the sum over $k$ includes all
particles, and hence also electrons, which can be created or destroyed in weak
nuclear reactions. $\Delta q$ is the change in the heat per baryon due to
external heating
sources. Let $\dot\epsilon_\text{ext}$ be an imposed external heating rate (per
baryon), then
\begin{align}
\Delta q = (\dot\epsilon_\text{ext} - \dot\epsilon_\nu)\Delta t, \label{eq:dq2}
\end{align}
where $\dot\epsilon_\nu$ is the neutrino heating/cooling rate of the
system on the environment given by \cref{eq:eps_nu}. Since we do not include
neutrinos in the internal state of the system, the neutrino heating/cooling
must be treated as an external source of heat. And because we define
$\dot\epsilon_\nu$ as the heating/cooling on the environment, it has a minus
sign in \cref{eq:dq1}. Combining \cref{eq:dq1} with \cref{eq:dq2} and solving
for $\Delta s$ yields
\begin{align}
 \Delta s &= \frac{\Delta q}{T} - \frac{1}{T}\sum_i \Delta Y_i (\mu_i +
Z_i\mu_{e^-}) \nn\\
 &= (\dot\epsilon_\text{ext} - \dot\epsilon_\nu)\frac{\Delta t}{T} - \sum_i
\Delta Y_i \left(\frac{m_i}{T} + \frac{Z_i\mu_{e^-}}{T} +
\ln\left[\frac{n_i}{G_i(T)}\left(\frac{2\pi}{m_i T}\right)^{3/2}\right]\right),
\label{eq:deltaS0}
\end{align}
where we used \cref{eq:mu} and switched to index $i$ that runs only over the
nuclides. So, we explicitly include the contribution from the $Z_i$ electrons
that come with nuclide $i$. To make the rest mass terms
in the sum closer to unity, we define the mass excess $\mathcal{M}_i$ as
\begin{align}
\mathcal{M}_i = m_i - A_i m_u,
\end{align}
where $A_i$ is the number of neutrons and protons of species $i$, and $m_u$ is
the atomic mass unit, defined such that the mass excess of \nc\ is
exactly 0 (i.e., $m_u = m_{\nc} / 12$). Since $Y_i = N_i/N_B$ (with $N_i$
being the number of particles of species $[i]$; see
\crefalt{eq:Y}), we find
\begin{align}
\sum_i Y_i A_i = \frac{1}{N_B} \sum_i N_i A_i = \frac{N_B}{N_B} = 1,
\label{eq:sumYA}
\end{align}
because species $[i]$ is made up of $A_i$ neutrons and protons, and hence $A_i$
baryons. Thus, we have $\sum_i Y_i(t) A_i = 1$ for all times $t$, which is
another way of saying that the total baryon number $N_B$ is conserved. Using
this, we find
\begin{align}
\sum_i \Delta Y_i m_i &= \sum_i \Delta Y_i (\mathcal{M}_i + m_u A_i) \nn\\
&= \sum_i \Delta Y_i\mathcal{M}_i + m_u\left(\sum_i Y_i(t+\Delta t) A_i -\sum_i
Y_i(t)A_i\right) \nn\\
&= \sum_i \Delta Y_i \mathcal{M}_i + m_u(1 - 1) = \sum_i \Delta Y_i
\mathcal{M}_i. \label{eq:mass_excess_cons}
\end{align}
Using the above and the fact that $n_i = Y_i n_B$, we can write
\cref{eq:deltaS0} as
\begin{align}
 \Delta s = (\dot\epsilon_\text{ext} - \dot\epsilon_\nu)\frac{\Delta t}{T} -
\sum_i \Delta Y_i \left(\frac{\mathcal{M}_i}{T} + \frac{Z_i\mu_{e^-}}{T} +
\ln\left[\frac{Y_i n_B}{G_i(T)}\left(\frac{2\pi}{m_i
T}\right)^{3/2}\right]\right). \label{eq:deltaS}
\end{align}
Note that the external heating is accounted for with a first-order forward
Euler method. We plan to improve this in the future when we implement
higher-order integration methods for the network itself.
With the above, \skynet\ can update the entropy after every time step and then
obtain the new temperature at the end of the time step from the \EOS{}.
Therefore, we only need to know the initial entropy (or temperature, from which
the entropy is determined). This changes the evolution \crefp{eq:root}
slightly, because we now have to use the entropy at the beginning of the time
step to estimate the temperature at the end of the time step. That is, we solve
\begin{align}
\vec 0 = \vec F(\vec x, T^\ast, \rho(t + \Delta t)), \label{eq:NR_heat}
\end{align}
where $T^\ast$ is given by the \EOS{} as
\begin{align}
T^\ast = \text{EOS}(s(t), \rho(t + \Delta t), \vec Y(t)).
\end{align}
\Cref{eq:NR_heat} is solved with the \NR{} method, as described in the
previous section. Note that $\Delta t$ is fixed during the \NR{} iterations,
which means that the temperature and density are also fixed. After the \NR{}
iterations have converged, we have found the new abundances $Y_i(t + \Delta
t)$ and then we can compute $\Delta s$ according to \cref{eq:deltaS} and
update the entropy as
\begin{align}
s(t + \Delta t) = s(t) + \Delta s.
\end{align}
Hence, we have a hybrid
implicit/explicit scheme where the abundances are evolved implicitly but the
entropy is evolved explicitly. One could also evolve the entropy implicitly
together with the abundances, which would require computing $\partial \dot
Y_i/\partial s$ and $\partial \dot s/\partial Y_i$ and adding these terms to the
Jacobian. We may extend \skynet\ in
the future to support such a fully implicit scheme, but for now, we have
achieved good results with the hybrid approach.

The energy released as nuclear binding energy due to nuclear reactions is
\begin{align}
\dot\epsilon_\text{nuc} = -\frac{1}{\Delta t}\sum_i \Delta Y_i m_i =
-\frac{1}{\Delta t}\sum_i \Delta Y_i \mathcal{M}_i,
\end{align}
where we used \cref{eq:mass_excess_cons} and $\Delta Y_i/\Delta t$ is an
approximation for $\dot Y_i$ over the time step. Note that the minus sign comes
from the fact that some rest mass (or mass excess) is converted into energy, and hence
the heating rate is positive if there is a net reduction in the total rest mass. Some
authors \citep[e.g.,][]{hix:99} treat $\dot\epsilon_\text{nuc}$ as an external
heat source. This is necessary if the \EOS{} does not depend on the entire
composition but only on $\bar A$ and $\bar Z$, the average mass and charge
numbers, for example. In order to compare the nuclear heating in \skynet\ to
other codes, \skynet\ computes and records a total heating rate
$\dot\epsilon_\text{tot}$, regardless of whether self-heating is enabled.
$\dot\epsilon_\text{tot}$ is computed as
\begin{align}
\dot\epsilon_\text{tot} = \frac{\Delta q}{\Delta t} + \dot\epsilon_\text{nuc} =
-\dot\epsilon_\nu + \dot\epsilon_\text{ext} -
\frac{1}{\Delta t}\sum_i \Delta Y_i \mathcal{M}_i. \label{eq:heating}
\end{align}
Note that the above has units of erg~s${}^{-1}$~baryon${}^{-1}$. To convert it
to the more commonly used units of erg~s${}^{-1}$~g${}^{-1}$, we simply multiply
$\dot\epsilon_\text{tot}$ by the Avogadro constant $N_A$.

Currently, \skynet\ records the total heating rate shown above. In reality,
this heating rate is composed of multiple components that are thermalized in
the material in different ways \citep[e.g.,][]{barnes:16}. For example, emitted
electrons and positrons, as well as the kinetic energy of fission fragments,
thermalize with very high efficiency, while only a small fraction of
the energy released as neutrinos might thermalize. In a future version of
\skynet, we plan to record the different heating rate components, so that
thermalization can be taken into account in kilonova light curve calculations,
for example.

\subsection{Convergence criteria and time stepping} \label{sec:dt}

The time step for the network evolution needs to be adjusted depending on how
well the \NR{} iteration \crefp{eq:NR} converges. All of the default values and
thresholds mentioned in this section are adjustable by the user. To check that
the \NR{} iterations have completely converged, a standard criterion is
\citep{numrep}
\begin{align}
\sum_{x^{(n+1)}_i \geq Y_\text{thr}} \left|\frac{x^{(n+1)}_i -
x^{(n)}_i}{x^{(n+1)}_i}\right| < \varepsilon_{\text{tol},\Delta x},
\label{eq:conv_dYbyY}
\end{align}
where \smash{$x_i^{(n+1)}$} is the $i$th component of the vector \smash{$\vec
x_{n+1}$}, and \smash{$\vec x = \vec Y(t + \Delta t)$} are the unknown abundances
at the end of the current time steps that we want to find. The sum only runs
over the indices $i$ for which \smash{$x^{(n+1)}_i
\geq Y_\text{thr}$} for some abundance threshold $Y_\text{thr}$, which we
usually set to \smash{$10^{-20}$}. The default value is
\smash{$\varepsilon_{\text{tol},\Delta x} = 10^{-6}$}. Although this convergence
criterion ensures that any
subsequent \NR{} iterations would no longer change the solution \smash{$\vec
x_{n+1}$}, we found that this criterion is too strict in practice.
Instead, \skynet\ typically uses mass conservation as a heuristic convergence
criterion (which is also used by \cite{hix:99}), which takes the form
\begin{align}
\left|1 - \sum_i x_i^{(n+1)} A_i\right| < \varepsilon_\text{tol,mass},
\label{eq:conv_mass}
\end{align}
where we usually use \smash{$\varepsilon_\text{tol,mass} = 10^{-10}$}. Note
that the sum now runs over all nuclear species, and there is no threshold for
\smash{$x_i^{(n+1)}$}. Since \smash{$x_i^{(n+1)} = Y_i(t + \Delta t)$}, this
convergence criterion is simply the conservation of the total baryon number. The user of
\skynet\ can choose to use \cref{eq:conv_dYbyY}, \cref{eq:conv_mass}, or both as
the convergence criterion for \cref{eq:NR}.

\begin{figure}
\centering
\includegraphics[width=\singlecol]{%
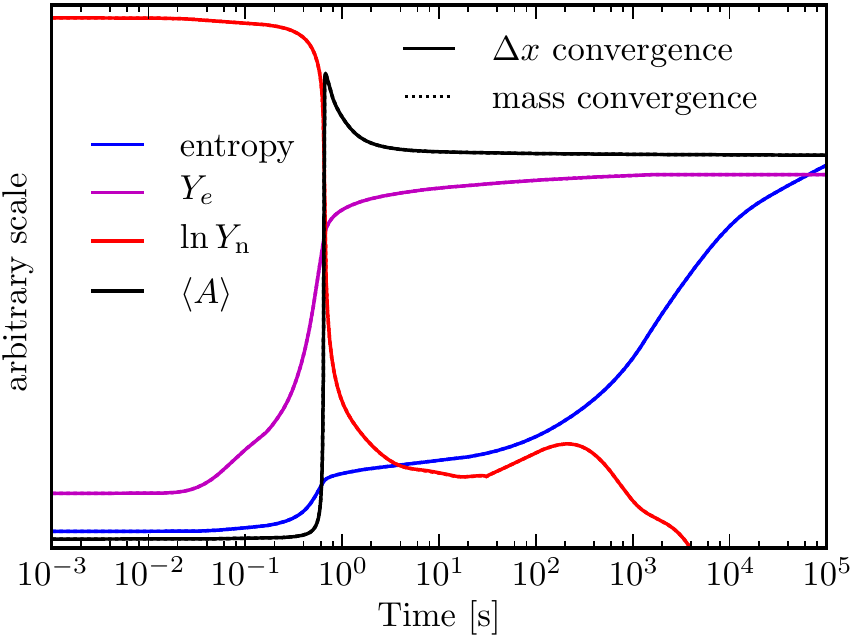}
\caption{Comparison of the two convergence criteria. The solid lines show
the entropy, electron fraction $Y_e$, logarithm of the neutron abundance
$Y_\text{n}$, and average mass number $\langle A\rangle$ as a function of time
using the $\Delta x$ convergence criterion \crefp{eq:conv_dYbyY} with
\smash{$\varepsilon_{\text{tol},\Delta x} = 10^{-6}$}. The dotted lines
plotted on top of the solid lines are the same quantities using the mass
conservation convergence criterion \crefp{eq:conv_mass} with
\smash{$\varepsilon_\text{tol,mass} =
10^{-10}$}. For all quantities, the two lines are exactly on top of each other
and so the dotted lines are not visible. All quantities have been scaled by an
arbitrary amount to fit on one figure. The networks evolve an r-process
starting at 6~GK with initial $Y_e = 0.1$, $s = 10\,\kb$ and an analytic
density profile described in \citet{lippuner:15} with expansion timescale
7.1~ms. The networks contain 7843 nuclides and 140,000 reactions.}
\label{fig:conv}
\end{figure}

\Cref{fig:conv} shows an r-process evolution with the two different convergence
criteria using \smash{$\varepsilon_{\text{tol},\Delta x} = 10^{-6}$} and
$\varepsilon_\text{tol,mass} = \smash{10^{-10}}$. These convergence thresholds
result in almost exactly the same time-step sizes, but if we made
\smash{$\varepsilon_{\text{tol},\Delta x}$} smaller, that would result in much
smaller time steps. However, using the $\Delta x$ convergence criterion requires
an average of 3.1 \NR{} iterations per time step, while mass conservation only
needs 1.2 \NR{} iterations per time step. Since the total
number of time steps
is almost the same, using mass conservation as the convergence criterion is
about 2.1 times faster for this particular case.
As \cref{fig:conv} shows, however, the
nucleosynthesis evolution is identical in the two cases. No differences are
visible in the entropy, electron fraction, neutron abundance, or average mass
number as a function of time. The maximum absolute difference in the final
abundances of the two cases (using the $\Delta x$ or mass conservation
convergence criterion) is about $1.6\times 10^{-7}$.

\begin{figure}
\centering
\includegraphics[width=1.2\singlecol]{%
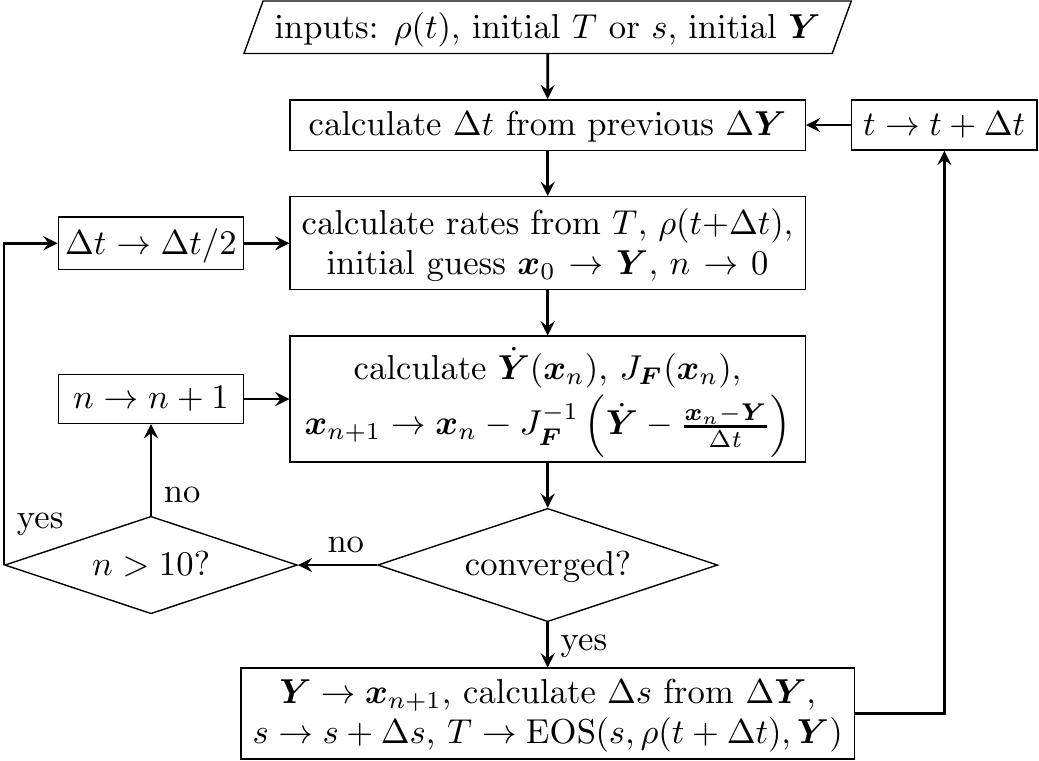}
\caption{Simplified overview of the adaptive time-stepping mechanism used in
\skynet. If the \NR{} iterations do not converge after 10 iterations, the time-step
size $\Delta t$ is cut in half and the time step is attempted again. There
are other conditions that can result in a failed step and a subsequent retry
with a smaller time-step size. See the text for details. After a successful
time step, the next $\Delta t$ is computed from the size of the abundance
changes $\Delta \vec Y$ \crefp{eq:new_dt}, and at that point $\Delta t$ can
increase or decrease.}
\label{fig:flowchart}
\end{figure}

\skynet\ adjusts the time-step size $\Delta t$ dynamically. Once the \NR{}
iterations have converged according to the chosen criterion, \skynet\ checks
that the composition did not change too much over the last time step. The
temperature and entropy are allowed to change by at most 1\%. If either of them
changes by more than this threshold, then the time
step is considered failed and $\Delta t$ is reduced by a factor of two,
and the whole step is attempted again with the reduced
time-step size. \skynet\ also considers the time step as failed
if the \NR{} iterations do not
converge after 10 iterations, or if the error measure
used for the \NR{} convergence criterion increases compared to the error of the
previous iteration, or if the error decreases by less than 10\%. In all of
those cases, $\Delta t$ is reduced and the time step is attempted again. A
simplified schematic of this mechanism is shown in \cref{fig:flowchart}.

\pagebreak

After a successful time step, \skynet\ attempts to increase the step size for
the next time step. \skynet\ tries to double $\Delta t$ after every successful
step, but this new time step can be limited if the abundance of a particular
nuclide changed by a large amount in the previous time step. If that is the
case, then the new time step is limited to the approximate step size necessary
to keep the abundance of the nuclide that changed the most over the last time
step from changing by more than 10\%. Hence, the new time step is computed as
\begin{align}
\Delta t^{(n+1)} = \min\left\{\Delta t_\text{max}, 2\Delta t^{(n)},\frac{10\%\,
\Delta t^{(n)}}{\max_i(\Delta Y_i/Y_i)}\right\}, \label{eq:new_dt}
\end{align}
where $\Delta t_\text{max}$ is the maximum allowed time step and $\Delta
t^{(n)}$ is the previous time step. Using this adaptive time-stepping
mechanism, we typically get a time-step size that grows exponentially with
time in freely expanding trajectories while keeping the error measure used for
the convergence criterion below its prescribed tolerance.

Very rarely, it is necessary to renormalize the composition. In that
case, every abundance is divided by the total mass, i.e.,
\begin{align}
Y_{i,\text{new}} = \frac{Y_i}{\sum_i A_iY_i},
\end{align}
and then the new composition satisfies $\sum_i A_iY_i = 1$ exactly. Although this
artificially injects or removes energy from the system, it is useful as a
last resort if the time-step size is kept small because the composition is far
away from mass conservation (but still within the error tolerance). After
renormalization, the evolution usually proceeds normally with a larger time
step than before. We renormalize if the time step falls below a certain limit
(usually $10^{-16}$), or if there are more than 25 time steps in a row that
tried to increase the step size but subsequently failed and had to keep the
step size constant. In such cases, it could be that the time step is small
because the mass conservation convergence criterion is preventing the time step
from increasing. If this is the case, then renormalizing the abundances usually
helps to increase the time step, because after renormalizing, mass conservation
in \cref{eq:conv_mass} is fulfilled exactly. But in some cases, for example
when trying to evolve the network near \NSE{} with reaction rates that are
inconsistent with \NSE{} (see \sref{sec:nse_consistency}), the time step is
small because the abundances are changing rapidly and so renormalizing the
composition does not help.

\subsection{NSE evolution mode} \label{sec:NSE_evolution}

If the abundances approach the \NSE{} composition, the forward and inverse
strong rates exactly balance (\sref{sec:inverse_rates}). In that case, all the
partial
derivatives in the Jacobian \crefp{eq:dydy} would be zero, resulting in a
singular Jacobian. The Jacobian is not exactly singular, however, because the
weak reactions (that are not in equilibrium with their inverses) contribute
non-zero derivatives to the Jacobian. Nevertheless, as the strong reactions
move into equilibrium, the network time step becomes very small as the Jacobian
becomes close to being numerically singular. To alleviate
this problem, \skynet\ automatically switches from a full network evolution to
an \NSE{} evolution scheme, if the strong nuclear reaction time scale becomes
shorter than the time scale over which the density changes, and if the
temperature is above some threshold (a user setting with a default value
of 7~GK). The full network is
turned back on when these conditions are no longer satisfied.
A similar approach is used by other groups \citep[e.g.][]{iwamoto:99,
brachwitz:00}.

If \skynet\ determines that switching to \NSE{} evolution is appropriate, it
computes the \NSE{} composition from the current internal energy, density, and
electron fraction. If the entropy and temperature of that \NSE{} composition
differs by less than 1\% (user setting) from the current network entropy and
temperature, then the switch to \NSE{} is allowed. Otherwise, the full network
evolution will continue and \skynet\ will try to switch to the \NSE{} evolution
mode again after the next step. A test of the \NSE{} evolution mode that
demonstrates its necessity and consistency is presented in
\sref{sec:test_nse_evolution}.

In the \NSE{} evolution mode, \skynet\ no longer evolves the abundances of
all nuclear species. Instead, \skynet\ only evolves the entropy $s$ and
electron fraction $Y_e$ of the composition, which can change due to weak
reactions, such as $\beta$-decays or neutrino interactions, which can change the
charge of nuclides and heat the material. Recall that the electron fraction is
$Y_e = \sum_i Z_i Y_i$, and so
\begin{align}
\dot Y_e = \sum_i Z_i \dot Y_i,  \label{eq:dotYe}
\end{align}
where $\dot Y_i$ is given by \cref{eq:dotYi} as a function of $T$, $\rho$,
and $\vec Y$. The temperature is given by the \EOS{} as a
function of $s$, $\rho$, and $\vec Y$. $\vec Y$ is given by \NSE{} as a function
of $s$, $\rho$, and $Y_e$ (see \sref{sec:NSE_T}). Thus, we have
\begin{align}
\vec Y &= \text{NSE}(s,\rho,Y_e), \\
T &= \text{EOS}(s, \rho, \vec Y), \\
\dot {\vec Y} &= [\text{weak reactions}](T, \rho, \vec Y), \\
\dot Y_e &= \sum_i Z_i \dot Y_i.
\end{align}
The rate of change of the entropy is obtained from dividing \cref{eq:deltaS}
by $\Delta t$, namely,
\begin{align}
\dot s &= \frac{\dot\epsilon_\text{ext}-\dot\epsilon_\nu}{T} - \sum_i \dot Y_i
\left(\frac{\mathcal{M}_i}{T} +  \frac{Z_i\mu_{e^-}}{T} +
\ln\left[\frac{Y_in_B}{G_i(T)}\left(\frac{2\pi}{m_iT}\right)^{3/2}
\right]\right) \nn\\
&= \dot s(\dot\epsilon_\nu, \dot\epsilon_\text{ext}, T, \dot{\vec Y}, \vec Y,
\rho) = \dot s(\dot\epsilon_\nu, \dot\epsilon_\text{ext}, s, \rho, Y_e).
\label{eq:dotS}
\end{align}
Since $\dot\epsilon_\nu(t)$, $\dot\epsilon_\text{ext}(t)$, and $\rho(t)$ are
known, we thus have two coupled \ODE{s} for $Y_e$ and $s$, which \skynet\
integrates with the Runge--Kutta--Fehlberg 4(5) method
\citep[e.g.,][\S5.5]{burden:15}. This is a fourth-order explicit integration
method that also
computes a fifth-order error estimate that is used to adaptively control the
integration time step. The heating rate can be calculated analogously to
\cref{eq:heating} as
\begin{align}
\dot\epsilon = -\dot\epsilon_\nu + \dot\epsilon_\text{ext} - \sum_i \dot Y_i
\mathcal{M}_i.
\end{align}
Note that in the \NSE{} evolution mode, we only evolve two variables and they are
changing on similar timescales because they are both influenced by the weak
reactions. In this case, however, even though the weak reactions span a large
range of timescales, this does not introduce any stiffness because we only deal
with the sum of the abundance derivatives in both \cref{eq:dotYe,eq:dotS}. Thus,
we can safely use an explicit integration method.

\section{Electron screening}
\label{sec:screening}

Nuclear reaction rates strongly depend on
the Coulomb interaction between the nuclides in the entrance channel
\citep[e.g.,][]{salpeter:54}. In conditions where nuclear burning can occur, the
nuclides are almost all fully ionized. Therefore,
nuclear reaction rates can be computed assuming that bare nuclei of charge $Z_i$
interact with each other. At high temperature, the Coulomb interaction
energy of the electrons with the nuclei is small compared to the thermal
energies of the particles and so the screening effect is negligible. But
if the density of the medium is sufficiently high and the
temperature sufficiently low, electron screening is likely important. In
these conditions, a nucleus will repel neighboring nuclei and
attract nearby electrons, thus creating an electron charge cloud around the
nucleus. This charge cloud partially screens or shields the nuclear charge
$Z_ie$, where $e$ is the elementary charge unit. Thus, the Coulomb repulsion
between the two positively charged nuclei is reduced by the screening effect,
which can enhance the nuclear reaction rates, which depend strongly on the
probability of Coulomb barrier penetration. Obviously, screening
corrections are only important for charged particle reactions. Neutron
capture reactions are unaffected by the polarization of the electron gas.

In this section, we present how electron screening is implemented in \skynet.
Our focus will be on writing down the equations that \skynet\ uses to compute
the screening corrections in a useful way with adequate justification. We will
not develop the screening theory from first principles but refer the reader to
the established literature on this subject \citep[e.g.,][to name
but a few]{salpeter:54,dewitt:73,graboske:73,itoh:79,ichimaru:84,brown:97,
bravo:99, yakovlev:06}.
For a handful of reactions, screening has been investigated experimentally
\citep[e.g.,][]{engstler:88, rolfs:95, chen:04, gatu:17}.
We use Gaussian cgs units throughout this section.

The strength of the electron screening effect depends mainly
on the ratio of the Coulomb interaction energy between a nucleus and the nearby
electrons to the thermal energy. If the thermal energy is large compared to the
Coulomb interaction energy, then the electron charge cloud around the nucleus
will be large and diffuse, providing less screening to the nuclear charge.
We define the ion density
\begin{align}
  n_I = \sum_i n_i = \sum_i Y_i n_B = n_B \sum_i Y_i.
\end{align}
The average interionic spacing is
\begin{align}
a = \left(\frac{3}{4\pi n_I}\right)^{1/3}.
\end{align}
Now, we define the dimensionless screening parameter $\Lambda_0$ as
\begin{align}
\Lambda_0 &= \sqrt{4\pi n_I}e^3 \beta^{3/2}, \label{eq:lam0}
\end{align}
where $\beta = 1/T$. For some average (dimensionless)
charge per ion $\zeta$ (defined later on in \crefalt{eq:zeta}), let
\begin{align}
\Lambda = \zeta^3 \Lambda_0 = \left(\frac{3^{1/3}\zeta^2
e^2}{aT}\right)^{3/2}. \label{eq:Lambda}
\end{align}
Thus, $\Lambda$ is a measure of the ratio of the average Coulomb interaction
energy $\zeta^2e^2/a$ to the average thermal energy $T$. If $\Lambda$ is large,
we expect the screening effect to be strong, and when $\Lambda$ is small,
screening should be weak. The screening corrections in these two regimes, as
well as the intermediate regime where $\Lambda \sim 1$, are the subject of the
following sections, after we introduce the general screening factor that
modifies the nuclear reaction rate.

\subsection{General screening factor}

For the two-body reaction $[1] + [2] \to [3]$, the total Coulomb potential can
be written as
\begin{align}
U_\text{tot}(r_{12}) = \frac{Z_1Z_2 e^2}{r_{12}} + U(r_{12}),
\end{align}
where $U(r_{12})$ is a potential correction to the bare Coulomb interaction
between the two nuclei due to screening, $r_{12}$ is the
separation of the two reactants, and $Z_1$ and $Z_2$ are the charge numbers.
\citet{salpeter:54} showed that the screening correction to the nuclear
reaction rate $\lambda_{12}$ is given by
\begin{align}
\lambda_{12} = e^{-U_0/T} \lambda_{12,\text{no-sc}} =
f_\text{sc} \lambda_{12,\text{no-sc}},
\end{align}
where $U_0 = U(r_{12} = 0)$, $\lambda_{12,\text{no-sc}}$ is the unscreened
reaction rate, and $f_\text{sc} = e^{-U_0/T}$ is the general screening factor.
Note that $f_\text{sc}$ is sometimes written as $\exp(H_{12}(0))$ in the
literature. An approximation for the screening factor $f_\text{sc}$ can be
found in terms of $Z_1$, $Z_2$, density, temperature, and other quantities determined by the
composition. However, it is advantageous to (equivalently) write down the screening factor in
terms of a chemical potential correction $\mu_\text{sc}(Z)$ that depends on the
charge of a single
nucleus. This way, we can apply the screening corrections in \NSE{} as well, and
we are not constrained to only correcting reaction rates with two reactants. In
this section, we show how to compute the reaction screening factor if we have
the chemical potential correction $\mu_\text{sc}(Z)$. In the following
sections, we will show how to compute $\mu_\text{sc}(Z)$ in different screening
regimes.

For the same two-body reaction $[1] + [2] \to [3]$, \citet{dewitt:73}
found that the screening factor can also be written as
\begin{align}
f_\text{sc} = \exp\left(\beta\mu_\text{sc}(Z_1) +
\beta\mu_\text{sc}(Z_2)-\beta\mu_\text{sc}(Z_3)\right), \label{eq:H12}
\end{align}
where $\beta = 1/T$, $\mu_\text{sc}(Z)$ is the correction to the chemical potential for the
addition of a charge $Z$ to the system when the electron gas is non-uniform, and $Z_3 = Z_1 + Z_2$ is the charge of
the product. It is straightforward to generalize this result to an arbitrary
(i.e.\ not just two-body) reaction $\alpha$ as
\begin{align}
f_\text{sc} =
\exp\left(\sum_{i\in\mathcal{R}_\alpha}N_i^\alpha \beta\mu_\text{sc}(Z_i)
-\beta\mu_\text{sc}(Z_\mathcal{R}^\alpha)\right),
\label{eq:screening}
\end{align}
where $\mathcal{R}_\alpha$ is the set of reactant species of the reaction and
$N_i^\alpha$ is the number of species $[i]$ destroyed or produced in the
reaction (see \sref{sec:kinetic} and \crefalt{eq:reac_species}).
We also define
\begin{align}
Z_\mathcal{R}^\alpha = \sum_{i\in\mathcal{R}_\alpha} N_i^\alpha Z_i.
\end{align}
The above expression for the screening factor $f_\text{sc}$ for reactions with
an arbitrary number of particles in the entrance channel holds for multistep
reactions. This can be shown by breaking up the multibody reaction into two-body
reactions, e.g., treating $[1] +
[2] + [3] \to X$ as $[1] + [2] \to [12]$ followed by $[12] + [3] \to X$, and
then calculating the overall screening factor.
To justify the general form of $f_\text{sc}$, consider a two-body reaction that
produces an arbitrary set of products $\mathcal{P}_\alpha$, i.e.,
\begin{align}
[1] + [2] \to \sum_{j\in\mathcal{P}_\alpha} N_j^\alpha [j],
\end{align}
where $Z_1 + Z_2 = \sum_{j\in\mathcal{P}_\alpha} N_j^\alpha Z_j$ due to charge
conservation. Let $\lambda_{\alpha,\text{no-sc}}$ and
$\lambda_{\alpha',\text{no-sc}}$ be the reaction rates of the forward and
inverse reactions without the screening corrections. The corrected forward rate
is
\begin{align}
\lambda_{\alpha} =
\lambda_{\alpha,\text{no-sc}} \exp\left(\beta\mu_\text{sc}(Z_1) + \beta
\mu_\text{sc}(Z_2) - \beta\mu_\text{sc}(Z_1+Z_2)\right).
\end{align}
\Cref{eq:mu} gives the chemical potential of nuclear species $[i]$ without the
screening correction, which we call $\mu_{i,\text{no-sc}}$. The corrected
chemical
potential is $\mu_i = \mu_{i,\text{no-sc}} + \mu_\text{sc}(Z_i)$. Note that the
chemical potential correction $\mu_\text{sc}(Z_i)$ enters on the same level as
the term $m_i$ in \cref{eq:mu} and therefore, $\mu_\text{sc}(Z_i)$ can be
absorbed into $Q_\alpha$, the rest mass difference between the reactants and
products, defined in \cref{eq:Qa}. $Q_\alpha$ thus becomes
\begin{align}
Q_\alpha = Q_{\alpha,\text{no-sc}} + \sum_{i\in\mathcal{R}_\alpha} N_i^\alpha
\mu_\text{sc}(Z_i) - \sum_{j\in\mathcal{P}_\alpha} N_j^\alpha
\mu_\text{sc}(Z_j) = Q_{\alpha,\text{no-sc}} + \mu_\text{sc}(Z_1) +
\mu_\text{sc}(Z_2) - \sum_{j\in\mathcal{P}_\alpha} N_j^\alpha
\mu_\text{sc}(Z_j).
\end{align}
Substituting the above corrected expressions for $\lambda_\alpha$ and
$Q_\alpha$ into \cref{eq:detailed_balance} yields
\begin{align}
\lambda_{\alpha'} &= \lambda_{\alpha',\text{no-sc}}
\exp\left(\beta\mu_\text{sc}(Z_1) + \beta\mu_\text{sc}(Z_2) -
\beta\mu_\text{sc}(Z_1+Z_2)\right)
\exp\left(\sum_{j\in\mathcal{P}_\alpha} N_j^\alpha \beta\mu_\text{sc}(Z_j)
-\beta\mu_\text{sc}(Z_1) -\beta\mu_\text{sc}(Z_2)\right) \nonumber\\
&= \lambda_{\alpha',\text{no-sc}}
\exp\left(\sum_{j\in\mathcal{P}_\alpha} N_j^\alpha \beta\mu_\text{sc}(Z_j)
-\beta\mu_\text{sc}(Z_1+Z_2)\right) = \lambda_{\alpha',\text{no-sc}}
\exp\left(\sum_{j\in\mathcal{P}_\alpha} N_j^\alpha \beta\mu_\text{sc}(Z_j)
-\beta\mu_\text{sc}(Z_\mathcal{P}^\alpha)\right),
\end{align}
where we used charge conservation, i.e., $Z_1 + Z_2 =
\sum_{j\in\mathcal{P}_\alpha} N_j^\alpha Z_j = Z_\mathcal{P}^\alpha$. Thus, the
screening
factor for the inverse reaction (whose set of reactants is
$\mathcal{P}_\alpha$) is indeed exactly the generalized screening factor
postulated in \cref{eq:screening}. Thus, we only need to know how to compute
$\mu_\text{sc}(Z)$, which is the subject of the following sections. Note that in
terms of the screened forward rate, we combine the above with
detailed balance \crefp{eq:detailed_balance} to obtain the screened inverse rate
\begin{align}
\lambda_{\alpha'} &= \lambda_\alpha(T,\rho) \exp\left(
\sum_{j\in\mathcal{P}_\alpha} N_j^\alpha\beta\mu_\text{sc}(Z_j) -
\sum_{i\in\mathcal{R}_\alpha} N_i^\alpha\beta\mu_\text{sc}(Z_i)
\right) \nonumber\\
&\phantom{=} {}\times e^{-Q_\alpha/T}\Gamma_\alpha(T) M_\alpha^{3/2}
\left(\frac{T}{2\pi} \right)^{ 3\Delta N_\alpha/2}
(\rho N_A)^{-\Delta N_\alpha},
\end{align}
where we used $Z_\mathcal{P}^\alpha = Z_\mathcal{R}^\alpha$ since strong
reactions conserve charge.

It now remains to find the screening chemical potential correction
$\mu_\text{sc}(Z)$ for a nuclide of charge $Z$. Thanks to \skynet's modularity,
arbitrary expressions for $\mu_\text{sc}(Z)$ can be plugged into the general
screening framework. In the following sections, we describe the current
implementation of $\mu_\text{sc}(Z)$ in \skynet.

\subsection{Weak screening}

The weak screening limit is the limiting case where the Coulomb interaction
energy
is much lower than the thermal energy, hence, where $\Lambda \ll 1$. In this
case, the electrostatic Poisson--Boltzmann equation describing the screening can
be solved approximately to find \citep{salpeter:54}
\begin{align}
-\frac{U_0}{T} = \frac{Z_1Z_2e^2}{\lambda_D T}, \label{eq:weak_sc_H}
\end{align}
where $\lambda_D$ is the Debye screening length. $\kappa_D = \lambda_D^{-1}$ is
called the Debye wave number, and it is given by \citep{brown:97}
\begin{align}
\kappa_D^2 = \sum_i \kappa_{D,i}^2,
\end{align}
where the Debye wave number of species $[i]$ is
\begin{align}
\kappa_{D,i}^2 = 4\pi (Z_ie)^2\int \frac{d^3p}{(2\pi)^3}\frac{\partial
f_i(p,\mu_i)}{\partial \mu_i},
\end{align}
where $Z_ie$ is the charge, $\mu_i$ is the chemical potential (defined in
\crefalt{eq:mu}), and $f_i(p,\mu_i)$ is the distribution function of species
$[i]$.

\skynet\ assumes that the ions (nuclides) are non-degenerate and
non-relativistic. Therefore, they obey Boltzmann statistics with $E(p) =
p^2/(2m_i) + m_i$, where $p$ is the momentum and $m_i$ is the mass of species
$[i]$. Hence,
\begin{align}
f_i(p,\mu_i) = \exp\left(-\beta(E(p) - \mu_i)\right) = \exp\left(\beta\mu_i -
\frac{p^2\beta}{2m_i}-\beta m_i\right),
\end{align}
and so
\begin{align}
\int \frac{d^3p}{(2\pi)^3}\frac{\partial f_i(p,\mu_i)}{\partial \mu_i} = \int
\frac{d^3p}{(2\pi)^3} \beta f_i(p,\mu_i) = \beta n_i,
\end{align}
and hence
\begin{align}
\kappa_{D,i}^2 = 4\pi Z_i^2e^2\beta n_i,
\end{align}
where $n_i$ is the number density of ion species $[i]$.

The electrons and positrons are allowed to be arbitrarily degenerate and arbitrarily
relativistic. They are both fermions and thus follow Fermi-Dirac statistics, so
\begin{align}
f_{e^\pm}(p,\mu_{e^\pm}) = \frac{1}{\exp\left(\beta(E(p) - \mu_{e^\pm})\right)
+ 1}, \label{eq:fepm}
\end{align}
where $E(p)$ is the total energy
\begin{align}
E(p) = \sqrt{m_e^2 + p^2}.
\end{align}
We find
\begin{align}
\frac{\partial f_{e^\pm}(p,\mu_{e^\pm})}{\partial \mu_{e^\pm}} =
\frac{\beta \exp\left(\beta(E(p) - \mu_{e^\pm})\right)}{[\exp\left(\beta(E(p) -
\mu_{e^\pm})\right) + 1]^2} = \beta f_{e^\pm} (1-f_{e^\pm}),
\end{align}
and so
\begin{align}
\kappa_{D,e^\pm}^2 = 4\pi e^2 \beta\int \frac{d^3 p}{(2\pi)^3} 2 f_{e^\pm}
(1-f_{e^\pm}),
\end{align}
where the extra factor of 2 in the phase-space integral comes from the fact
that fermions have two spin states, and we used $Z_{e^\pm} = \pm 1$.

The Debye length is thus
\begin{align}
\lambda_D = \kappa_D^{-1} = \sqrt{\frac{1}{4\pi e^2\beta n_I \zeta^2}},
\end{align}
where
\begin{align}
\zeta^2 = \frac{n_B}{n_I} \left[\sum_i Z_i^2 Y_i + \frac{2}{n_B}\int
\frac{d^3p}{(2\pi)^3}\left(f_{e^-}(1-f_{e^-}) +
f_{e^+}(1-f_{e^+})\right)\right]. \label{eq:zeta}
\end{align}
Note that the factor of $n_B/n_I$ in $\zeta$ occurs because we
define $\zeta$ as the RMS charge per ion rather than per nucleon.
The distribution functions $f_{e^\pm}$ \crefp{eq:fepm} depend on the electron
and positron chemical potentials $\mu_{e^\pm}$, which are given by the \EOS{}
(\namecrefs{eq:mue}~\ref{eq:mue} and \ref{eq:mup}). \skynet\ uses the adaptive
QAG integration routines provided by the GNU Scientific Library\footnote{\url{%
https://www.gnu.org/software/gsl/manual/html_node/Numerical-Integration.html}}
to evaluate the above integral numerically.
The two-body screening potential in \cref{eq:weak_sc_H} becomes
\begin{align}
-\frac{U_0}{T} = \frac{Z_1Z_2e^2}{\lambda_D T} = Z_1Z_2\zeta e^3\sqrt{4\pi
n_I} \beta^{3/2} = Z_1Z_2\zeta\Lambda_0,
\end{align}
and so the screening factor is
\begin{align}
f_\text{sc} = e^{-U_0/T} = \exp(-Z_1Z_2\zeta\Lambda_0). \label{eq:f2}
\end{align}
This is consistent with the result from \citet{dewitt:73}, who showed that in
the weak screening case, the chemical potential correction due to screening is
\begin{align}
\beta\mu_\text{sc,weak}(Z) = -\frac{1}{2}Z^2\zeta \Lambda_0. \label{eq:mu_weak}
\end{align}
Combining the above with the expression for the general screening factor
\crefp{eq:screening} yields
\begin{align}
f_\text{sc} &= \exp\left(\beta\mu_\text{sc,weak}(Z_1) +
\beta\mu_\text{sc,weak}(Z_2) -\beta\mu_\text{sc,weak}(Z_1 + Z_2)\right) \nn\\
&= \exp\left(-\frac{1}{2}\zeta\Lambda_0 \left[Z_1^2 + Z_2^2 - (Z_1 +
Z_2)^2\right]\right) \nn\\
&= \exp(-Z_1Z_2\zeta\Lambda_0),
\end{align}
as shown in \cref{eq:f2}.

\subsection{Strong and intermediate screening}

In the strong screening limit, the Poisson--Boltzmann equation governing
screening has to be solved numerically. \citet{dewitt:73} found that
\begin{align}
\beta\mu_\text{sc,strong}(Z) = -\frac{Z}{\bar
Z}\left(Z^{2/3}\bar Z^{4/3} \Gamma_0\left(c_0+c_1\left(\frac{\bar
Z}{Z}\right)^{1/3}+c_2\left(\frac{\bar Z}{Z}\right)^{2/3}\right) + d_0 +
d_1\left(\frac{\bar Z}{Z}\right)^{1/3}\right), \label{eq:strong_screening1}
\end{align}
where $c_0 = 9/10$, $c_1 = 0.2843$, $c_2 = -0.054$, $d_0 = -9/16$, and $d_1 =
0.4600$. The parameter $\Gamma_0$ is
\begin{align}
\Gamma_0 = \Lambda_0^{2/3}/3^{1/3},
\end{align}
and the applicable average charge per ion is the arithmetic mean $\bar Z$,
instead of the RMS $\zeta$, which is given by
\begin{align}
  \bar Z = \frac{\sum_i Z_iY_i}{\sum_i Y_i}.
\end{align}
We can write \cref{eq:strong_screening1} in terms of $\Lambda_0$ as
\begin{align}
\beta\mu_\text{sc,strong}(Z) = -\Lambda_0^{2/3} \left(0.6240\, Z^{5/3}\bar
Z^{1/3} + 0.1971\, Z^{4/3}\bar
Z^{2/3} - 0.0374\,Z\bar Z\right) + \frac{9}{16}\frac{Z}{\bar Z} -
0.4600\left(\frac{Z}{\bar Z}\right)^{2/3}, \label{eq:mu_strong}
\end{align}
which is valid when $\Gamma = \bar Z^2\Gamma_0 \gg 1$. Alternatively, if $\zeta
\sim \bar Z$, then $\Gamma \sim \Lambda^{2/3}$ and so the strong screening limit
applies if $\Lambda \gg 1$. The strong screening limit is also applicable if
$\Lambda \ll 1$ but the charge $Z$ is such that $Z\zeta^2\Lambda_0 \gg 1$. Then
strong screening applies for charge $Z$ \citep{dewitt:73}.

The intermediate screening regime is where
$\Lambda = \zeta^3\Lambda_0 \sim 1$, in which case \citet{dewitt:73} found
\begin{align}
\beta\mu_\text{sc,intermediate}(Z) = -0.380\,\Lambda_0^b \eta_b Z^{b+1},
\label{eq:mu_int}
\end{align}
where $b = 0.860$ and
\begin{align}
\eta_b = \frac{\sum_i Z_i^{3b-1}n_i/n_I}{\zeta^{3b-2}\bar Z^{2-2b}} =
\frac{n_B}{n_I}\frac{\sum_i Z_i^{3b-1}Y_i}{\zeta^{3b-2}\bar Z^{2-2b}}.
\end{align}

\subsection{Combining the different screening regimes}

\Cref{eq:mu_weak,eq:mu_strong,eq:mu_int} give the chemical
potential corrections in the limits of weak, strong, and intermediate screening,
respectively. To smoothly transition between these three regimes, we need a
single parameter that determines which regime is applicable and a function that
smoothly interpolates between the regimes based on the said parameter.

Weak screening applies if $\Lambda = \zeta^3\Lambda_0 \ll 1$ and strong
screening if $\Lambda \gg 1$ or $\Lambda \ll 1$ but $Z\zeta^2\Lambda_0 =
\Lambda Z/\zeta \gg 1$. We thus define the dimensionless parameter
\begin{align}
p(Z) = \Lambda + \Lambda\frac{Z}{\zeta} = (\zeta + Z)\zeta^2\Lambda_0.
\end{align}
Note that $p(Z) \ll 1$ if and only if $\Lambda \ll 1$ and $\Lambda Z/\zeta \ll
1$, in which case weak screening applies. If and only if $\Lambda \gg 1$ or
$\Lambda Z/\zeta \gg 1$ is $p(Z) \gg 1$, in which case we are in the strong
screening regime. If $p(Z) \sim 1$, then intermediate screening applies.
To ensure a smooth transition of $\beta\mu_\text{sc}$ from one regime to
another, we will compute the screening correction as a weighted sum of the
corrections computed in the different regimes. Hence, we compute
\begin{align}
  \beta\mu_\text{sc}(Z) = f_w \beta\mu_\text{sc,weak}(Z) + f_s
\beta\mu_\text{sc,strong}(Z) + f_i \beta\mu_\text{sc,intermediate}(Z),
\label{eq:mu_screen}
\end{align}
where each coefficient $f_j$ is between 0 and 1 and defined as
\begin{align}
f_w(p) &= \frac{1}{2}\left[\tanh\left(-2\ln p - \ln25\right) + 1\right],
\\
f_s(p) &= \frac{1}{2}\left[\tanh\left(2\ln p - \ln25\right) +
1\right], \\
f_i(p) &= \frac{1}{2}\left[\tanh\left(2\ln p + \ln25\right) +
\tanh\left(-2\ln p + \ln25)\right)\right].
\end{align}
Note that $f_w(p) + f_s(p) + f_i(p) = 1$ for all values of $p$. The transition
from weak to intermediate screening occurs at $p(Z) = 1/\sqrt{25} = 0.2$,
and the transition from intermediate to strong screening occurs at $p(Z) = \sqrt{25}
= 5$.

\begin{figure}
\centering
\includegraphics[width=\singlecol]{%
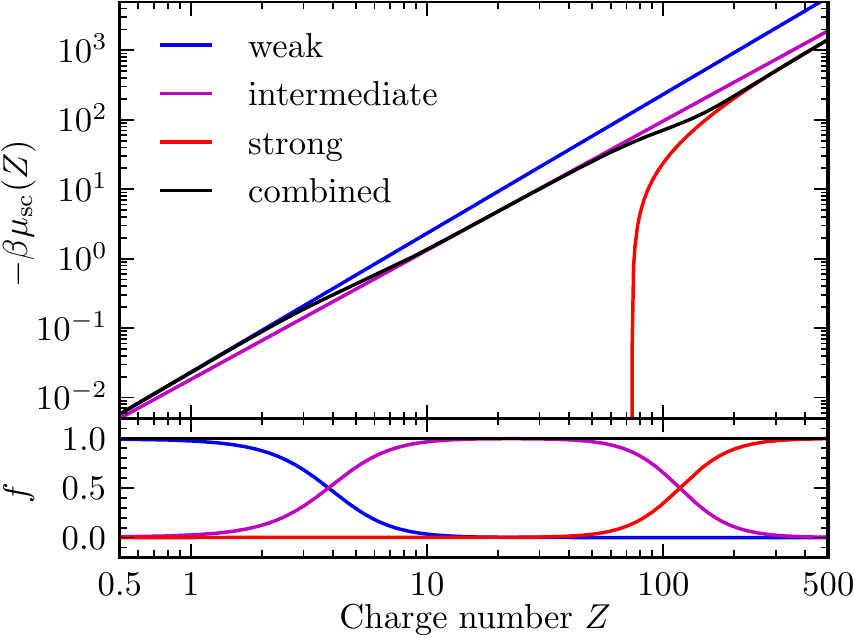}
\caption{Screening correction to the chemical potential for a test particle
with charge $Z$. The upper panel shows the corrections from the weak,
intermediate,
and strong regimes, as well as the combined correction. The bottom panel shows
the functions $f_w$, $f_i$, and $f_s$ that are used to weight the chemical
potential corrections due to weak, intermediate, and strong screening. These
screening corrections are computed at $T = 3$~GK and $\rho = 2 \times 10^9\
\text{g cm}^{-3}$ with a composition consisting of 59\% neutrons, 40\% protons,
and 1\% \smce{U^{238}} (by mass).}
\label{fig:screen_vs_Z}
\end{figure}

\Cref{fig:screen_vs_Z} demonstrates the transition between the different
screening regimes. We choose a composition consisting of 59\% neutrons, 40\%
protons, and 1\% \smce{U^{238}} (by mass) at $T = 3$~GK and $\rho = 2 \times
10^9\ \text{g cm}^{-3}$. In this composition, all three screening regimes occur
as the charge number of a test particle ranges from $Z = 1$ to $Z \sim 100$. In
order to show
the transitions more clearly, we plot the chemical potential corrections from
the different regimes for $Z$ ranging from 0.5 to 500 in
\cref{fig:screen_vs_Z}. Such charge numbers are not expected to occur in
reality, but the expressions for the chemical potential corrections are valid
nonetheless. The chosen composition results in $\Lambda_0 = 0.05089$, $\zeta =
0.9104$, $\bar Z = 0.4079$, $\eta_b = 0.6214$, and $\Lambda = 0.03840$.

\begin{figure}
\centering
\includegraphics[width=\singlecol]{%
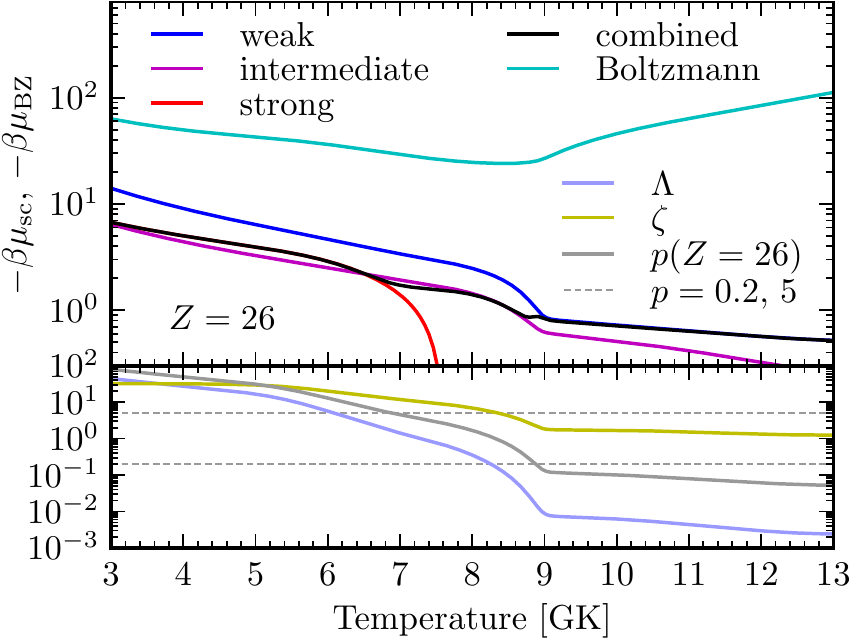}
\caption{Screening correction to the chemical potential as a function of
temperature for a fixed charge $Z = 26$. The composition is computed from
\NSE{} with screening corrections with the given temperature, $\rho = 10^8\
\text{g cm}^{-3}$, and $Y_e = 0.4$. The upper panel shows the screening
corrections from the weak, intermediate, and strong screening regimes, as well
as the combined correction. The Boltzmann chemical potential $\mu_{BZ}$ of
\smce{^{56}Fe} (\crefalt{eq:mu} without the rest mass) is also shown for
comparison. The lower panel shows the screening parameter $\Lambda$
\crefp{eq:Lambda} and average charge per ion $\zeta$ \crefp{eq:zeta} computed
from the composition, and the $p(Z)$ parameter used to transition between the
different regimes for $Z = 26$. The dashed gray lines show the values of $p$ at
which the transitions from strong to intermediate screening ($p = 5$) and from
intermediate to weak screening ($p = 0.2$) happen.}
\label{fig:screen_vs_T}
\end{figure}

Whereas \cref{fig:screen_vs_Z} shows the screening correction for different ion
charges in a fixed composition, \cref{fig:screen_vs_T} shows the screening
correction for a fixed charge ($Z = 26$) in different compositions determined
by the temperature $T$. Also shown
is the Boltzmann chemical potential (\crefalt{eq:mu} without the rest mass) for
\smce{^{56}Fe}. The composition is determined from \NSE{} with screening
corrections (\sref{sec:nse_screen}) using a temperature ranging from 3 to 13~GK,
\smash{$\rho = 10^8\ \text{g cm}^{-3}$}, and $Y_e = 0.4$. At low temperatures
($T \lesssim 7$~GK), strong screening is applicable since the thermal energy is
small and Coulomb interactions dominate. Electron screening provides a 10\%
correction over the unscreened Boltzmann chemical potential. Intermediate
screening is applicable between 7 and 9~GK. At 9~GK, \nhe\
nuclei are broken up and free neutrons and protons start to dominate the
composition. Thus, $\zeta$ remains roughly constant at 1, and so $\Lambda$ and
$p(26)$ also stop changing rapidly because $\Lambda_0$ does not depend strongly
on the temperature. $T \sim 9$~GK is also where weak screening sets in, because
the thermal energy now overcomes the Coulomb interaction energy, which has been
reduced by the smaller average charge per ion provided by the free protons. In
that regime, the screening effect is at the 4\% to 0.5\% level.

\Cref{fig:screen_vs_T} also shows that our screening corrections are indeed
approximate. The weak and intermediate screening corrections never meet, so any
scheme to transition between them is necessarily approximate and arbitrary to
some degree. But considering the fact that our transition scheme needs to work
robustly for a wide range of compositions and ion charges, it seems to do
reasonably well in interpolating between the different (somewhat disjoint)
screening regimes. Although progress has been made in improving screening
calculations in various regimes, a unifying theory for screening across all
regimes is still elusive \citep[e.g.,][]{itoh:77, shaviv:96, shaviv:00,
chugunov:07}.

\subsection{NSE with screening} \label{sec:nse_screen}

As was noted above, in addition to screening nuclear reactions, electronic
correlations also change the free energetic cost of adding or removing a charged
particle from the medium. Therefore, the chemical potentials of the nuclides have an
additional correction due to screening, i.e.,
$\mu_i = \mu_{i,\text{no-sc}} + \mu_\text{sc}(Z_i)$, where
$\mu_{i,\text{no-sc}}$ is the unscreened chemical potential given by
\cref{eq:mu} and $\mu_\text{sc}(Z_i)$ is the screening correction given by
\cref{eq:mu_screen} with $Z_i$ being the charge number of nuclide $i$.
\aref{app:calc_NSE} presents in detail how \NSE{} is computed. Here, we briefly
discuss how the method shown in \aref{app:calc_NSE} is modified to take chemical
potential corrections into account.
\cref{eq:muhat} becomes
\begin{align}
\hat\mu_i = \mu_{i,\text{no-sc}} + \mu_\text{sc}(Z_i) - m_i - \text{BE}_i,
\end{align}
which means \cref{eq:nse_Y} gives
\begin{align}
Y_i = e^{\eta_i - \beta\mu_\text{sc}(Z_i) + \beta\text{BE}_i}
\frac{G_i(T)}{n_B} \left(\frac{m_i T}{2\pi}\right)^{3/2}, \label{eq:nse_Y_scr}
\end{align}
where $\beta = 1/T$. Note that $\mu_\text{sc}(Z_i)$ depends on all $Y_i$
because the screening corrections depend on different types of average ion
charges ($\bar Z$, $\zeta$, and $\eta_b$).
Computing \NSE{} involves an \NR{} iteration that requires partial derivatives
of $Y_i$ (see \aref{app:calc_NSE}).
Thus, the screening corrections in \cref{eq:nse_Y_scr} introduce a large number of complicated partial
derivatives to the Jacobian, which we will not attempt to write down. We
experimented with using numerical derivatives to compute the Jacobian, with
limited success. Another complication is that \cref{eq:nse_Y_scr} itself
depends on all $Y_i$ on the right-hand side and thus has to be solved
iteratively.

We find that it is much more robust to keep the screening corrections fixed
during the \NR{} iterations. This does not introduce any additional derivatives
in the Jacobian. \NSE{} is computed exactly as shown in \aref{app:calc_NSE},
with the only difference that \cref{eq:nse_Y_scr} is used to compute
$Y_i$ from $\eta_i$, but the terms $\beta\mu_\text{sc}(Z_i)$ are constant
throughout the \NR{} iterations. In this case, to obtain an \NSE{} composition
that is self-consistent with the screening corrections it is based on, the
\NSE{} computation itself needs to be iterated.

We start by computing the
\NSE{} composition without screening (i.e., $\mu_\text{sc}(Z_i) = 0$). We
denote the resulting composition by \smash{$\vec Y^{(0)}$}. Then, we compute
$\mu_\text{sc}(Z_i)$ based on \smash{$\vec Y^{(0)}$} and use these as the
constant screening corrections for the next \NSE{} computation that yields
\smash{$\vec Y^{(1)}$}. From this new composition we compute new screening
corrections, which are used to compute \smash{$\vec Y^{(2)}$} and so on. This
iteration stops once \smash{$\max_i(|Y_i^{(n+1)} - Y_i^{(n)}|) < 10^{-12}$} or
$n$ reaches 20 (both of these criteria can be changed by the user). This method
of iteratively updating the screening corrections and computing \NSE{} with them
being fixed is not guaranteed to converge (but neither is the \NR{} method
itself). However, in practice, we find that this method works very well and
converges quite quickly in a large region of parameter space.

\begin{figure*}
\centering
\includegraphics[width=\linewidth]{%
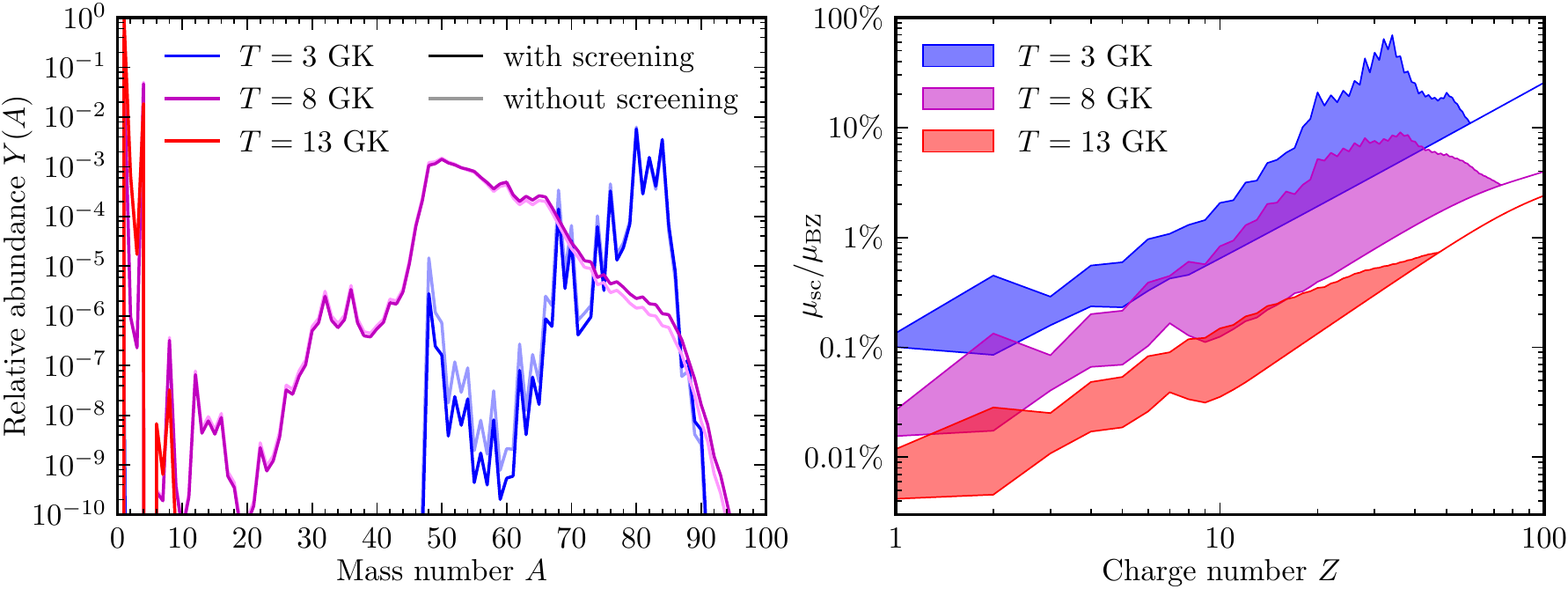}
\caption{\NSE{} compositions with and without screening at different
temperatures. In all cases, $\rho = 10^8\ \text{g cm}^{-3}$ and $Y_e = 0.4$.
\textbf{Left:} Abundances as a function of mass number $A$ of the compositions
with and without screening. Screening pushes the abundance distribution to
slightly higher masses. The effect at $T = 3$~GK is clearly visible. At $T =
8$~GK the screening effect is quite small and at $T = 13$~GK it is practically
absent. \textbf{Right:} The ratio of the screening chemical potential
correction to the Boltzmann chemical potential (\crefalt{eq:mu} without the rest
mass). The bands show the range of this ratio for all isotopes with the same
charge number $Z$. The screening correction can be as large as 70\% in the $T =
3$~GK case, and as low as 0.004\% for $T = 13$~GK. At high $Z$ where the bands
collapse, all isotopes of the same element have the same screening to Boltzmann
chemical potential ratio because their abundances are all extremely small.}
\label{fig:nse_screen}
\end{figure*}

\Cref{fig:nse_screen} shows the \NSE{} abundance distribution as a function of
mass number $A$ for three different temperatures. For all temperatures, $\rho =
10^8\ \text{g cm}^{-3}$ and $Y_e = 0.4$, so this is the same composition as the
one shown in \cref{fig:screen_vs_T}. To show the impact of the screening
corrections, the \NSE{} compositions with and without screening are shown in
the left panel. Screening is strongest for $T = 3$~GK and the effect of
screening is to reduce the abundances below $A \sim 80$ and slightly increase
them above that mass number. Screening is weaker at $T = 8$~GK, but still
enhances the high-mass abundances above $A \sim 60$. Finally, for $T = 13$~GK,
screening has virtually no effect. In the right panel of \cref{fig:nse_screen},
we show the ratio of the screening chemical potential $\mu_\text{sc}$ to the
Boltzmann chemical potential $\mu_\text{BZ}$ (\crefalt{eq:mu} without the rest
mass). $\mu_\text{sc}$ depends on the composition and the charge, hence it is
the same for all isotopes of a given element. But the Boltzmann chemical
potential also depends on the mass, partition function, and the abundance of a
given isotope. Thus, for a given charge number $Z$, the ratio
$\mu_\text{sc}/\mu_\text{BZ}$ varies for the isotopes of that charge. In
\cref{fig:nse_screen}, we show the range of the chemical potential ratios as
colored bands. The bands collapse to a line at large $Z$, because there the
abundances of all isotopes are essentially zero. In the strongest screening
case ($T = 3$~GK), the screening effect ranges from 0.1\% to 70\%. For $T =
8$~GK, it ranges from about 0.02\% to 10\%, and for $T = 13$~GK, the screening
effect is much less than 1\% except for very large $Z$.

\section{Implementation details}
\label{sec:impl}

The main design goals of \skynet\ are usability and flexibility.  Since \skynet\ is built in
a modular fashion, different physics implementations can easily be switched out
or new physics can be added, making \skynet\ very flexible (see the next sections
for details). To achieve the modularity, \skynet\ is written in object-oriented
C++ and makes use of some C++11 features. \skynet\ contains a small amount of
Fortran code to provide a minimal interface for \skynet\ to be called from
Fortran. CMake (\url{http://www.cmake.org}) provides a
cross-platform, compiler-independent build system for \skynet\ that
automatically finds the required external libraries. CMake also provides an
automated testing facility and \skynet\ comes with a suite of tests that check
basic functionality and correctness of \skynet.

To make it easy to use, \skynet\ comes with Python bindings that make it
possible to use all parts of \skynet\ from an interactive Python shell or a
Python script. Therefore, one can use standard Python libraries like
\emph{NumPy} (\url{http://www.numpy.org}) to read in and manipulate input
data, like the list of nuclides to
be evolved, the initial composition, density vs.\ time history, etc., and these
data can be passed to \skynet\ using standard Python data structures. This
means that one does not have to deal with C++ to run \skynet. But of course,
\skynet\ can also be used from a C++ or Fortran application. One can even run
multiple copies of \skynet\ in parallel within Python, using Python's
\texttt{multiprocessing} module---a facility we use extensively whenever
post-processing nucleosynthesis on many tracer particles form
hydrodynamical simulations \citep[e.g.,][]{lippuner:17a, roberts:16b}. An example of how
to run \skynet\ in parallel with Python is included with the \skynet\ source
code available at \skyneturl. The \skynet\ Python bindings are provided by
SWIG (\url{http://www.swig.org}) and using
Python is the most convenient and most flexible way to run \skynet.

\subsection{Modularity and extendability}

\skynet\ is a modular library of different C++ classes rather than a monolithic
program. Some of the most important classes in \skynet\ are the various reaction
library classes that contain different types of nuclear reactions (see next
section), a nuclide library class that contains all nuclear data, and a reaction
network class that implements the actual nuclear reaction network. There are
various other types of classes that implement specific functionalities. For
example, there are different function interpolation classes, \ODE{}
integrators, and general numerical method classes (bisection, line search). On
the physics side, there are different classes that are responsible for
different pieces of physics. The \NSE{} class computes \NSE{} given an
electron fraction and two of the following properties: temperature, density,
entropy, or internal energy. There are also separate classes that are responsible
for the \EOS{} and screening corrections.

Since \skynet\ is built in an object-oriented fashion, different parts of the
code are separated from each other and only interact via well-defined
interfaces. This makes \skynet\ extremely modular because the implementation
of a certain class can be changed or extended, without having to modify the
rest of the code. For example, one could easily extend the nuclide library
class to support reading nuclear data from a different file format. Since
all of the nuclear data are handled by this one class, only this class has to be
modified to support the new file format. Furthermore, some classes are
implemented as abstract base classes, meaning they only specify the interface
for a particular physics module without tying it to a specific implementation.
Examples of this are the \EOS{} class and the screening corrections class. For
both
of these, \skynet\ currently has one implementation, namely the extended Timmes
\EOS{} described in \sref{sec:eos} and the screening corrections discussed in
\sref{sec:screening}. One can easily add a new \EOS{} class that implements a
different \EOS{} but has the same interface as the abstract \EOS{} base class.
This new \EOS{} class then plugs into the existing \skynet\ framework. In a
similar way, one can add additional screening implementations to \skynet.

The various classes provided in \skynet\ can be used individually through the
Python bindings. For instance, one can use the \NSE{} class in Python to
compute \NSE{} in various conditions or use the nuclide library class to
access the nuclear data and partition functions from Python.

\subsection{Nuclear reaction libraries}
\label{sec:reac_types}

\skynet\ supports different types of nuclear reactions. Reactions of the same
type or from the same data source are grouped into reaction library classes.
The network class contains an arbitrary list of reaction library classes that
collectively contain all of the reactions that are evolved in the network. The
reaction library classes have a common interface that allows the network to be
agnostic as to how the reaction rate is determined. Via this interface, the
network can tell the reaction libraries to recompute the reaction rates for a
given thermodynamic state (temperature, density, electron fraction, electron
degeneracy parameter, etc.) to get the contributions to all \smash{$\dot Y_i$}
from the reactions in the network and to get the contributions to
\smash{$\partial \dot Y_i/\partial Y_j$}. This makes \skynet\ extremely
flexible because many different types of reactions can be evolved at the same
time, and furthermore, the data for reactions of the same type can be split
across multiple files, allowing the user to quickly switch out certain
reactions. Finally, thanks to the abstract interface of reaction library
classes, it is easy to add new types of reactions to \skynet.

\pagebreak[4]

The following reaction types of nuclear reactions are currently implemented in
\skynet.

\begin{itemize}
\item \textbf{Constant:} These reactions have a constant rate that does not
depend on any properties of the thermodynamic state.
\item \textbf{REACLIB:} These are reactions that come from the REACLIB database
\citep{cyburt:10}. The rates of these reactions are given by parametric
fitting formulae that depend on temperature and density.
\item \textbf{Tabulated:} This reaction library contains
tabulated $\beta$-decay (both $\beta^-$ and $\beta^+$) and electron/positron
capture rates based on \citet{fuller:82} and \citet{langanke:01}. The rates
are tabulated as a function of temperature and $Y_e\rho$.
\item \textbf{Neutrino interactions:} These are neutrino emission and
absorption reactions on free neutrons and protons. The rates are calculated
according to \cref{eq:lec,eq:lpc,eq:lnue,eq:lnuebar} given the electron
neutrino and electron antineutrino distribution functions.
\item \textbf{Arbitrary rate functions:} This reaction library contains
reactions whose rates are given by arbitrary, user-specified functions. This
can be used to quickly test a new or modified reaction rate that can depend on
various thermodynamic quantities and also time.
\end{itemize}

Since the different reaction types and rate sources can be used concurrently in
\skynet, care must be taken to ensure that no reaction rate is contained
multiple times in the network, since that would effectively multiply the
reaction rate by the number of times it occurs. \skynet\ provides a facility to
remove all reactions in one reaction library that also occur in another library.
However, in some cases, there are supposed to be multiple rates for the same
reaction. In this case, the total reaction rate is the sum of the individual
rates. REACLIB uses this mechanism to capture the different resonant and
non-resonant parts of a reaction rate with its limited fitting formula.

\section{Code verification and tests}
\label{sec:tests}

In order to verify the correctness of \skynet, we compare its results to the
results of other existing reaction network codes, specifically \winnet\
\citep{winteler:13} and \xnet\ \citep{hix:99}, and also to results published
in the literature. The scripts and input files to reproduce these code tests
are distributed with \skynet\ in the directory \code{examples/code\_tests}.
\skynet\ also has a test suite that contains simple code tests, regression
tests, and tests that compare very simple networks to analytic solutions. The
main purpose of that test suite is to ensure that changes to the code do not
break the functionality or correctness of \skynet.

\subsection{Nuclear statistical equilibrium} \label{sec:nse_test}

To verify the \NSE{} solver in \skynet, we perform a consistency test and
comparison to literature results of the \NSE{} abundances computed by \skynet.
We use nuclear masses and partition functions distributed with REACLIB
\citep{cyburt:10}, which contains experimental data where available and
finite-range droplet macroscopic model \citep[FRDM; see, e.g.,][]{moller:16}
data otherwise. The temperature-dependent partition functions are from
\citet{rauscher:00}.

\subsubsection{Consistency test}
\label{sec:nse_consistency}

In this section, we verify that the abundances computed with the \NSE{} solver
in \skynet\ are consistent with the strong reactions. We perform a test
evolution starting with purely free neutrons and protons and let only strong
reactions take place. We use the strong reaction rates from REACLIB and the
default fission rates distributed with \skynet. The network contains 7824
nuclear species, ranging from free neutron and protons to \smce{^{337}Cn} ($Z =
112$). We keep the temperature constant at $T = 7$~GK and the
density $\rho = 10^9\ \text{g cm}^{-3}$ is also fixed. We pick these values to
ensure that the composition achieves \NSE{} within a reasonable amount of time.
We set $Y_e = 0.4$, so
the initial composition is $Y_\text{n} = 0.6$ and $Y_\text{p} = 0.4$, and since
the network does not contain any weak reactions, the electron fraction remains
constant at $Y_e = 0.4$. Screening corrections are enabled for both the \NSE{}
solver and the network evolution. At every step, we compare the network
abundances to the \NSE{} abundances and compute the error $\Delta Y$ as
\begin{align}
\Delta Y = \max_i \left|Y_i^\text{network} - Y_i^\text{NSE}\right|.
\end{align}
We perform two
different network evolutions: one where the strong inverse rates are computed
from detailed balance (\sref{sec:inverse_rates}) and another one where the
inverse rates from REACLIB are used. We never use inverse fission reactions.

\begin{figure}
\centering
\includegraphics[width=\singlecol]{%
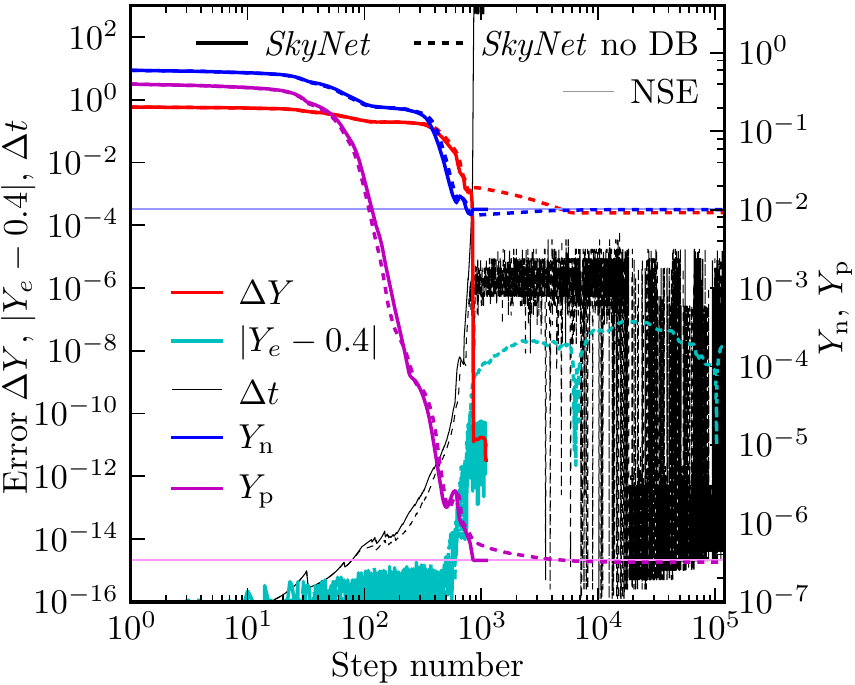}
\caption{Evolving pure neutrons and protons to \NSE{} with a fixed temperature
$T = 7$~GK, density $\rho = 10^9\ \text{g cm}^{-3}$, and $Y_e = 0.4$. The
evolution only includes strong reactions, because weak reactions would change
the electron fraction. The horizontal lines show the
neutron and proton
abundances of the \NSE{} composition. The \skynet\ evolutions are done with
screening correction turned on. The solid lines are the quantities from the
\skynet\ evolution where the inverse rates are computed from detailed balance,
while the dashed lines are from \skynet\ without detailed balance, i.e.,
the inverse rates are taken from REACLIB. The error $\Delta Y$ is the maximum
abundance difference between the network and the \NSE{} composition. We stop
the
network evolution without detailed balance after 130,000
time steps, corresponding to 0.053~s of physical time, because the time
step $\Delta t$ remains very small and the error $\Delta Y$
appears to converge to a $\text{few}\times 10^{-4}$. In contrast, the evolution
with detailed balance only requires about 850 steps ($\sim 1$~s physical time)
to reach $\Delta Y \sim 10^{-11}$. We stop that evolution after 1100 steps
when it reached $t = 10^{10}$~s.}
\label{fig:nse_consistency}
\end{figure}

The results are shown in \cref{fig:nse_consistency}. After evolving the
network using detailed balance to compute the inverse
rates for 850 time steps, corresponding to $t = 1$~s, the network reaches
the \NSE{} composition. The error between \NSE{}
and the network composition is $\Delta Y \sim 10^{-11}$ and the
deviation of $Y_e$ from 0.4 is on the same level. We note that this is
comparable to the mass conservation limit of $10^{-10}$ that \skynet\ uses as
the \NR{} iteration convergence criterion. The neutron and proton
abundances also match the values from the \NSE{} composition with very high
precision. In the subsequent 250 time steps, the network reaches $t =
10^{10}$~s and $\Delta Y$ decreases another order of magnitude. This
demonstrates that the \NSE{} solver in \skynet\ and the
implementation of detailed balance for the inverse rates are consistent. The
\NSE{} compositions computed with \skynet\ are indeed the compositions that the
network produces if the strong reactions are in equilibrium.

Furthermore, \cref{fig:nse_consistency} also shows that the inverse rates
provided in
REACLIB are not completely consistent with the \NSE{} composition that is
computed from the nuclear data (masses and partition functions) distributed
together with REACLIB. If the REACLIB inverse rates are used (``\skynet\ no
DB'' in \cref{fig:nse_consistency}), the network
evolution is extremely slow after about 900 steps. We stopped the network
evolution after 130,000 steps at $t \sim 0.053$~s, when it became clear
that $\Delta Y$ converged to
$2.5\times 10^{-4}$. The
evolution without detailed balance becomes very slow because the inverse rates
from REACLIB try to push the composition into a certain equilibrium
configuration, but chemical potential balance predicts a different equilibrium
composition. This makes the evolution very difficult and
keeps the time step between $10^{-15}$ and $10^{-5}$~s. Computing the
inverse rates from detailed balance so
that the inverse rates exactly cancel the forward rates when the chemical
potentials balance is therefore necessary for the network
evolution to be consistent with \NSE{} (see \sref{sec:inverse_rates}).

A detailed investigation of the reverse rates in REACLIB reveals that a
significant fraction of them were computed with nuclear masses different from
the mass model distributed with REACLIB, which we use in \skynet\ to compute the
inverse rates from detailed balance and to compute \NSE{}.
We observe this for the latest REACLIB version (REACLIB V2.2 from 2016-11-14 and
also newer pre-release version that is not publicly available yet) as well as
previous REACLIB versions. For 74\% of the
inverse rates, the $Q$ values used in the REACLIB inverse rates differ by 1\%
or less compared to the $Q$ values computed from our mass model \crefp{eq:Qa}.
About 19\% of the $Q$ values agree within 1\% to 10\%, 5\% within 10\% to 30\%, and
almost all the $Q$ values agree within a factor of 2. However, there are some
rare cases where the $Q$ value used in REACLIB differs by up to two orders of
magnitude from the one given by the mass model used in \skynet. Fortunately,
the impact of the discrepancy between the reverse REACLIB rates and the mass
model on the final abundances of a network evolution is very small, as we
shall see in \sref{sec:test_net_evolution}. Nevertheless, the test presented in
this section highlights the advantage of using detailed balance with the mass
model used in the network to compute inverse rates. This is especially
important to keep the time step from becoming very small when the composition
moves into \NSE{}. Some of the inconsistencies in the inverse rates will be
corrected in a future REACLIB version \citep{schatz_private:17}. However,
because REACLIB is a collection of rates from different sources that use
different mass models to compute the rates and their inverses, it seems
unlikely that all inverse rates in REACLIB will ever be consistent with a
single mass model. We therefore recommend to always directly compute inverse
rates from detailed balance with the masses used in the network when consistency
with \NSE{} is desired.

\subsubsection{Comparison with published results}

\begin{figure*}
\centering
\includegraphics[width=\linewidth]{%
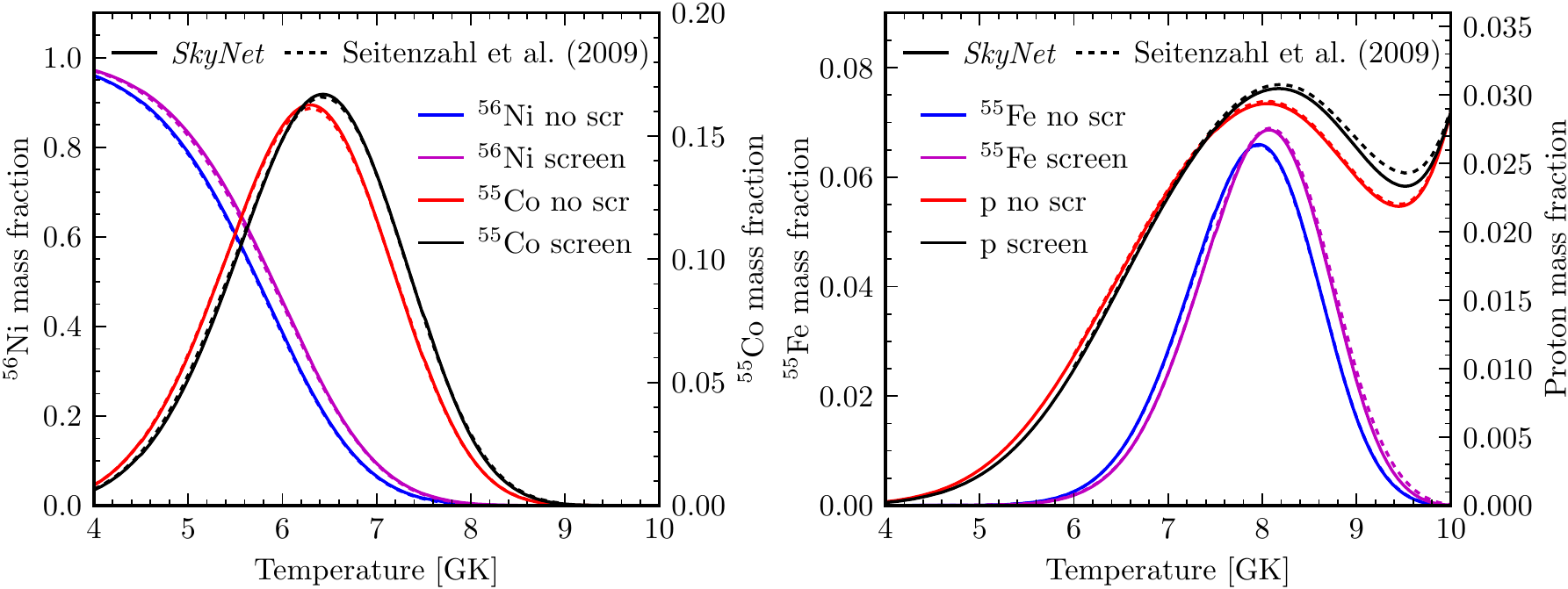}
\caption{\smce{^{56}Ni}, \smce{^{55}Co}, \smce{^{55}Fe}, and proton mass
fractions in an \NSE{} composition with varying temperatures and fixed $\rho =
5\times 10^8\ \text{g cm}^{-3}$ and $Y_e = 0.5$. The composition includes 443
species and \NSE{} is computed with and without screening corrections. The
results are compared to those published in \citet[][data used with permission]{seitenzahl:09}.
We see excellent agreement between \skynet\ and the published
results. The differences are slightly enhanced when screening is turned on, due
to the different screening implementations.}
\label{fig:test_nse}
\end{figure*}

We compare the \skynet\ \NSE{} solver to the \NSE{} results with and without
electron screening by \citet{seitenzahl:09}. The composition includes 443
nuclides ranging from free neutrons and protons to multiple isotopes of krypton
\citep[see Figure~1 in][]{seitenzahl:09}. We compute \NSE{} with and without
screening at a fixed density of $\rho = 5 \times 10^8\ \text{g cm}^{-3}$ and
fixed $Y_e = 0.5$ for temperatures ranging from 4 to 10~GK. The results are
shown in \cref{fig:test_nse} along with the results from
\citet[][Figures 3 to 6, data used with permission]{seitenzahl:09}.

We find excellent agreement between the \skynet\ results and those presented in
\citet{seitenzahl:09}. When screening corrections are included, the deviation
between the two results is slightly larger, because \citet{seitenzahl:09} use a
different screening implementation. The effect of the different screening
implementation is most pronounced for the proton mass fraction at $T \gtrsim
9$~GK, but even then, the difference is less than 5\%. \citet{seitenzahl:09}
use a fit for the screening corrections that has a significant correction even
in the weak screening regime at $T \gtrsim 8$~GK. On the other hand, the
weak screening correction in \skynet\ becomes much smaller at those
temperatures. Hence, the differences in the \NSE{} mass fractions are due to the
increased disagreement between the different screening corrections as the
temperature increases. Furthermore, \citet{seitenzahl:09}
use different nuclear masses, so that could account for the small differences
between our results and theirs when screening is turned off.

\subsection{Network evolution}
\label{sec:test_net_evolution}

In this section, we present comparisons of nucleosynthesis evolutions with
\skynet\ and other nuclear reaction networks. We compare \skynet\ to \winnet\
and \xnet. \winnet\ was originally developed at the University of Basel by
\citet{winteler:13} based on the earlier \emph{BasNet} by
\citet{thielemann:11}. \winnet\ has been used by various authors for r-process
nucleosynthesis calculations in core-collapse supernovae and neutron star
mergers, and to investigate the impact of nuclear physics on the r-process
\citep[e.g.,][]{korobkin:12b,winteler:12,eichler:15, martin:15,martin:16}.
\xnet\ was developed at Oak Ridge National Laboratories by \citet{hix:99} and
has been used for r-process nucleosynthesis in accretion disk outflows and
neutron star mergers, and for explosive nucleosynthesis in type I X-ray bursts
and core-collapse supernovae \citep[e.g.,][]{surman:06,fisker:08,roberts:11,
harris:17}.

Since nuclear physics data, such as nuclear masses, partition functions, and
nuclear reaction rates, have a significant influence on the nucleosynthesis
calculations, we take care in ensuring that exactly the same nuclear physics
input data are used for all of the different codes that we consider. However, this
means that we are restricted to using the greatest common denominator of
nuclear physics data sources that can be used by all codes. For the comparisons
in this section, we use the strong and weak reaction rates distributed in
REACLIB \citep{cyburt:10}, neutron-induced fission reactions with symmetric
fission fragments from \cite{panov:10}, and spontaneous fission rates
calculated from the approximation of \cite{frankel:47} using the spontaneous
fission barriers of \cite{mamdouh:01}. In future versions of \skynet, we
plan to add additional fission reactions and fission fragment distributions.
The nuclear masses and partition
functions are again the ones distributed with REACLIB, as in
\sref{sec:nse_test}.

\subsubsection{Neutron-rich r-process}
\label{sec:test_r-process}

We run an r-process nucleosynthesis calculation in a neutron-rich environment
with all three networks. We use 7836 nuclear species and about 93,000
reactions. The density history is a trajectory from the ejecta of a black
hole--neutron star
merger \citep{roberts:16b}. The initial composition is \NSE{} with $T =
6.1$~GK, $\rho = 7.4 \times 10^{9}\ \text{g cm}^{-3}$, and $Y_e = 0.07$. We run
all combinations of screening and self-heating turned on and off. For each
case, we perform two separate \skynet\ evolutions: one where the inverse rates
are computed from detailed balance, and another where the inverse rates
from REACLIB are used. We consider these two cases because \skynet\ is usually
run with inverse rates computed from detailed balance, but \winnet\ and \xnet\
use the inverse rates from REACLIB, and so we also run \skynet\ with those
inverse rates for a more direct comparison.

The self-heating method currently implemented in \xnet\ only applies in the
case of constant density \citep{harris_private:17}. Hence, for this r-process
computation with an evolving density, we cannot use the self-heating capability
of \xnet.
Instead, to compare \xnet\ to \skynet\ and \winnet\ when self-heating is turned
on, we use the \skynet\ temperature history (from the \skynet\ run without
detailed balance) in \xnet. However, the temperature provided to \xnet\ has to
be limited to a lower bound of 0.01~GK. At temperatures lower than that, the
reaction rate fits from REACLIB no longer apply and some of the rates blow up.
\skynet\ internally also uses a lower bound 0.01~GK for the REACLIB reactions,
but the network temperature used in the \EOS{} is allowed to drop below that
bound, until the lower limit of the \EOS{} is reached at around $4 \times
10^5$~K (this lower bound is due to the tabulated electron/positron part of the
Timmes \EOS{}).

\begin{figure*}
\centering
\includegraphics[width=\linewidth]{%
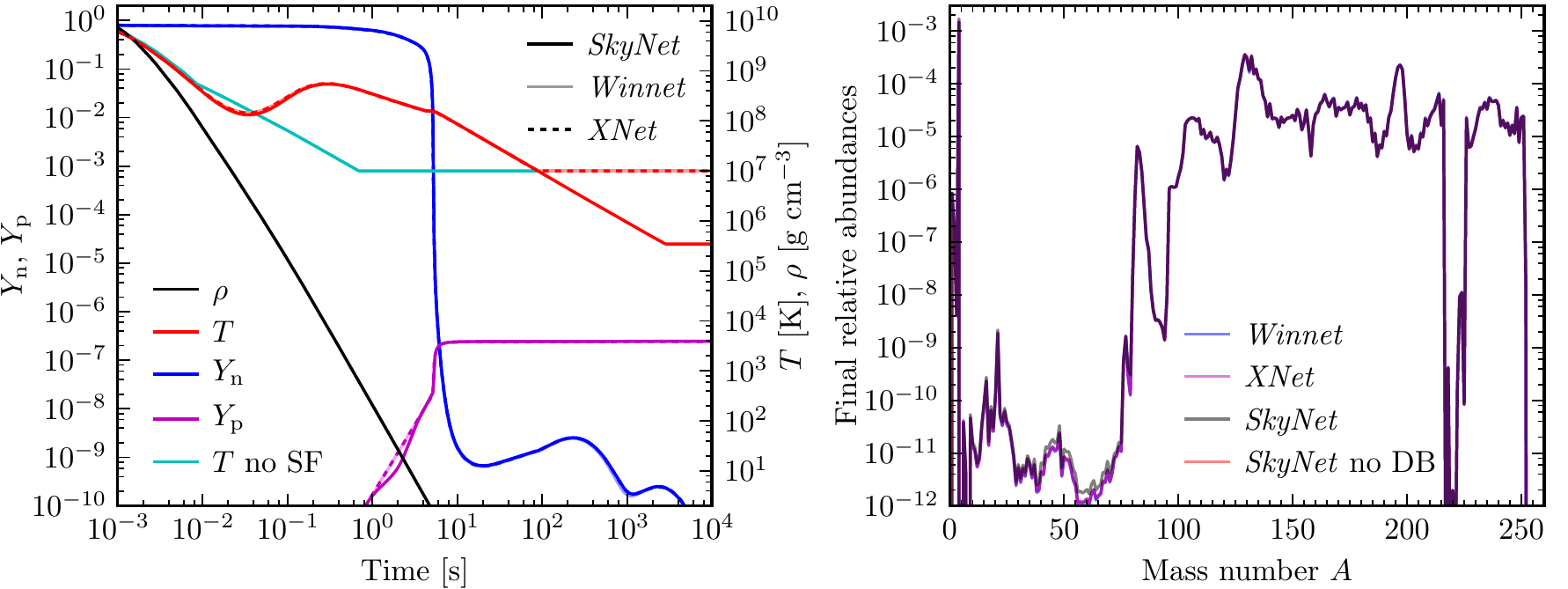}
\caption{Neutron-rich r-process calculation with three different reaction
networks: \skynet, \winnet, and \xnet. Screening corrections are turned on.
\skynet\ and \winnet\ evolve the temperature, but \xnet's temperature was fixed
to that computed by \skynet\ (with a lower bound of 0.01~GK).
\textbf{Left:} Prescribed density $\rho$ as a function of time, resulting
temperature $T$, neutron abundance $Y_\text{n}$, and proton abundance
$Y_\text{p}$. For comparison, the temperature without self-heating (SF) from
the trajectory is also
shown to illustrate the importance of self-heating. The solid dark lines show
the \skynet\ results, the solid light lines are the \winnet\ results, and the
dashed lines are the \xnet\ results. The three networks agree extremely well
with each other, with the lower temperature floor in \skynet\ being the largest
difference. The small deviation in \smash{$Y_\text{p}$} at $t \sim 2$~s is
because \skynet\ uses detailed balance to compute the inverse rates. Detailed
balance also accounts for the small temperature difference around $t = 0.01 -
0.1$~s. \textbf{Right:} Final
abundances as a function of mass number $A$ after $5\times 10^8$~s. Showing two
\skynet\ results: with detailed balance (DB) and without. We again see
excellent agreement between the networks, and the small differences around $A =
50$ are again because of detailed balance. \skynet\ without detailed balance
matches \winnet\ and \xnet\ in that region.}
\label{fig:test_r-process}
\end{figure*}

\newcommand{\0}{\phantom{0}}
\newcommand{\p}{\phantom{.}}

\begin{table}
\caption{Errors between the final mass-summed abundances between the different
networks \crefp{eq:net_error}.}
\label{tab:test_networks}
\centering
\begin{tabular}{@{}lllcccccc@{}}
\hline
Test case & Screening & Self-heating &  S -- SnoDB & S -- W & SnoDB -- W  &
S -- X & SnoDB -- X & X -- W \\
\hline
Neutron-rich r-process                                 & yes & yes/no$^\ast$ & \03.2\0  & \05.8\0  & 3.4\0 & \03.2\0  & 0.066\0\0    & 3.4\0  \\
(\sref{sec:test_r-process}, \cref{fig:test_r-process}) & yes & no            & 33\p\0\0 & 35\p\0\0 & 4.5\0 & 32\p\0\0 & 1.8\0\0\0\0  & 3.3\0  \\
                                                       & no  & yes/no$^\ast$ & \03.4\0  & \05.9\0  & 3.4\0 & \03.4\0  & 0.042\0\0    & 3.4\0  \\
                                                       & no  & no            & 33\p\0\0 & 36\p\0\0 & 3.3\0 & 33\p\0\0 & 0.10\0\0\0   & 3.2\0  \\
\noalign{\smallskip}
Explosive X-ray burst                                  & yes & yes/no$^\ast$ & \00.39   & \02.3\0  & 2.1\0 & \01.3\0  & 1.1\0\0\0\0  & 1.0\0  \\
(\sref{sec:test_xray}, \cref{fig:test_xray})           & yes & no            & 13\p\0\0 & 14\p\0\0 & 4.0\0 & 11\p\0\0 & 3.5\0\0\0\0  & 3.3\0  \\
                                                       & no  & yes/no$^\ast$ & \00.38   & \02.3\0  & 2.0\0 & \00.47   & 0.086\0\0    & 2.0\0  \\
                                                       & no  & no            & 17\p\0\0 & 18\p\0\0 & 1.5\0 & 18\p\0\0 & 1.2\0\0\0\0  & 0.40   \\
\noalign{\smallskip}
Hydrostatic C/O burn                                   & yes & yes           & \00.64   & \00.68   & 0.57  & \03.9\0  & 4.1\0\0\0\0  & 3.9\0  \\
(\sref{sec:test_co_burn}, \cref{fig:test_co_burn})     & yes & no            & \00.10   & \00.22   & 0.18  & \00.16   & 0.22\0\0\0   & 0.13   \\
                                                       & no  & yes           & \00.68   & \01.5\0  & 1.9\0 & \04.0\0  & 4.4\0\0\0\0  & 2.5\0  \\
                                                       & no  & no            & \00.10   & \00.27   & 0.17  & \00.10   & 0.00036      & 0.17   \\
\hline \\[-4pt]
\multicolumn{9}{@{}p{0.86\textwidth}@{}}{\textbf{Notes.} S: \skynet\ with detailed balance, SnoDB: \skynet\ without detailed
balance, W: \winnet, X: \xnet. The error measures the average fractional
difference between the final abundances. The numbers shown are in percent. The three
networks generally agree very well with each other. The error between \skynet\
without detailed balance (SnoDB), \winnet\ (W), and \xnet\ (X) are usually of
similar magnitude and on the few percent level. We also see that using
detailed balance for the inverse rates has a big impact in the first two test
cases, especially when self-heating is turned off. Since \winnet\ and \xnet\ do
not use detailed balance, the error is bigger when they are compared to
\skynet\ with detailed balance (S).} \\
\multicolumn{9}{@{}p{0.86\textwidth}@{}}{$^\ast$ Self-heating is
turned on in \skynet\ and \winnet, but not in \xnet, because its self-heating
method does not apply in these test cases. Instead, \xnet\ uses the
temperature computed by the self-heating \skynet\ run without detailed balance
(SnoDB).} \\
\end{tabular}
\end{table}

\Cref{fig:test_r-process} shows the results of running the neutron-rich
r-process with the three reaction networks with screening and self-heating
turned on. We find excellent agreement between the different networks with the
temperature evolution of \skynet\ and \winnet\ being virtually
indistinguishable (for \xnet, we prescribe the \skynet\ temperature). The final
abundances (at $t = 5 \times 10^8$~s) also agree very well. To compare the
results of the three networks quantitatively, we compute the numeric error
between the final mass-summed abundances as
\begin{align}
\text{error} = \frac{\displaystyle\sum_{A=1}^{A_\text{max}}
\left|Y_\text{net1}(A) - Y_\text{net2}(A)\right|}
{\displaystyle\sum_{A=1}^{A_\text{max}} \frac{Y_\text{net1}(A) +
Y_\text{net2}(A)}{2}}, \label{eq:net_error}
\end{align}
where the mass-summed abundance $Y(A)$ is given by
\begin{align}
Y(A) = \sum_{\substack{i\ \text{where}\\A_i = A}} Y_i.
\end{align}
This error measure is the average absolute difference between the abundance
results divided by the average abundances. This effectively measures the
fractional error in the final abundances, averaged over all mass numbers $A$.
However, we compute the quotient of the sums rather than the sum of quotients,
because the latter would be dominated by tiny abundances \smash{($\sim
10^{-20}$)}
that may differ by a factor of several between the two networks. This would
result in a large overall error, but abundance differences at the
\smash{$10^{-20}$}
level are not important, even if it is by a factor of several.

\Cref{tab:test_networks} shows the errors in percent between the different
networks. Since this is a neutron-rich environment, we expect that screening
plays no important role in the nucleosynthesis evolution. The fact that the
errors between the
different networks are almost the same regardless whether screening is turned
on or off confirms that screening is not important in this case. Furthermore,
the error between \skynet\ without detailed balance and
\winnet\ or \xnet\ are comparable to the errors between \winnet\ and \xnet. For
example, with screening and self-heating turned on (first row in
\cref{tab:test_networks}), the error between \skynet\ without detailed balance
and \winnet\ is 3.4\%, while the error between \winnet\ and \xnet\ is 3.4\%,
and the error between \xnet\ and \skynet\ without detailed balance is 0.066\%.
\skynet\ is closer to \xnet\ in this case because \xnet\ uses the same
temperature as \skynet\ whereas \winnet\ evolves its own temperature.
This demonstrates that \skynet\ produces results that are compatible with
\winnet\ and \xnet. The errors between \skynet\ and \winnet\ or \xnet\ are
larger if detailed balance is used to compute the inverse rates in \skynet. For
example, again for the first row in \cref{tab:test_networks}, the error between
\winnet\ and \skynet\ with detailed balance is 5.8\%, but only 3.4\% when
compared to \skynet\ without detailed balance.
This is not surprising, because \skynet\ is effectively
evolving slightly different reaction rates when inverse rates are computed
from detailed balance. It does illustrate, however, that
using detailed balance, which produces inverse rates that are consistent
with the nuclear masses and partition functions, has a measurable impact and
might be a reasonable standard practice. In the self-heating runs, \skynet\ is
much
closer to \xnet\ than \winnet, because \xnet\ uses the temperature from \skynet.

\subsubsection{X-ray burst}
\label{sec:test_xray}

\begin{figure*}
\centering
\includegraphics[width=\linewidth]{%
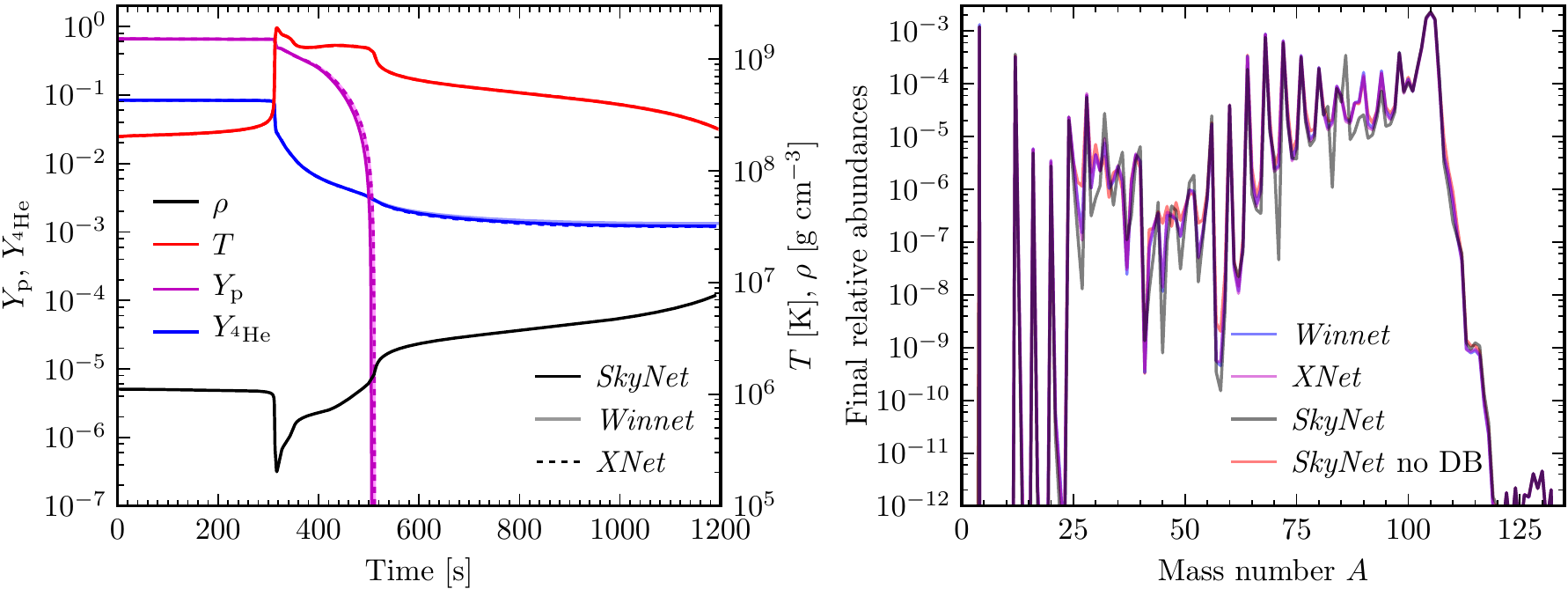}
\caption{Explosive nucleosynthesis in an X-ray burst with three different
reaction networks: \skynet, \winnet, and \xnet. Screening corrections are
included, but self-heating is turned off for the computations shown in this
figure so that the impact of screening can be presented. We use the temperature
and density history
from \citet{schatz:01}. \textbf{Left:} Prescribed density $\rho$ and
temperature $T$ as a function of time, resulting proton abundance $Y_\text{p}$,
and helium abundance $Y_{^4\text{He}}$. All three networks agree very well with
each other. \textbf{Right:} Final abundances as a
function of mass number $A$ at $t = 1242.6$~s. Showing two \skynet\
results: with detailed balance (DB) and without. \skynet\ with detailed balance
produces final abundances at some specific values of $A$ that differ by up to
two orders of magnitude from the other networks. This shows the importance of
using detailed balance to compute inverse rates.}
\label{fig:test_xray}
\end{figure*}

In \cref{fig:test_xray}, we show a comparison between the three networks for a
different type of trajectory. This trajectory captures the situation of
unstable hydrogen burning on the surface of a neutron star, which produces a
type I X-ray burst, presented in \citet{schatz:01}. The density and temperature
histories as well as the initial and final abundances were graciously provided
by \citet{schatz:01}. The temperature starts at 0.2~GK and peaks at 1.9~GK
during the burst. The density starts at $1.1\times 10^6\ \text{g cm}^{-3}$. The
initial composition is 66.0\% hydrogen (by mass), 33.6\% helium, and 0.4\%
heavier elements, mostly oxygen.
For this test, we use a small network containing
only 686 species going up to \smce{^{136}Xe} and 8400 reactions.

We again see good agreement in \cref{fig:test_xray} between the
three networks in the proton and helium abundance evolution (left panel of
\cref{fig:test_xray}). But there are a handful of mass numbers at which
\skynet\ with detailed balance produces final abundances that deviate from the
other networks by up to two orders of magnitude (right panel of
\cref{fig:test_xray}). This indicates that it is vital to compute the inverse
reaction rates correctly with detailed balance in
this scenario. However, the average fractional abundance errors shown in
\cref{tab:test_networks} are still at the few percent level, and the differences
between \skynet\ and \winnet\ or \xnet\ are comparable to the differences
between \winnet\ and \xnet. We note that the errors between \skynet\
without detailed balance and \winnet\ or \xnet\ are smaller if screening is
turned off. This indicates that screening is somewhat important in this case
and we expect the discrepancy between the different codes to increase if
screening is turned on due to the different screening implementations. The
effect is especially noticeable when comparing \skynet\ to \xnet\ with
self-heating turned on, because in this case \xnet\ uses the temperature
computed from \skynet, but if screening is switched off, the error between the
two codes decreases from 1.1\% to 0.086\%.

\skynet's screening implementation is presented in \sref{sec:screening}.
\winnet\ uses a single fit of the two-body screening factor by
\citet{chugunov:07} across all screening regimes. \xnet\ computes the two-body
screening function provided by \citet{graboske:73} for weak and intermediate
screening and the one provided by \citet{dewitt:99} for strong
screening. \xnet\ then uses a selection rule to select one of the three
screening regimes without interpolating between them.

\pagebreak

\subsubsection{Hydrostatic carbon-oxygen burning}
\label{sec:test_co_burn}

\begin{figure*}
\centering
\includegraphics[width=\linewidth]{%
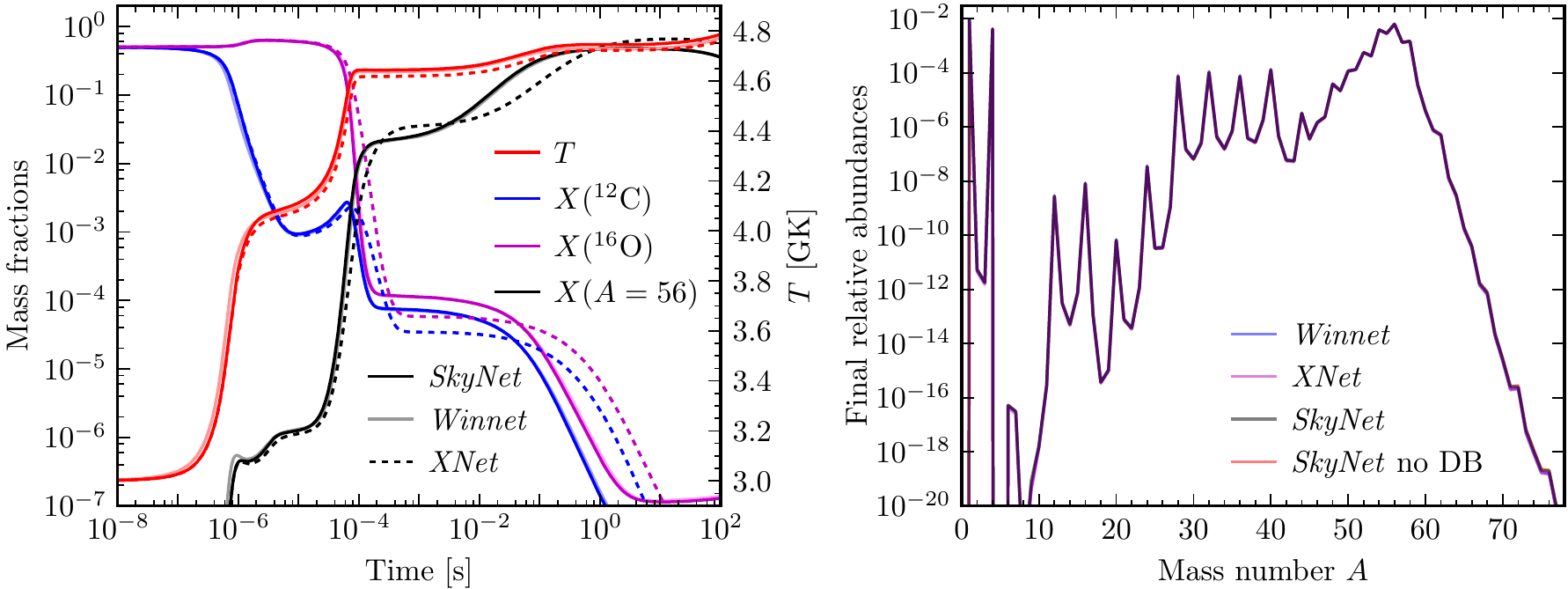}
\caption{Hydrostatic burning of carbon and oxygen with three different
reaction networks: \skynet, \winnet, and \xnet. The density is fixed at $\rho
= 10^7\ \text{g cm}^{-3}$ and the evolution starts with $T = 3$~GK and half
\nc, half \no\ (by mass). Screening and self-heating
are enabled in all three networks. \textbf{Left:} Resulting temperature
$T$, \nc\ and \no\ mass fractions, and the sum of the mass
fractions of all isotopes with $A = 56$. We clearly see the difference in
the self-heating implementations. \skynet\ and \winnet\
evolve the entropy and compute the temperature from it, which takes
composition changes into account. \xnet\ evolves the temperature directly
solely based on the released nuclear binding energy, which becomes small after
$t \sim 10^{-4}$~s and so the temperature stops changing in \xnet.
\textbf{Right:} Final abundances as a function of mass number $A$ at $t =
100$~s. Showing two \skynet\ results: with detailed balance (DB) and without.
All codes produce extremely similar results.}
\label{fig:test_co_burn}
\end{figure*}

In order to compare the self-heating methods implemented in the three networks,
we perform a hydrostatic burn at constant density. We keep $\rho = 10^7\
\text{g cm}^{-3}$ fixed and start with $T = 3$~GK and the initial composition
consists of half \nc\ and half \no\ (by mass). As a baseline
comparison, we also perform non-self-heating runs, where we keep the
temperature fixed at 3~GK. The increase in the errors between the networks
with self-heating enabled compared to with it disabled must then be due to the
difference in the self-heating implementations in the codes. We use a mid-size
network containing all nuclides from the full network
(\sref{sec:test_r-process}) with $A \leq 100$. This
results in a network with 1530 species and 20,000 reactions.

\Cref{fig:test_co_burn} shows the results of this test case with
self-heating and screening turned on. We find good qualitative agreement
between the three networks, but quantitatively, the differences between \xnet\
and the other two codes are much larger
than in the previous test cases. These discrepancies come from the different
self-heating implementations in the three codes. \winnet\ uses the same
self-heating method as \skynet\ (described in \sref{sec:self-heating}), but
\winnet\ does not include the entropy change due to the electron chemical
potential. Also, \winnet\
uses the original Timmes \EOS{}, which computes the entropy with a single
representative heavy ion species, while \skynet\ computes the entropy by
considering all species in the network separately (\sref{sec:eos}).
However, these two differences have virtually no impact on the temperature
evolution or the final abundances produced by \winnet\ and \skynet.
\xnet, on the other hand, uses a different self-heating method. It
evolves the temperature directly using \citep{harris_private:17}
\begin{align}
\frac{dT}{dt} = \frac{\dot \epsilon_\text{nuc}}{c_V} = -\frac{1}{c_V}\sum_i \dot
Y_i \mathcal{M}_i,
\end{align}
where $c_V$ is the specific heat capacity at constant volume (provided by the
Timmes \EOS{}), $\dot Y_i = dY_i/dt$ is the abundance time derivative of
species $i$, and $\mathcal{M}_i$ is the mass excess. As can be seen in the left
panel of \cref{fig:test_co_burn}, this method is comparable to the methods in
\skynet\ and \winnet. Note that \xnet\ is using exactly the same \EOS{} as
\winnet. We see only very small differences in the final abundances of the
three networks in the right panel of \cref{fig:test_co_burn}.

The differences in the self-heating implementations are also apparent
in \cref{tab:test_networks}. When self-heating is turned on, \xnet\ differs
from \skynet\ and \winnet\ by about 4\%, and \winnet\ differs from
\skynet\ by about 1\%. However, when
self-heating is disabled, the three networks agree at the 0.2\% level, and if
screening is turned off, \skynet\ without detailed balance and \xnet\ agree to
an astounding precision of 0.0004\%.

\subsection{NSE evolution test}
\label{sec:test_nse_evolution}

To ensure that the \NSE{} evolution mode in \skynet\ produces the correct
results, we perform a test that evolves a trajectory with and without the \NSE{}
evolution mode. Of course, the trajectory must experience some heating that
forces the composition into \NSE{} at some point during the evolution,
otherwise the \NSE{} evolution mode
would not be triggered. We first attempt this test with a trajectory from a
neutron star merger accretion disk outflow simulation \citep{lippuner:17a}.
That trajectory experiences late-time fallback, which causes a spike in the
density that results in late-time heating and forces the composition into
\NSE{}. Although \skynet\ is able to evolve this trajectory without issues using
the \NSE{} evolution mode, when the \NSE{} evolution mode is turned off, \skynet\
gets stuck with a time step of $\sim 10^{-16}$~s for at least 350,000 steps at
the time when the heating occurs. Thus, we
cannot use this trajectory for this test, since we cannot evolve it without the
\NSE{} evolution mode. However, this trajectory serves as an illustration of
the necessity of the \NSE{} evolution mode in order to evolve certain
trajectories.

\begin{figure*}
\centering
\includegraphics[width=\linewidth]{%
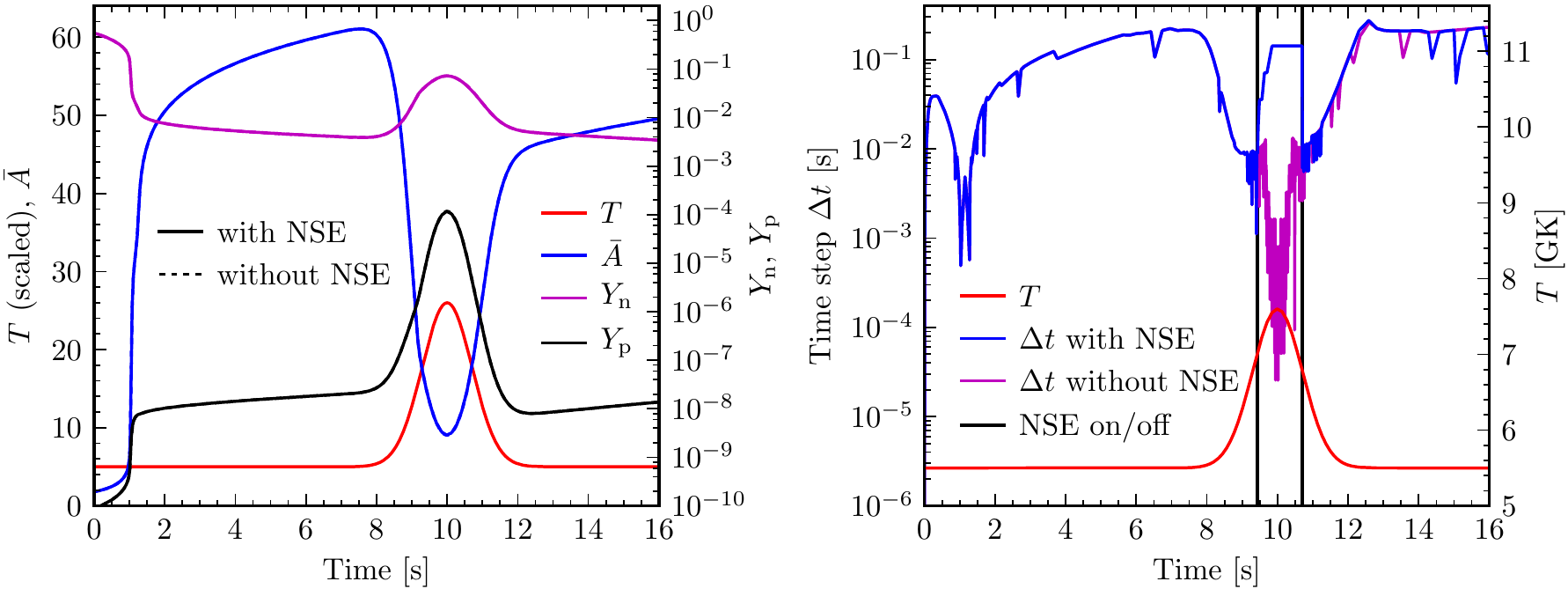}
\caption{Test of the \skynet\ \NSE{} evolution mode. \skynet\ is evolved with a
constant density of $\rho = 10^8\ \text{g cm}^{-3}$ and a constant temperature
of 5.5~GK except for a bump up to 7.6~GK at 10~s. Self-heating and screening
corrections are turned off and inverse rates are computed from detailed
balance. The initial composition is \NSE{} with $Y_e = 0.1$. \textbf{Left:} The
temperature $T$, average mass number $\bar A$, and neutron and proton
abundances $Y_\text{n}$ and $Y_\text{p}$ as a function of time. In the first
second or so, neutrons are captured onto seed nuclei and heavy nuclei around
the first r-process peak are synthesized with $\bar A$ increasing to about 60.
Then, at 8~s as the temperature begins to rise, the heavy nuclei are destroyed
again as the material is being forced into \NSE{}. This increases $Y_\text{n}$
and $Y_\text{p}$ and the average nuclear mass drops to around 10. As the
temperature returns back to 5.5~GK, the iron peak is formed and $\bar A$
increases to 56 at $T\sim 50$~s, eventually settling around 54. Only the first
16 seconds are shown to highlight the region around $t = 10$~s. We see
absolutely no differences between the evolution with and without the \NSE{}
evolution mode. \textbf{Right:}
The imposed temperature $T$ and resulting time-step size $\Delta t$ as a
function of time showing that using the \NSE{} evolution mode allows \skynet\ to
evolve more quickly. The vertical black lines indicate the times at which
\skynet\ decides to turn the \NSE{} evolution mode on and off. As the temperature
increases, the composition moves into \NSE{} and the time step decreases
because the reaction rates become faster. With the \NSE{} evolution mode, \skynet\
can turn off the strong reactions that are all in equilibrium now and hence do
not change the composition. The composition is only evolved under the influence
of weak reactions that change the electron fraction and entropy. Without the \NSE{}
evolution mode, the time step keeps decreasing to a $\text{few}\times 10^{-5}$,
until the temperature decreases again.}
\label{fig:test_nse_evolution}
\end{figure*}

Since it is challenging to evolve a trajectory without the \NSE{} evolution mode
that moves into \NSE{} during the evolution, we use an artificial trajectory
that has a temperature peak that is less than 8~GK. In practice, we found that
a trajectory with a peak of 7.6~GK can be evolved without the \NSE{} evolution
mode, but anything hotter becomes problematic. So, we use an imposed temperature
history that starts out at 5.5~GK and remains constant at that value, except for
a short peak up to 7.6~GK at 10~s. We keep the density fixed at $\rho = 10^8\
\text{g cm}^{-3}$. We
evolve \skynet\ without screening or self-heating, but with inverse rates from
detailed balance, as these are important for consistency with \NSE{}
(\sref{sec:nse_consistency}). We use the same full network (7836 species and
93,000 reactions) as in \sref{sec:test_r-process}. The initial composition is
\NSE{} with $Y_e = 0.1$. \Cref{fig:test_nse_evolution} shows the results of
this test. From the left panel, we see that the free neutrons are captured in
the first second to make heavy nuclei, raising the average mass number $\bar A$
to about 60. The nuclei synthesized are mainly around the first r-process peak
and the composition decays toward stability until the temperature starts to
rise at around 8~s. The rising temperature forces the material into \NSE{},
which liberates neutrons and protons from nuclei, quickly reducing $\bar A$ to
around 10. When the temperature drops back to 5.5~GK, the free neutrons and
protons are absorbed into nuclei again and the iron peak is formed. In the left
panel of \cref{fig:test_nse_evolution}, there are no visible differences
between the \skynet\ evolution with and without \NSE{} evolution mode. They
produce exactly the same results. The right panel of
\cref{fig:test_nse_evolution} contains the time step $\Delta t$ and temperature
as
a function of time. We see that the time steps with and without the \NSE{}
evolution mode are the same until \skynet\ turns on the \NSE{} evolution mode when
the temperature is sufficiently high. With the \NSE{} evolution mode, \skynet\
evolves with $\Delta t \sim 10^{-1}$~s until the temperature drops
again and the \NSE{} evolution mode is turned off. Without the \NSE{} evolution mode,
however, the time step drops to a $\text{few} \times 10^{-5}$~s. After the
temperature has returned to 5.5~GK, the time steps in both cases are virtually
identical again. Not shown are the final abundances of the two \skynet\
evolutions, but we find no discernible differences down to a level of
$10^{-30}$. The error according to \cref{eq:net_error}
between the two \skynet\ evolutions is 0.0098\%, indicating that the \NSE{}
evolution mode in \skynet\ produces correct results, i.e., exactly the same
results as would be obtained with the full network, but that the \NSE{} evolution mode
prevents the time step from getting stuck at a very small value.

\section{Summary and future work}
\label{sec:summary}

We presented the new nuclear reaction network \skynet\ and the physics
that it currently implements. Details are provided of how the abundance
evolution equations implemented in \skynet\ are derived from kinetic theory. We
discuss how inverse reaction rates are computed with detailed balance and how
this is related to \NSE{}. A detailed description is given of the numerical
methods used in \skynet\ for the network integration and the self-heating
evolution that accounts for heating due to nuclear reactions. Further, we show
how \skynet\ automatically transitions between evolving the full network and
evolving only the entropy and electron fraction under the influence of weak
reactions when the composition is in \NSE{} and all the strong reactions are in
equilibrium. A general treatment of electron screening that
computes the screening factors for arbitrary strong reactions from chemical
potential corrections that only depend on the nucleus charge and the
composition is presented. We then show in detail how \skynet\ computes these chemical
potential corrections in the weak, strong, and intermediate screening regimes
and how it smoothly transitions between the regimes. These screening
corrections are also taken into account when computing \NSE{} compositions.

After providing some code implementation details that highlight
the modularity and expandability of \skynet, we present comprehensive code
tests and comparisons. The \NSE{} compositions computed by \skynet\ are shown
to be consistent with the evolved strong reactions, but only if detailed
balance is used to compute the inverse rates. Furthermore, we show that
\skynet's \NSE{} results are compatible with results in the published
literature. \skynet\ is compared to two other nuclear reaction networks in
three different cases: a neutron-rich r-process, a proton-rich explosive X-ray
burst, and hydrostatic carbon/oxygen burning. All three tests are conducted with
and without electron screening and with and without self-heating. We find that
all three networks agree with each other at the few percent level in most
cases, although there are some situations where the disagreement is larger due
to somewhat different implementations of the physics in the codes. Finally, in
the
appendices, we discuss the physics of ideal Boltzmann gases and how this is
implemented in the \skynet\ \EOS{} that accounts for all nuclear species
individually. Technical details of how \NSE{} is calculated in \skynet\ and how
the neutrino interactions are implemented are also presented in the appendices.

We hope that \skynet\ will be a useful tool for the nuclear astrophysics
community to compute nucleosynthesis in various scenarios. We also hope that
the theoretical and experimental nuclear physics communities will find \skynet\
useful as a low-barrier entry point to running nucleosynthesis models. \skynet\
can be used for
testing the impact of newly measured or calculated reaction rates or nuclear
properties, or to conduct sensitivity studies in order to determine which
nuclides should be the focus of future experiments. \skynet\ is available as an
open-source software at \skyneturl. We value any feedback, whether it be bug
reports, feature requests, or code contributions.

In the future, we plan to extend the electron screening implementation in
\skynet\ in a way that screening corrections can be consistently accounted for
in the \EOS{} as well. We are also thinking of investigating different
screening prescriptions and making them available in \skynet. Currently, only
the first-order backward Euler method is implemented in \skynet, but we intend
to add higher-order integration methods in the future. \skynet\ is limited to
evolving a one-zone model right now (i.e., there is only one density and
temperature). We plan to add support for multiple zones in \skynet\ and couple
it to existing hydrodynamical simulations. Finally, we plan to investigate
offloading some or most of the computations in \skynet\ to GPUs, which could
significantly speed up the network evolution and might be necessary to
efficiently evolve large hydrodynamical simulations that are coupled to \skynet.

\acknowledgments

We thank J.\ Austin Harris and W.\ Raph Hix for giving us access to
\xnet, for assisting us with running the code comparison tests with \xnet, and
for helpful discussion of the test results. We are also very
grateful to Moritz Reichert, Dirk Martin, Oleg Korobkin, and Marius Eichler for
giving us access to \winnet\ and for helping us run the code comparisons with
it. We thank Hendrik Schatz for providing the X-ray burst trajectories and Ivo
Seitenzahl for allowing us to use his \NSE{} data. We are very grateful to
Christian D.\ Ott for carefully reading the manuscript and providing helpful
comments and suggestions. We also thank our referee, Friedel Thielemann, for
the thoughtful comments and additional reference suggestions.
L.R.\ thanks W.\ Raph Hix and Stan
Woosley for many useful discussions concerning nuclear reaction networks and
nucleosynthesis. We are indebted to the authors of
the sparse linear solver package PARDISO \citep{schenk:04, schenk:06,
karypis:98} for making their code available for free for academic uses. This
work was supported in part by the Sherman Fairchild Foundation and NSF awards
NSF CAREER PHY-1151197, TCAN AST-1333520, and AST-1205732. The calculations were
performed on the \emph{Zwicky} compute cluster at Caltech, supported by NSF
under MRI award PHY-0960291 and by the Sherman Fairchild Foundation. This work
benefited from access to the NSF XSEDE computing network under allocation
TG-PHY100033 and to NSF/NCSA Blue Waters under NSF PRAC award ACI-1440083. This
work was supported in part by NSF grant PHY-1430152 (JINA Center for the
Evolution of the Elements).

\appendix

\section{Equation of state (EOS)}
\label{app:eos}

\skynet\ requires an \EOS{} in order to relate different thermodynamic
quantities, such as temperature, entropy, chemical potential, etc., to each
other. Ions behave as non-relativistic, non-degenerate particles in the majority
of situations where nuclear burning occurs and so the
\EOS{} in \skynet\ treats all ions as non-interacting, non-degenerate,
non-relativistic ideal Boltzmann gases. Electrons and positrons, on the other
hand, can be both degenerate and relativistic. An important exception to the
assumption that the ions are non-interacting is electron screening, which is
discussed in detail in \sref{sec:screening}.

In this section, we present a brief summary of the most relevant properties
of ideal Boltzmann gases and introduce the notation used in this paper. We also
describe the \EOS{} implemented in \skynet.

\subsection{Ideal Boltzmann gas}

The grand partition function $\mathcal{Z}$ of a Maxwell--Boltzmann gas is given
by \citep[e.g.,][\S9.D.3]{reichl:80}
\begin{align}
\ln\mathcal{Z} = \sum_k e^{-\beta(\epsilon_k-\mu)},
\end{align}
where the sum runs over all single-particle states $k$ with energy
$\epsilon_k$, $\beta = 1/T$, and $\mu$ is the chemical potential. Let the
particles be non-relativistic with rest mass $m$ and kinetic energy $E
= p^2/(2m)$. Also, let the particles have internal states with excitation
energies $\Delta_l$ (with respect to the ground state $l = 0$, so $\Delta_0
= 0$) and spins $J_l$, so that the multiplicity factor is $2J_l + 1$. Thus, a
state $k$ is described by the internal state label $l$ and momentum $p$, and
the energy is given by
\begin{align}
\epsilon_k = \epsilon(l,p) = \frac{p^2}{2m} + m + \Delta_l.
\end{align}
Recalling that the momentum phase-space volume element is $V/h^3 d^3p$, we find
\begin{align}
\ln\mathcal{Z} &= \sum_l (2J_l + 1) \frac{V}{h^3} \int d^3p\,
e^{-\beta(p^2/(2m)+m+\Delta_l-\mu)} \nn\\
&= 4\pi \frac{V}{(2\pi)^3} e^{\beta(\mu-m)} \sum_l (2J_l + 1) e^{-\beta
\Delta_l} \int_0^\infty dp\, p^2 e^{-\beta p^2/(2m)} \nn\\
&= \frac{V}{2\pi^2} e^{\beta(\mu-m)} G(T) \sqrt{\frac{\pi}{2}}
\left(\frac{m}{\beta}\right)^{3/2} = VG(T)
\left(\frac{mT}{2\pi}\right)^{3/2}e^{\beta(\mu-m)},
\end{align}
where $(2J_l+1)$ is the multiplicity of the internal state $l$, we used $\int
d^3p = 4\pi\int_0^\infty dp\,p^2$ and $h = 2\pi\hbar = 2\pi$ since $\hbar =
1$, and we define the internal partition function
\begin{align}
G(T) = \sum_l (2J_l+1)e^{-\beta\Delta_l}. \label{eq:part_func}
\end{align}
Note that the internal partition function is sometimes given normalized to the
ground-state spin factor, i.e.,
\begin{align}
G(T) = (2J_0 + 1) g(T),
\end{align}
where $J_0$ is the ground-state spin of the nuclide and $g(T)$ is a tabulated
function \citep[e.g.,][]{rauscher:00}.

The grand potential $\Omega$ is defined as
\begin{align}
\Omega(T,V,\mu) = -T\ln\mathcal{Z} = - VG(T)
\left(\frac{m}{2\pi}\right)^{3/2} T^{5/2}e^{\beta(\mu-m)},
\end{align}
and the particle number $N$, pressure $P$, entropy $S$, and internal energy
$U$ are given by
\citep[e.g.,][\S9.B.3]{reichl:80}
\begin{align}
N &= -\left(\frac{\partial \Omega}{\partial \mu}\right)_{V,T} \\
P &= -\left(\frac{\partial \Omega}{\partial \mu}\right)_{T,\mu} \\
S &= -\left(\frac{\partial \Omega}{\partial T}\right)_{V,\mu} \\
U &= \Omega + TS + \mu N.
\end{align}
We find
\begin{align}
N = -\beta \Omega = VG(T)\left(\frac{mT}{2\pi}\right)^{3/2}e^{\beta(\mu-m)},
\end{align}
which we can solve for the chemical potential $\mu$ to find
\begin{align}
\mu = m + T\ln\left[\frac{n}{G(T)} \left(\frac{2\pi}{mT}\right)^{3/2}\right],
\label{eq:mu}
\end{align}
where $n = N/V$ is the number density. The pressure $P$ becomes
\begin{align}
P = -\frac{\Omega}{V} = \frac{NT}{V} = nT,
\end{align}
since $\Omega = -NT$. For the entropy $S$ we obtain
\begin{align}
S = -\frac{\partial G(T)}{\partial T} \frac{\Omega}{G(T)}
-\frac{5}{2}\frac{\Omega}{T} - \Omega \left(-\frac{\mu-m}{T^2}\right) =
-\frac{5}{2}\frac{\Omega}{T} - \frac{\Omega}{T} \left(-\frac{\mu-m}{T}\right)
-\frac{\Omega}{T}\frac{\partial \ln G(T)}{\partial \ln T},
\end{align}
and since $N = -\Omega/T$, the specific entropy per particle $s = S/N$ is
\begin{align}
s = \frac{5}{2} + \ln\left[\frac{G(T)}{n} \left(\frac{mT}{2\pi}\right)^{3/2}
\right] +  \frac{\partial \ln G(T)}{\partial\ln T}, \label{eq:s}
\end{align}
where we used \cref{eq:mu}. Finally, the internal energy per particle is
\begin{align}
u = \frac{U}{N} = \frac{\Omega}{N} + Ts + \mu = -\Omega\frac{T}{\Omega}
+\frac{5}{2}T -(\mu-m) + T \frac{\partial \ln G(T)}{\partial\ln T} + \mu
= \frac{3}{2} T + m + T \frac{\partial \ln G(T)}{\partial\ln T}. \label{eq:u}
\end{align}

\subsection{Modified Timmes EOS} \label{sec:eos}

In the previous section, we found the most relevant thermodynamic properties of
a non-interacting, non-relativistic, non-degenerate Boltzmann gas. In this
section, we describe the complete \EOS{} implemented in \skynet. \skynet\ uses a
modified Timmes \EOS{} developed in \citet{timmes:99} and \citet{timmes:00}.
The Timmes \EOS{} consists of three independent parts: a photon gas, an
arbitrarily degenerate and relativistic electron/positron gas, and a
non-degenerate, non-relativistic Boltzmann gas for the heavy ions
\citep{timmes:99}. The electron/positron part is implemented via
table interpolation of the Helmholtz free energy \citep{timmes:00}.

For the photon gas and electron/positron gas, the code from the original author
of the Timmes \EOS{}, which is available at
\url{http://cococubed.asu.edu/codes/eos/helmholtz.tbz}, is used. That code also provides
the electron/positron chemical potential $\eta_{e^-,\text{Timmes}} =
\mu_{e^-,\text{Timmes}}/T$, which we need to compute neutrino interactions
(\aref{app:nu_reac}) and electron screening corrections (\sref{sec:screening}).
Note that the electron/positron chemical potential $\mu_{e^-,\text{Timmes}}$ in
the Timmes \EOS{} is defined with the electron rest mass subtracted out
\citep[\S2]{timmes:99}. The positron chemical potential is
\begin{align}
\mu_{e^+,\text{Timmes}} = -\mu_{e^-,\text{Timmes}} - 2m_e.
\end{align}
So, the electron and positron chemical potentials that include the rest masses
are
\begin{align}
\mu_{e^-} &= \mu_{e^-,\text{Timmes}} + m_e = T\eta_{e^-,\text{Timmes}} + m_e,
\label{eq:mue} \\
\mu_{e^+} &= \mu_{e^+,\text{Timmes}} + m_e = -T\eta_{e^-,\text{Timmes}} - m_e =
-\mu_{e^-}. \label{eq:mup}
\end{align}

For the heavy ions, the Timmes \EOS{} implementation uses a single representative
ion species with mass $\bar A$ and charge $\bar Z$, which are the average mass
and charge of all nuclides, respectively. Since \skynet\ has the
complete composition information at all times, we decided to extend the
original Timmes \EOS{} to take into account all ion species individually.
Furthermore, we use the expressions derived in the previous section for the ion
quantities, which take the internal nuclear partition functions into account.

The overall specific entropy of the system is computed in units of
$k_B\,\text{baryon}^{-1}$ as
\begin{align}
s_\text{tot} = \frac{s_{e^\pm,\text{Timmes}} + s_{\gamma,\text{Timmes}}}{k_B
N_A} + s_\text{ions}, \label{eq:stot}
\end{align}
where $s_{e^\pm,\text{Timmes}}$ and $s_{\gamma,\text{Timmes}}$ are the
electron/positron and photon entropies provided by the Timmes \EOS{},
respectively. We divide them by $k_B N_A$, where $N_A \approx 6.022 \times
10^{23}\ \text{baryon g}^{-1}$ is the Avogadro constant, because the Timmes
\EOS{} returns the entropies in units of $\text{erg g}^{-1}\,\text{K}^{-1}$.
The specific entropy of the ions $s_\text{ions}$ is calculated by \skynet\
itself according to \cref{eq:s} as
\begin{align}
s_\text{ions} = \sum_i \frac{N_i s_i}{N_B} = \sum_i Y_i s_i = \sum_i Y_i
\left(\frac{5}{2} + \ln\left[\frac{G_i(T)}{n_i}
\left(\frac{m_iT}{2\pi}\right)^{3/2}
\right] +  \frac{\partial \ln G_i(T)}{\partial\ln T}\right), \label{eq:sion}
\end{align}
where the sum runs over all nuclear species labeled by $i$, $N_i$ is the number
of particles of species $i$, $N_B$ is the total number of baryons, and $s_i$ is
the entropy per particle of species $i$ given by \cref{eq:s}. Recall that the
abundance $Y_i$ is \crefp{eq:Y}
\begin{align}
Y_i \equiv \frac{n_i}{n_B} = \frac{N_i/V}{N_B/V} = \frac{N_i}{N_B},
\end{align}
where $N_i$ and $N_B$ are the total number of particles species $i$ and the total
number of baryons, respectively, and $V$ is the volume. Thus, the abundance
$Y_i$ is the fraction of particles of species $i$ compared to the total number
of baryons. Note that $N_i s_i$ is the total entropy contribution of species
$i$ and so $N_i s_i / N_B$ is the entropy per baryon contribution of species
$i$. Also note that
\begin{align}
n_B = \rho N_A,
\end{align}
where $\rho$ is the mass density.

In \sref{sec:NSE_T}, we require the partial derivative of the entropy with
respect to temperature. From \cref{eq:s}, we get
\begin{align}
\frac{\partial s}{\partial T} = \frac{1}{k_B N_A} \left(\frac{\partial
s_{e^\pm,\text{Timmes}}}{\partial T} + \frac{\partial
s_{\gamma,\text{Timmes}}}{\partial T}\right) + \frac{\partial
s_\text{ions}}{\partial T}. \label{eq:dsdt}
\end{align}
The first two partial derivatives are provided by the Timmes \EOS{}, and from
\cref{eq:sion} we find
\begin{align}
\frac{\partial s_\text{ions}}{\partial T} &= \sum_i Y_i \left(0 +
\frac{1}{G_i(T)}\frac{\partial G_i(T)}{\partial T} +
\frac{3}{2}\frac{1}{T} + \frac{\partial}{\partial
T}\left[\frac{\partial \ln G_i(T)}{\partial \ln T}\right]\right) \nn\\
&= \sum_i \frac{Y_i}{T} \left(\frac{3}{2} + \frac{\partial \ln
G_i(T)}{\partial\ln T} + \frac{\partial^2\ln G_i(T)}{\partial(\ln T)^2}\right),
\end{align}
since $\partial f / \partial \ln T = T \partial f\partial T$. In the current
\skynet\ implementation, however, we ignore the second derivative of $\ln
G_i(T)$, because the partition functions we currently have available do not
have continuous second derivatives. This will be fixed in a future version of
\skynet.

Similar to \cref{eq:stot}, the specific internal energy is computed as
\begin{align}
u_\text{tot} = u_{e^\pm,\text{Timmes}} + u_{\gamma,\text{Timmes}} +
u_\text{ions}
\end{align}
in units of $\text{erg g}^{-1}$. The ion internal energy is computed as
\begin{align}
u_\text{ions} = N_A\sum_i Y_i \left[T\left(\frac{3}{2} +
\frac{\partial\ln G(T)}{\partial\ln T}\right) - \text{BE}_i - Z_i(m_n -
m_p)\right] + N_AY_e m_e, \label{eq:u_ions}
\end{align}
where $m_n$ and $m_p$ are the neutron and proton mass, respectively, $Z_i$ is
the charge number of nuclide $i$, and $\text{BE}_i$ is its binding energy. Note
that the sum in the above expression gives the internal energy per baryon, so we
multiply by $N_A$ to convert this to the internal energy per gram. We need to add
the electron rest mass because it is not accounted for in
$u_{e^\pm,\text{Timmes}}$ \citep{timmes:99}. The
binding energy $\text{BE}_i$ is defined as
\begin{align}
\text{BE}_i = N_i m_n + Z_i m_p - m_i, \label{eq:BE}
\end{align}
with $N_i$ being the number of neutrons of nuclide $i$ (not to be confused with
the same symbol used above for the total number of particles of species $i$ in
the composition) and $m_i$ being the rest mass of nuclide $i$. Thus, the term
$-\text{BE}_i - Z_i(m_n - m_p)$ in \cref{eq:u_ions} is
\begin{align}
-\text{BE}_i - Z_i(m_n - m_p) = m_i - N_i m_n - Z_i m_p - Z_i m_n + Z_i m_p =
m_i - A_im_n,
\end{align}
where $A_i = Z_i + N_i$ is the mass number of nuclide $i$. Thus,
\cref{eq:u_ions} differs from the expression for the specific internal energy
of a single species \crefp{eq:u} only by the subtraction of $A_im_n$ from the
particle rest mass $m_i$. Thus, the specific internal energy we calculate is
relative to the neutron rest mass, which has the advantage that the numerical
value of the specific internal energy is not too large but may be comparable to
the thermal energy. However, if we ignored the rest mass altogether, the
internal energy would not be conserved under nuclear reactions. Nuclear
reactions change particles from one species to another that have different
binding energies, and so we have to account for the binding energy and mass
difference between neutrons and protons, as we do in \cref{eq:u_ions}. Other
\EOS{s} use the same convention \citep[e.g.,][]{lseos:91}. Note that we have
\begin{align}
u_\text{ions} = N_A\sum_i Y_i (u_i - A_i m_n) = N_A \sum_i Y_i u_i - N_A m_n
\sum_i Y_i A_i,
\end{align}
where $u_i$ is given by \cref{eq:u}. But since $\sum_i Y_i A_i = 1$
\crefp{eq:sumYA}, we find $u_\text{ions} = N_A \sum_i Y_i u_i - N_A m_n$, which
means
our definition of specific internal energy only differs by a constant ($N_A
m_n c^2 \approx 9.065 \times 10^{20}\,\text{erg g}^{-1}$) from the specific
internal energy we would calculate by using \cref{eq:u} directly.

Currently, the electron screening corrections implemented in \skynet\
(\sref{sec:screening}) are not yet included in the modified Timmes \EOS{}.
Since screening is implemented as a correction to the ion chemical potential
\crefp{eq:mu}, it is not straightforward to propagate those corrections to the
other thermodynamic quantities of the ions, let alone the electron/positron
gas. We plan to incorporate screening into the \EOS{} in a future version of
\skynet.

\section{Calculating NSE}
\label{app:calc_NSE}

In this section, we show in detail how \NSE{} is computed in \skynet\ given a
temperature and density (\sref{sec:NSE}) and with an unknown temperature
(\sref{sec:NSE_T}).

\subsection{From temperature and density}
\label{sec:NSE}

The \NSE{} evolution mode requires a robust method for calculating \NSE{}.
We have a list of nuclides for which we want to calculate the \NSE{}
composition, given a temperature $T$, density $\rho$, and electron fraction
$Y_e$. Recall that \NSE{} is governed by \cref{eq:NSE}:
\begin{equation}
\mu_i = Z_i \mu_p + N_i \mu_n, \label{eq:NSE2}
\end{equation}
where $\mu_i$, $\mu_p$, and $\mu_n$ are the chemical potentials of the nuclide
$i$, protons, and neutrons, respectively.
To make the values of the chemical potentials closer to unity, we introduce a
renormalized chemical potential $\hat \mu$ given by
\begin{align}
  \hat \mu_i = \mu_i - m_i - \mathrm{BE}_i, \label{eq:muhat}
\end{align}
where $\mathrm{BE}_i$ is the binding energy of the nuclide $i$ defined in
\cref{eq:BE}. Recall that
\begin{align}
  \mathrm{BE}_i = Z_i m_p + N_i m_n - m_i, \label{eq:BE2}
\end{align}
where the proton and neutron masses $m_p$ and $m_n$ are generally chosen such
that the binding energies of
the neutron and proton are exactly zero. \Cref{eq:NSE2} now becomes
\begin{align}
  & \hat\mu_i + m_i + \mathrm{BE_i} = Z_i \hat\mu_p + Z_im_p
+ N_i\hat\mu_n + N_i m_n \nonumber\\
\Leftrightarrow\ \ & \hat\mu_i = Z_i\hat\mu_p + N_i\hat\mu_n,
\end{align}
where we used the definition of $\mathrm{BE}_i$, \cref{eq:BE2}, and the fact
that $\mathrm{BE}_p = \mathrm{BE}_n = 0$.

It is always possible to choose any two chemical potentials as the basis
vectors and express all other chemical potentials in terms of those two.  In
terms of the species $l$ and $m$ that have $Z_l$ and $Z_m$ protons and $N_l$ and
$N_m$ neutrons, we find
\begin{align}
& \begin{bmatrix}
Z_l & N_l \\ Z_m & N_m
\end{bmatrix}
\cdot
\begin{bmatrix}
\hat \mu_p \\ \hat \mu_n
\end{bmatrix}
=
\begin{bmatrix}
\hat \mu_l \\ \hat \mu_m
\end{bmatrix} \nn\\
\Leftrightarrow\ \ &
\begin{bmatrix}
\hat \mu_p \\ \hat \mu_n
\end{bmatrix}
= \frac{1}{Z_lN_m - Z_mN_l}
\begin{bmatrix}
N_m & -N_l \\ -Z_m & Z_l
\end{bmatrix}
\cdot
\begin{bmatrix}
\hat \mu_l \\ \hat \mu_m
\end{bmatrix}.
\end{align}
And so
\begin{align}
\hat\mu_i &= Z_i\hat\mu_p + N_i\hat\mu_n \nn\\
&= \frac{1}{Z_lN_m-Z_mN_l} \left(Z_i(N_m\hat\mu_l - N_l\hat\mu_m) +
   N_i(-Z_m\hat\mu_l + Z_l\hat\mu_m)\right) \nn\\
&= \frac{Z_i N_m - N_i Z_m}{Z_l N_m - N_lZ_m} \hat\mu_l
      + \frac{Z_i N_l - N_i Z_l}{Z_m N_l - N_mZ_l} \hat\mu_m \nonumber\\
&= \alpha_i \hat\mu_l + \beta_i \hat\mu_m,
\end{align}
where
\begin{align}
\alpha_i = \frac{Z_i N_m - N_i Z_m}{Z_l N_m - N_lZ_m} \qquad \text{and} \qquad
\beta_i = \frac{Z_i N_l - N_i Z_l}{Z_m N_l - N_mZ_l}.
\end{align}
Clearly, the species $l$ and $m$ must be chosen to have different
proton fractions $Z/N$.  However, they do not have to be nuclei in the
network. We have had reasonable success with $Z_l = 0$, $N_l = 1$ and
$Z_m = -1$, $N_m = 1$.

Using \cref{eq:mu}, the abundance $Y_i = n_i/n_B$ in terms of the
chemical potential $\hat\mu_i$ is given by
\begin{align}
  & \hat\mu_i = -\mathrm{BE}_i +
T\ln\left[\frac{n_i}{G_i(T)}\left(\frac{2\pi}{m_iT}\right)^{3/2}\right]
\nonumber\\
\Leftrightarrow\ \ & e^{(\hat\mu_i+\mathrm{BE}_i)/T} =
\frac{Y_in_B}{G_i(T)}\left(\frac{2\pi}{m_iT}\right)^{3/2} \nonumber\\
\Leftrightarrow\ \ & Y_i = e^{(\hat\mu_i+\mathrm{BE}_i)/T}
\frac{G_i(T)}{n_B}
\left(\frac{m_i T}{2\pi}\right)^{3/2} \nonumber\\
&\phantom{Y_i} = e^{\eta_i+\mathrm{BE}_i/T} \frac{G_i(T)}{n_B}
\left(\frac{m_i T}{2\pi}\right)^{3/2}, \label{eq:nse_Y}
\end{align}
where we define $\eta_i = \hat\mu_i / T$.

This system of equations is subject to baryon
number conservation and charge conservation,
\begin{align}
f_A &= 1   - \sum_i A_i Y_i = 0, \label{eq:fA} \\
f_Z &= Y_e - \sum_i Z_i Y_i = 0. \label{eq:fZ}
\end{align}
These equations can be zeroed by a two-dimensional \NR{} iteration,
where the Jacobian is given by
\begin{equation}
J_\text{NSE} = \begin{bmatrix}
\rule[-11pt]{0pt}{27pt}\dfrac{\partial f_A}{\partial \eta_l} & \dfrac{\partial
f_A}{\partial
\eta_m} \\
\rule[-11pt]{0pt}{27pt}\dfrac{\partial f_Z}{\partial \eta_l} & \dfrac{\partial
f_Z}{\partial
\eta_m}
\end{bmatrix},
\end{equation}
so that the chemical potential updates are given by
\begin{equation}
\Delta \vec\eta = -J_\text{NSE}^{-1} \cdot \begin{bmatrix} f_A \\ f_Z
\end{bmatrix}.
\end{equation}
From \cref{eq:fA,eq:fZ}, we find
\begin{align}
  \frac{\partial f_A}{\partial \eta_l} = -\sum_i A_i \frac{\partial
Y_i}{\partial \eta_l} = -\sum_i A_i Y_i \alpha_i,
\end{align}
and similarly
\begin{align}
\frac{\partial f_A}{\partial \eta_m} = -\sum_i A_i Y_i\beta_i, \\
\frac{\partial f_Z}{\partial \eta_l} = -\sum_i Z_i Y_i\alpha_i, \\
\frac{\partial f_Z}{\partial \eta_m} = -\sum_i Z_i Y_i\beta_i.
\end{align}

The trickiest part about calculating \NSE{} is choosing the basis nuclides $l$
and $m$ and the starting
guess for $\vec\eta = (\eta_l,\eta_m)$.  If one of the basis nuclides is
$(Z,N)=(-1,1)$, then the corresponding $\eta$ can be set to zero in
most cases.
To choose the second basis nuclide, compute the chemical potential of the most
bound nuclide if it had a mass fraction of 1 and compare this to the chemical
potential of the neutron if it had mass fraction 1. The second basis nuclide
will be the nuclide corresponding to the larger of those two chemical
potentials and the starting guess for that $\eta$ comes from that chemical
potential. Although this method may not work in every case, we have found it to
work robustly in a large region of parameter space.

\subsection{With an unknown temperature}
\label{sec:NSE_T}

In the previous section, we described how to calculate \NSE{} given a
temperature $T$, density $\rho$ (from which we get the baryon number density
$n_B = N_A\rho$), and electron fraction $Y_e$. However, there are cases where
the temperature is unknown but the entropy $s_0$ is given instead. In that case,
we have an additional unknown variable $T$, and the additional constraint
equation
\begin{align}
f_s = \frac{s(T,\eta_l,\eta_m)}{s_0} - 1 = 0, \label{eq:fs}
\end{align}
where $s_0$ is the given target entropy and $s(T,\eta_l,\eta_m)$ is the entropy
given by the \EOS{} from the current guess for $T$,
$\eta_l$, and $\eta_m$. The Jacobian becomes
\begin{align}
J_\text{NSE} = \begin{bmatrix}
\rule[-11pt]{0pt}{27pt}\dfrac{\partial f_A}{\partial \eta_l} & \dfrac{\partial
f_A}{\partial
\eta_m} & \dfrac{\partial f_A}{\partial T} \\
\rule[-11pt]{0pt}{27pt}\dfrac{\partial f_Z}{\partial \eta_l} & \dfrac{\partial
f_Z}{\partial
\eta_m} & \dfrac{\partial f_Z}{\partial T} \\
\rule[-11pt]{0pt}{27pt}\dfrac{\partial f_s}{\partial \eta_l} & \dfrac{\partial
f_s}{\partial
\eta_m} & \dfrac{\partial f_s}{\partial T}
\end{bmatrix}.
\end{align}
Note that \cref{eq:nse_Y} gives
\begin{align}
\frac{\partial Y_i}{\partial T} = \frac{Y_i}{T}
\left(\frac{3}{2}-\frac{\text{BE}_i}{T} + \frac{\partial \ln
G_i(T)}{\partial \ln T}\right) \label{eq:dYidT}
\end{align}
and so
\begin{align}
\frac{\partial f_A}{\partial T} &= -\sum_i A_i\frac{Y_i}{T}
\left(\frac{3}{2}-\frac{\text{BE}_i}{T} + \frac{\partial \ln
G_i(T)}{\partial \ln T}\right), \\
\frac{\partial f_Z}{\partial T} &= -\sum_i Z_i\frac{Y_i}{T}
\left(\frac{3}{2}-\frac{\text{BE}_i}{T} + \frac{\partial \ln
G_i(T)}{\partial \ln T}\right).
\end{align}
Recall that the entropy is calculated as \crefp{eq:stot}
\begin{align}
s = s_{e^\pm} + s_\gamma + s_\text{ions}.
\end{align}
The EOS provides $\partial s/\partial T$ \crefp{eq:dsdt}, so
\begin{align}
\frac{\partial f_s}{\partial T} = \frac{1}{s_0}\frac{\partial
s}{\partial T} + \frac{1}{s_0}\sum_i \frac{\partial
s}{\partial Y_i}\frac{\partial Y_i}{\partial T}.
\end{align}
The electron entropy $s_{e^\pm}$ and photon entropy $s_\gamma$ only depend on
the temperature and electron fraction. Thus, they do not depend on the
composition $\vec Y$ and thus also not on $\eta_l$ and $\eta_m$. The
ion entropy is calculated as \crefp{eq:sion}
\begin{align}
s_\text{ion} &= \sum_i Y_i \left(\frac{5}{2} + \ln\left[\frac{G_i(T)}{n_B Y_i}
\left(\frac{m_iT}{2\pi}\right)^{3/2}\right] + \frac{\partial \ln
G_i(T)}{\partial \ln T}\right),
\end{align}
so
\begin{align}
\frac{\partial s}{\partial Y_i} &= \frac{5}{2} +
\ln\left[\frac{G_i(T)}{n_B Y_i}
\left(\frac{m_iT}{2\pi}\right)^{3/2}\right] + \frac{\partial \ln
G_i(T)}{\partial \ln T} + Y_i\left(-\frac{1}{Y_i}\right) \nonumber\\
&=\frac{3}{2} + \ln\left[\frac{G_i(T)}{n_B Y_i}
\left(\frac{m_iT}{2\pi}\right)^{3/2}\right] + \frac{\partial \ln
G_i(T)}{\partial \ln T}.
\end{align}
From \cref{eq:nse_Y}, we get
\begin{align}
\ln\left[\frac{G_i(T)}{n_B Y_i}
\left(\frac{m_iT}{2\pi}\right)^{3/2}\right] = -\left(\eta_i +
\frac{\text{BE}_i}{T}\right).
\end{align}
Combining the above with \cref{eq:dYidT} yields
\begin{align}
\frac{\partial f_s}{\partial T} = \frac{1}{s_0}\frac{\partial
s}{\partial T} + \frac{1}{s_0}\sum_i \left(\frac{3}{2}
-\left(\eta_i + \frac{\text{BE}_i}{T}\right) + \frac{\partial \ln
G_i(T)}{\partial \ln T}\right)\frac{Y_i}{T}
\left(\frac{3}{2}-\frac{\text{BE}_i}{T} + \frac{\partial \ln
G_i(T)}{\partial \ln T}\right),
\end{align}
where the first partial derivative is provided by the \EOS{} \crefp{eq:dsdt}.
We also find
\begin{align}
\frac{\partial f_s}{\partial\eta_l} &= \frac{1}{s_0}\sum_i\frac{\partial
s}{\partial Y_i} \frac{\partial Y_i}{\partial \eta_l}
= \frac{1}{s_0}\sum_i\left(\frac{3}{2} -
\left(\eta_i+\frac{\text{BE}_i}{T}\right) + \frac{\partial \ln
G_i(T)}{\partial
\ln T}\right)Y_i\alpha_i
\end{align}
and similarly
\begin{align}
\frac{\partial f_s}{\partial\eta_m} = \frac{1}{s_0}\sum_i
\left(\frac{3}{2} -
\left(\eta_i+\frac{\text{BE}_i}{T}\right) + \frac{\partial \ln
G_i(T)}{\partial
\ln T}\right)Y_i\beta_i.
\end{align}

Unfortunately, the \NR{} iterations with three variables are much less stable
than if the temperature is fixed, unless a good initial guess for the
temperature is available. For this reason, if \NSE{} is computed from a given
entropy and density, \skynet\ first uses the bisection method
\citep[e.g.,][\S2.1]{burden:15} to find a good guess for the
temperature. The bisection attempts to find a guess temperature such that
\cref{eq:fs} is close to zero. Then, the \NR{} iterations are performed as
described in this section, which may lead to a temperature that satisfies the
three constraint equations better than the guess temperature found by the
bisection method. However, it can also happen that the bisection already found
the best temperature.

If \NSE{} needs to be calculated from a given temperature, entropy, and
electron fraction, then the bisection method is used to find the density that
produces the desired entropy. Similarly, if the internal energy, density, and
electron fraction are given, \skynet\ uses the bisection method to find the
temperature that produces an \NSE{} distribution with the desired internal
energy.

\section{Neutrino interaction reactions} \label{app:nu_reac}

The rate for a two-particle charged current weak interaction is given by
(see Equation \ref{eq:kinetic_rates})
\begin{equation}
\lambda_w = \frac{1}{n_2}
g_1 \int \frac{d^3 k_1}{(2 \pi)^3}
g_2 \int \frac{d^3 k_2}{(2 \pi)^3}
\int \frac{d^3 k_3}{(2 \pi)^3}
\int \frac{d^3 k_4}{(2 \pi)^3}
\delta^{(4)}(k_1^\mu + k_2^\mu - k_3^\mu - k_4^\mu)
\scriptr_w
f_1 f_2 (1 - f_3)(1 - f_4),
\end{equation}
where particles [1] and [3] are the incoming and outgoing neutrino and lepton,
and particles [2] and [4] are the incoming and outgoing nucleons. Under the
astrophysical
conditions relevant for reaction networks \citep[see, e.g.,][]{reddy:98}, we have
\begin{align}
\scriptr_w \approx (2 \pi)^4 G_F^2(g_V^2 + 3 g_A^2)(1 + h_{wm} E_3),
\end{align}
where $G_F$ is the Fermi coupling constant, $g_V$ and $g_A$ are the vector and
axial vector couplings of the weak current to the nucleons, and the
energy-dependent correction factor $(1 + h_{wm}E_3)$ comes from weak magnetism
and recoil corrections \citep{horowitz:02}.  Due to the large mass of the
nucleons relative
to the energy scale of neutrinos emitted from sites undergoing nuclear burning,
we can assume there is no momentum transfer from the nucleons to the leptons.
Also neglecting final state nucleon blocking, we then have
\begin{equation}
\lambda_w =
G_F^2 (g_V^2 + 3 g_A^2)
g_1 \int \frac{d^3 k_1}{(2 \pi)^3}
\int \frac{d^3 k_3}{(2 \pi)^3}
f_1 (1 - f_3)
2 \pi \delta(E_1 - E_3 + q_0)
(1 + h_{wm} E_3),
\end{equation}
where $q_0$ is the energy difference between the incoming and outgoing nucleons.
The angular integrals in momentum space are trivially integrated. We have $E_i d
E_i = k_i dk_i$ and the delta function gets rid of the integral over $k_3$.
This gives
\begin{equation}
\lambda_w =
G_F^2 (g_V^2 + 3 g_A^2)
\frac{g_1}{2 \pi^3}
\int_0^\infty d E_3  k_1 E_1 k_3 E_3
(1 + h_{wm} E_3)
\, \theta(E_1-m_1)
f_1 (1 - f_3),
\end{equation}
where $E_1 = E_3 - q_0$, $k_1 = \sqrt{E_1^2 - m_1^2}$, and $\theta(x) = 1$ if
$x>0$ and 0 otherwise.

\skynet\ contains neutrino interactions on free nucleons. Currently, the
following reactions are implemented:
\begin{align}
\lambda_\text{ec}\!:\ & \text{p} + e^- \to \text{n} + \nu_e, \\
\lambda_\text{pc}\!:\ & \text{n} + e^+ \to \text{p} + \bar\nu_e, \\
\lambda_{\nu_e}\!:\ &   \text{n} + \nu_e \to \text{p} + e^-, \\
\lambda_{\bar\nu_e}\!:\ & \text{p} + \bar\nu_e \to \text{n} + e^+.
\end{align}
Thus, we compute
\begin{align}
\lambda_\text{ec} &= C \int_{\omega_\text{ec}}^\infty dE\, E
\sqrt{E^2-m_e^2} (E - Q_\text{ec})^2 (1 + E
h_{wm})f_e(E,\mu_e)(1-f_{\nu_e}(E-Q_\text{ec},\mu_{\nu_e})), \label{eq:lec} \\
\lambda_\text{pc} &= C \int_{\omega_\text{pc}}^\infty dE\, E
\sqrt{E^2-m_e^2} (E - Q_\text{pc})^2 (1 + E
h_{wm})f_p(E,-\mu_e)(1-f_{\bar\nu_e}(E-Q_\text{pc},\mu_{\bar\nu_e})),
\label{eq:lpc} \\
\lambda_{\nu_e} &= C \int_{\omega_\text{ec}}^\infty dE\, E
\sqrt{E^2-m_e^2} (E - Q_\text{ec})^2 (1 + E
h_{wm})(1-f_e(E,\mu_e))f_{\nu_e}(E-Q_\text{ec},\mu_{\nu_e}), \label{eq:lnue} \\
\lambda_{\bar\nu_e} &= C \int_{\omega_\text{pc}}^\infty dE\, E
\sqrt{E^2-m_e^2} (E - Q_\text{pc})^2 (1 + E
h_{wm})(1-f_p(E,-\mu_e))f_{\bar\nu_e}(E-Q_\text{pc},\mu_{\bar\nu_e}),
\label{eq:lnuebar}
\end{align}
where $Q_\text{ec} = -Q_\text{pc} = m_n - m_p = 1.29333\,\text{MeV}$,
$\omega_x = \max(m_e,Q_x)$ for $x=\text{ec}$ or $x=\text{pc}$, $f_e$ and $f_p$
are the electron and
positron distribution functions with the electron chemical
potential $\mu_e$ given by \cref{eq:mue}, and $f_{\nu_e}$ and $f_{\bar\nu_e}$
are the electron neutrino and antineutrino distributions functions, which are
usually Fermi-Dirac with $\mu_{\nu_e} = \mu_{\bar\nu_e} = 0$, but could be set
to something else, too. We also use the Fermi-Dirac distribution for the
electrons and positrons, hence
\begin{align}
f_{e,p}(E,\mu) = \frac{1}{e^{\beta(E-\mu)}+1},
\end{align}
where $\beta = 1/T$. The factor $C$ is given by \citep{arcones:10}
\begin{align}
C = \frac{B\ln 2}{K m_e^5},
\end{align}
where the matrix element is $B = g_V^2 + 3g_A^2 = 5.76$ and $K = 6144$~s.

Note that the factor $(E-Q_x)$ in \cref{eq:lec,eq:lpc,eq:lnue,eq:lnuebar} is the
(anti) neutrino energy. We neglect nucleon recoils so that $E_{\nu_e} = E + m_p
- m_n = E - Q_\text{ec}$ for electron capture and neutrino absorption, and for
positron capture and antineutrino absorption, we get $E_{\bar\nu_e} = E + m_n -
m_p = E + Q_\text{ec} = E - Q_\text{pc}$. Therefore, to compute the neutrino
heating or cooling rates due to these reactions, we simply multiply the
integrand by another factor of $(E-Q_x)$ and we get
\begin{align}
\dot\epsilon_\text{ec} &= C \int_{\omega_\text{ec}}^\infty dE\, E
\sqrt{E^2-m_e^2} (E - Q_\text{ec})^3 (1 + E
h_{wm})f_e(E,\mu_e)(1-f_{\nu_e}(E-Q_\text{ec},\mu_{\nu_e})), \label{eq:eec} \\
\dot\epsilon_\text{pc} &= C \int_{\omega_\text{pc}}^\infty dE\, E
\sqrt{E^2-m_e^2} (E - Q_\text{pc})^3 (1 + E
h_{wm})f_p(E,-\mu_e)(1-f_{\bar\nu_e}(E-Q_\text{pc},\mu_{\bar\nu_e})),
\label{eq:epc} \\
\dot\epsilon_{\nu_e} &= -C \int_{\omega_\text{ec}}^\infty dE\, E
\sqrt{E^2-m_e^2} (E - Q_\text{ec})^3 (1 + E
h_{wm})(1-f_e(E,\mu_e))f_{\nu_e}(E-Q_\text{ec},\mu_{\nu_e}), \label{eq:enue} \\
\dot\epsilon_{\bar\nu_e} &= -C \int_{\omega_\text{pc}}^\infty dE\, E
\sqrt{E^2-m_e^2} (E - Q_\text{pc})^3 (1 + E
h_{wm})(1-f_p(E,-\mu_e))f_{\bar\nu_e}(E-Q_\text{pc},\mu_{\bar\nu_e}),
\label{eq:enuebar},
\end{align}
where the negative sign for the (anti) neutrino absorption reactions comes from
the fact that in those reactions, the neutrino energies are absorbed, and they
thus provide cooling instead of heating. The total neutrino heating/cooling
rate is thus
\begin{align}
\dot\epsilon_\nu = \dot\epsilon_\text{ec} + \dot\epsilon_\text{pc} +
\dot\epsilon_{\nu_e} + \dot\epsilon_{\bar\nu_e}. \label{eq:eps_nu}
\end{align}

The integrals shown in this section are evaluated numerically in \skynet\ using
the adaptive QAG integration routines provided by the GNU Scientific
Library\footnote{\href{%
https://www.gnu.org/software/gsl/manual/html_node/Numerical-Integration.html}{%
\nolinkurl{https://www.gnu.org/software/gsl/manual/html_node/}\\%
\nolinkurl{Numerical-Integration.html}}}.
As usual, all of the reaction rates and heating rates need to be multiplied by the
product of reactant abundances. So, the electron capture and antineutrino
absorption rates are multiplied by $Y_p$ and the positron capture and neutrino
absorption rates are multiplied by $Y_n$.

Alternatively, instead of computing the rates from the integrals provided here,
one can also specify the rates $\lambda_\text{ec}$, $\lambda_\text{pc}$,
$\lambda_{\nu_e}$, and $\lambda_{\bar\nu_e}$ directly as a function of time,
and \skynet\ will use these externally given rates. In that case, the neutrino
heating/cooling rate also has to be specified directly. This capability is
useful if the (anti) neutrino absorption and emission rates are computed in a
hydrodynamical simulation and the same rates should be used for the
nucleosynthesis calculations in \skynet\ for consistency.

\addcontentsline{toc}{section}{References}
\bibliographystyle{apj}
\bibliography{apj-jour,%
bibliography/nucleosynthesis_references.bib,%
bibliography/eos_references.bib,%
bibliography/methods_references.bib,%
bibliography/misc_references.bib,%
bibliography/nu_interactions_references.bib,%
bibliography/sn_theory_references.bib,%
bibliography/stellarevolution_references.bib,%
bibliography/typeIa_theory_references.bib,%
bibliography/xray_burst_references.bib,%
bibliography/pns_cooling_references.bib,%
bibliography/NSNS_NSBH_references.bib,%
private_comm.bib}

\end{document}